\documentclass[manuscript]{aastex}
\usepackage{rotating}
\usepackage{amssymb,amsmath,natbib,graphicx}
\usepackage{caption}

\begin{document}
\thispagestyle{empty}

\parbox{2.5cm}{\includegraphics{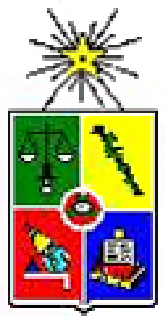}}
\parbox{14cm}{\bf UNIVERSIDAD DE CHILE \\ FACULTAD DE CIENCIAS F\'ISICAS Y 
MATEM\'ATICAS\\ DEPARTAMENTO DE ASTRONOM\'IA} 
\vspace{5 ex}

\begin{center}{\bf DETERMINACI\'ON DE LA DISTANCIA A 12 SUPERNOVAS DE TIPO 
II MEDIANTE EL M\'ETODO DE LA FOT\'OSFERA EN EXPANSI\'ON} \end{center}

\vspace{3 ex}

\begin{center}{\bf TESIS PARA OPTAR AL GRADO DE MAG\'ISTER EN \\
CIENCIAS, MENCI\'ON ASTRONOM\'IA} \end{center}

\vspace{3 ex}

\begin{center} {\bf MAT\'IAS IGNACIO JONES FERN\'ANDEZ} \end{center}

\vspace{3 ex}

\begin{center} {PROFESOR GU\'IA:\\
MARIO HAMUY WACKENHUT

\vspace{2 ex}
MIEMBROS DE LA COMISI\'ON:\\
JOS\'E MAZA SANCHO \\
PAULINA LIRA TEILLERY\\
GAST\'ON FOLATELLI \\
ALEJANDRO CLOCCHIATTI

\vspace{2 ex}
SANTIAGO DE CHILE\\
JUNIO 2008     
} \end{center}

\newpage
\thispagestyle{empty}
\begin{center}{\bf DEDICATION} \end{center}

\vspace{20 ex}
\hspace{8.5 cm} {\it A mi familia, Padre, Madre y hermanos}    
\vspace{2 ex}
\hspace{10.5 cm} {\it y a mi amada Carolina}

\newpage

\thispagestyle{empty}
\begin{center}{\bf RESUMEN} \end{center}

\vspace{2 ex}

Hemos usado fotometr\'ia y espectroscop\'ia temprana de 12 Supernovas de Tipo II plateau (SNs IIP)
para derivar sus distancias mediante el M\'etodo de la Fot\'osfera en Expansi\'on (EPM). Hemos
realizado este estudio usando dos sets de modelos de atm\'osfera de Supernovas de Tipo II (SNs II),
obtenidos de \citet{E96} y \citet{D05b}, tres sets de filtros ({\it \{BV\},\{BVI\},\{VI\}}) 
y dos m\'etodos para la determinaci\'on de la extinci\'on en la galaxia hu\'esped, con lo cual
hemos construido 12 diagramas de Hubble. 
Usando el set de filtros $\{VI\}$ y los modelos de \citet{D05b} hemos obtenido una disperis\'ion
en el diagrama de Hubble de \mbox{$\sigma_{\mu}$ = 0.32 {\it mag}} y su correspondiente constante de Hubble de $H_0$ = 52.4 $\pm$ 4.3 $km$ $s^{-1}$ $Mpc^{-1}$.
Adem\'as aplicamos el EPM a la SN IIP SN 1999em. Con el set de filtros $\{VI\}$ y los modelos de
\citet{D05b} hemos derivado una distancia a \'esta de 13.9 $\pm$ 1.4 Mpc, lo cual concuerda
con la distancia de Cefeida de 11.7 $\pm$ 1.0 Mpc a la galaxia hu\'esped de \'esta Supernova
(NGC 1637).
 
\newpage

\thispagestyle{empty}
\tableofcontents
\thispagestyle{empty}

\newpage

\thispagestyle{empty}

{\bf \begin{center} List of Tables \end{center}}
\vspace{3 ex}
\noindent \newline
2.1  \hspace{0.5 cm} Telescopes and instruments \hspace{0.1 cm} . . . . . . . . . . . . . . . . . . . . . . . . . . . . . .   
     \hspace{0.3 cm} 66\\
2.1  \hspace{0.5 cm} Telescopes and instruments (continued) \hspace{0.1 cm} . . . . . . . . . . . . . . . . . . . . . . .
     \hspace{0.3 cm} 67\\
2.2  \hspace{0.5 cm} SNe redshifts \hspace{0.1 cm} . . . . . . . . . . . . . . . . . . . . . . . . . . . . . . . . . . . . . . .
     \hspace{0.3 cm} 68\\
\vspace{1 ex}
\noindent \newline
3.1 \hspace{0.5 cm} Dilution factors coefficients \hspace{0.1 cm} . . . . . . . . . . . . . . . . . . . . . . . . . . . . . .
    \hspace{0.3 cm} 69\\
3.2 \hspace{0.5 cm} Spectroscopic velocities \hspace{0.1 cm} . . . . . . . . . . . . . . . . . . . . . . . . . . . . . . . . .
    \hspace{0.3 cm} 70\\
3.2 \hspace{0.5 cm} Spectroscopic velocities (continued) \hspace{0.1 cm} . . . . . . . . . . . . . . . . . . . . . . . . .
    \hspace{0.3 cm} 71\\
3.2 \hspace{0.5 cm} Spectroscopic velocities (continued) \hspace{0.1 cm} . . . . . . . . . . . . . . . . . . . . . . . . .
    \hspace{0.3 cm} 72\\
3.2 \hspace{0.5 cm} Spectroscopic velocities (continued) \hspace{0.1 cm} . . . . . . . . . . . . . . . . . . . . . . . . .
    \hspace{0.3 cm} 73\\
3.2 \hspace{0.5 cm} Spectroscopic velocities (continued) \hspace{0.1 cm} . . . . . . . . . . . . . . . . . . . . . . . . .
    \hspace{0.3 cm} 74\\
3.2 \hspace{0.5 cm} Spectroscopic velocities (continued) \hspace{0.1 cm} . . . . . . . . . . . . . . . . . . . . . . . . .
    \hspace{0.3 cm} 75\\
3.2 \hspace{0.5 cm} Spectroscopic velocities (continued) \hspace{0.1 cm} . . . . . . . . . . . . . . . . . . . . . . . . .
    \hspace{0.3 cm} 76\\
3.2 \hspace{0.5 cm} Spectroscopic velocities (continued) \hspace{0.1 cm} . . . . . . . . . . . . . . . . . . . . . . . . .
    \hspace{0.3 cm} 77\\
3.3 \hspace{0.5 cm} Photospheric velocity conversion coefficients \hspace{0.1 cm} . . . . . . . . . . . . . . . . . . . .
    \hspace{0.3 cm} 78\\
3.4 \hspace{0.5 cm} Host galaxy ad Galactic extinction \hspace{0.1 cm} . . . . . . . . . . . . . . . . . . . . . . . . . .
    \hspace{0.3 cm} 79\\ 
3.5 \hspace{0.5 cm} EPM distances \hspace{0.1 cm} . . . . . . . . . . . . . . . . . . . . . . . . . . . . . . . . . . . . . .
    \hspace{0.3 cm} 80\\
3.6 \hspace{0.5 cm} Error sources \hspace{0.1 cm} . . . . . . . . . . . . . . . . . . . . . . . . . . . . . . . . . . . . . . .
    \hspace{0.3 cm} 81\\
3.7 \hspace{0.5 cm} EPM quantities derived for SN 1992ba \hspace{0.1 cm} . . . . . . . . . . . . . . . . . . . . . . . .
    \hspace{0.3 cm} 82\\
3.8 \hspace{0.5 cm} EPM quantities derived for SN 1999br \hspace{0.1 cm} . . . . . . . . . . . . . . . . . . . . . . . .
    \hspace{0.3 cm} 83\\
3.9 \hspace{0.5 cm} EPM quantities derived for SN 1999em \hspace{0.1 cm} . . . . . . . . . . . . . . . . . . . . . . . 
    \hspace{0.3 cm} 84\\
3.9 \hspace{0.5 cm} EPM quantities derived for SN 1999em (continued) \hspace{0.1 cm} . . . . . . . . . . . . . . . . 
    \hspace{0.3 cm} 85\\
3.10 \hspace{0.3 cm} EPM quantities derived for SN 1999gi \hspace{0.1 cm} . . . . . . . . . . . . . . . . . . . . . . . .
     \hspace{0.3 cm} 86\\
3.11 \hspace{0.3 cm} EPM quantities derived for SN 2002gw \hspace{0.1 cm} . . . . . . . . . . . . . . . . . . . . . . . 
     \hspace{0.3 cm} 87\\
3.12 \hspace{0.3 cm} EPM quantities derived for SN 2003T  \hspace{0.1 cm} . . . . . . . . . . . . . . . . . . . . . . . .
     \hspace{0.3 cm} 88\\
3.13 \hspace{0.3 cm} EPM quantities derived for SN 2003bl \hspace{0.1 cm} . . . . . . . . . . . . . . . . . . . . . . . .
     \hspace{0.3 cm} 89\\
3.14 \hspace{0.3 cm} EPM quantities derived for SN 2003bn \hspace{0.1 cm} . . . . . . . . . . . . . . . . . . . . . . . .
     \hspace{0.3 cm} 90\\
3.15 \hspace{0.3 cm} EPM quantities derived for SN 2003ef \hspace{0.1 cm} . . . . . . . . . . . . . . . . . . . . . . . .
     \hspace{0.3 cm} 91\\
3.16 \hspace{0.3 cm} EPM quantities derived for SN 2003hl \hspace{0.1 cm} . . . . . . . . . . . . . . . . . . . . . . . .
     \hspace{0.3 cm} 92\\
3.17 \hspace{0.3 cm} EPM quantities derived for SN 2003hn \hspace{0.1 cm} . . . . . . . . . . . . . . . . . . . . . . . .
     \hspace{0.3 cm} 93\\
3.18 \hspace{0.3 cm} EPM quantities derived for SN 2003iq \hspace{0.1 cm} . . . . . . . . . . . . . . . . . . . . . . . .
     \hspace{0.3 cm} 94\\
\vspace{1 ex}
\noindent \newline
4.1 \hspace{0.5 cm} Summary of $H_0$ values \hspace{0.1 cm} . . . . . . . . . . . . . . . . . . . . . . . . . . . . . . . . . 
    \hspace{0.3 cm} 95\\
4.2 \hspace{0.5 cm} Summary of dispersions in the Hubble diagrams \hspace{0.1 cm} . . . . . . . . . . . . . . . . . . . .
    \hspace{0.3 cm} 96\\

\thispagestyle{empty}

\newpage

\thispagestyle{empty}

{\bf \begin{center} List of Figures \end{center}} 

\vspace{3 ex}

%\begin{enumerate}            
%\setcounter{enumi}{1}
%\item hoka
%\end{enumerate}
\noindent \newline 
2.1  \hspace{0.5 cm} Light curves (part 1) \hspace{0.1 cm} . . . . . . . . . . . . . . . . . . . . . . . . . . . . . . . . . . 
     \hspace{0.3 cm} 35\\
2.2  \hspace{0.5 cm} Light curves (part 2) \hspace{0.1 cm} . . . . . . . . . . . . . . . . . . . . . . . . . . . . . . . . . . 
     \hspace{0.3 cm} 36\\
2.3  \hspace{0.5 cm} Light curves (part 3) \hspace{0.1 cm} . . . . . . . . . . . . . . . . . . . . . . . . . . . . . . . . . .
     \hspace{0.3 cm} 37 \\
\vspace{1 ex}
\noindent \newline            
3.1  \hspace{0.5 cm} Dilution factors \hspace{0.1 cm}  . . . . . . . . . . . . . . . . . . . . . . . . . . . . . . . . . . . . . .
     \hspace{0.3 cm} 38\\
3.2  \hspace{0.5 cm} Line velocity evolution (part 1) \hspace{0.1 cm}  . . . . . . . . . . . . . . . . . . . . . . . . . . . . 
     \hspace{0.3 cm} 39\\
3.3  \hspace{0.5 cm} Line velocity evolution (part 2) \hspace{0.1 cm}  . . . . . . . . . . . . . . . . . . . . . . . . . . . . 
     \hspace{0.3 cm} 40\\
3.4  \hspace{0.5 cm} Line velocity evolution (part 3) \hspace{0.1 cm}  . . . . . . . . . . . . . . . . . . . . . . . . . . . . 
     \hspace{0.3 cm} 41\\
3.5  \hspace{0.5 cm} Photospheric velocity conversion \hspace{0.1 cm}  . . . . . . . . . . . . . . . . . . . . . . . . . . .
     \hspace{0.3 cm} 42\\
3.6  \hspace{0.5 cm} Ratio between the $H_\alpha$ and $H_\beta$ velocity \hspace{0.1 cm} . . . . . . . . . . . . . . . . . . . . . . . . 
     \hspace{0.3 cm} 43\\
3.7  \hspace{0.5 cm} Comparison between the {\it DES} and {\it OLI} reddening \hspace{0.1 cm} . . . . . . . . . . . . . . . . . 
     \hspace{0.3 cm} 44\\
3.8  \hspace{0.5 cm} Full EPM solution for SN 1999em \hspace{0.1 cm} . . . . . . . . . . . . . . . . . . . . . . . . . . .   
     \hspace{0.3 cm} 45\\
3.9  \hspace{0.5 cm} EPM solution for SN 1992ba \hspace{0.1 cm} . . . . . . . . . . . . . . . . . . . . . . . . . . . . . . 
     \hspace{0.3 cm} 46\\
3.10 \hspace{0.3 cm} EPM solution for SN 1999br \hspace{0.1 cm} . . . . . . . . . . . . . . . . . . . . . . . . . . . . . .
     \hspace{0.3 cm} 47\\
3.11 \hspace{0.3 cm} EPM solution for SN 1999em \hspace{0.1 cm} . . . . . . . . . . . . . . . . . . . . . . . . . . . . .
     \hspace{0.3 cm} 48\\
3.12 \hspace{0.3 cm} EPM solution for SN 1999gi \hspace{0.1 cm} . . . . . . . . . . . . . . . . . . . . . . . . . . . . . .
     \hspace{0.3 cm} 49\\
3.13 \hspace{0.3 cm} EPM solution for SN 2002gw \hspace{0.1 cm} . . . . . . . . . . . . . . . . . . . . . . . . . . . . . .
     \hspace{0.3 cm} 50\\       
3.14 \hspace{0.3 cm} EPM solution for SN 2003T \hspace{0.1 cm}  . . . . . . . . . . . . . . . . . . . . . . . . . . . . . .
     \hspace{0.3 cm} 51\\
3.15 \hspace{0.3 cm} EPM solution for SN 2003bl \hspace{0.1 cm} . . . . . . . . . . . . . . . . . . . . . . . . . . . . . .
     \hspace{0.3 cm} 52\\
3.16 \hspace{0.3 cm} EPM solution for SN 2003bn \hspace{0.1 cm} . . . . . . . . . . . . . . . . . . . . . . . . . . . . . .
     \hspace{0.3 cm} 53\\
3.17 \hspace{0.3 cm} EPM solution for SN 2003ef \hspace{0.1 cm} . . . . . . . . . . . . . . . . . . . . . . . . . . . . . .
     \hspace{0.3 cm} 54\\
3.18 \hspace{0.3 cm} EPM solution for SN 2003hl \hspace{0.1 cm} . . . . . . . . . . . . . . . . . . . . . . . . . . . . . .
     \hspace{0.3 cm} 55\\
3.19 \hspace{0.3 cm} EPM solution for SN 2003hn \hspace{0.1 cm} . . . . . . . . . . . . . . . . . . . . . . . . . . . . . .
     \hspace{0.3 cm} 56\\
3.20 \hspace{0.3 cm} EPM solution for SN 2003iq \hspace{0.1 cm} . . . . . . . . . . . . . . . . . . . . . . . . . . . . . .
     \hspace{0.3 cm} 57 \\
\vspace{1 ex}
\noindent \newline
4.1  \hspace{0.5 cm} EPM distances as a function of the host galaxy extinction \hspace{0.1 cm} . . . . . . . . . . . .
     \hspace{0.3 cm} 58\\
4.2  \hspace{0.5 cm} Hubble diagrams using the $\{BV\}$ filter subset and {\it OLI} \hspace{0.1 cm} . . . . . . . . . . . . . .
     \hspace{0.3 cm} 59\\
4.3  \hspace{0.5 cm} Hubble diagrams using the $\{BVI\}$ filter subset and {\it OLI} \hspace{0.1 cm}  . . . . . . . . . . . . .
     \hspace{0.3 cm} 60\\
4.4  \hspace{0.5 cm} Hubble diagrams using the $\{VI\}$ filter subset and {\it OLI} \hspace{0.1 cm}  . . . . . . . . . . . . . .
     \hspace{0.3 cm} 61\\
4.5  \hspace{0.5 cm} Hubble diagrams using the $\{BV\}$ filter subset and {\it DES} \hspace{0.1 cm} . . . . . . . . . . . . .
     \hspace{0.3 cm} 62\\
4.6  \hspace{0.5 cm} Hubble diagrams using the $\{BVI\}$ filter subset and {\it DES} \hspace{0.1 cm} . . . . . . . . . . . . .
     \hspace{0.3 cm} 63\\
4.7  \hspace{0.5 cm} Hubble diagrams using the $\{VI\}$ filter subset and {\it DES} \hspace{0.1 cm}   . . . . . . . . . . . . . .
     \hspace{0.3 cm} 64\\
4.8  \hspace{0.5 cm} Corrected {\it E96} and {\it D05} distances \hspace{0.1 cm} . . . . . . . . . . . . . . . . . . . . . . . . . . . 
     \hspace{0.3 cm} 65\\

\thispagestyle{empty}

\newpage
\setcounter{page}{2}
\thispagestyle{empty}

\begin{center}{\bf ABSTRACT} \end{center}

\vspace{2 ex}

We used early time photometry and spectroscopy of 12 Type II plateau Supernovae (SNe IIP) to
derive their distances using the Expanding Photosphere Method (EPM).
We performed this study using two sets of Type II supernovae (SNe II) atmosphere models
from \citet{E96} and \citet{D05b}, three filter subsets ({\it \{BV\},\{BVI\},\{VI\}})
and two methods for the host galaxy extinctions, which led to 12 Hubble diagrams.
Using the $\{VI\}$ filter subset and the \citet{D05b} models we obtained a dispersion in the
Hubble diagram of \mbox{$\sigma_{\mu}$ = 0.32 {\it mag}}
and a Hubble constant of $H_0$ = 52.4 $\pm$ 4.3 $km$ $s^{-1}$ $Mpc^{-1}$.
We also applied the EPM analysis to the well-observed SN IIP SN 1999em. With the $\{VI\}$
filter subset and the \citet{D05b} models we derived a distance of 13.9 $\pm$ 1.4 Mpc, which is in
agreement with the Cepheid distance of 11.7 $\pm$ 1.0 Mpc to the SN 1999em host galaxy
(NGC 1637).

\newpage
\section{Introduction}
\setcounter{page}{3}

%The SNe correspond to the explosive end of the lives of some kind of stars. 
%The energy realesed in the explosion range tipycally between 0.1 to 10 foe ($10^{53}erg$).
%These events are believed to be driven by two different mechanism: gravitational core 
%collapse and thermonuclear explosion. 
%Also these events has been classified in two classes based on the presence (Type II) or not 
%(Type I) of hydrogen lines in their spectra.

Type II supernovae (SNe II) are believed to be produced by the gravitational collapse of 
massive stars ($M$ $>$ $8M_{\sun}$), that at the moment of the explosion have most of their 
hydrogen envelope intact. The energy released in the explosion is typically $10^{53}$ 
{\it erg} (mainly in the form of neutrinos), and the luminosity of the SN during the first few months after 
explosion can be comparable to the total luminosity of its host galaxy.
These objects have been classified based on their light curves into
Type IIP (plateau) and Type IIL (linear) \citep{Pat94}. The former present a nearly 
constant luminosity during the photospheric phase ($\sim$ 100 days after explosion), while 
the latter show a slow decline in luminosity during that phase. However, there are 
some SN II events, such as the SN 1987A, that show peculiar photometric properties.        
Also, further studies of SNe II spectra, have revealed the existence of a new 
subclass, characterized by the presence of narrow spectral lines, called SNe IIn. 
\newline \indent
Due to their high intrinsic luminosities, SNe II have great potential as extragalactic
distance indicators. To date, several methods have been proposed to derive distances to 
SNe II, but two are the most commonly used: the Expanding Photosphere Method (EPM) 
\citep{Kir74} and the Standardized Candle Method (SCM) \citep{Ham02}. The former is a
geometrical technique that relates the physical radius and the angular radius of a SN in
order to derive its distance, and has been applied to several SNe to derive the Hubble 
constant \citep{SKE92}. The EPM is independent of the extragalactic distance ladder,
and therefore does not need any external calibration.
The SCM, is based on the observational relation between expansion 
velocity and luminosity of the SNe. Recently, this method has been applied to a sample of 
high redshift SNe \citep{Nug06}.                                            
Other methods have been used also to determine distances to SNe II such as the Spectral-fitting 
Expanding Atmosphere Method (SEAM) \citep{Bar04} or the Plateau-Tail relation proposed
by \citet{Nad03}. \newline \indent
In this work we apply the EPM using early spectroscopy and photometry of 12 \mbox{SNe IIP} in order to 
derive their distances. We apply the method using two sets of SNe II atmospheres models, 
from \citet{E96} and \citet{D05a}, three filter subsets ($\{BV\},\{BVI\},\{VI\}$) and two methods 
for the host galaxy extinctions, which leads to 12 Hubble diagrams. This work is divided
as follows: $\S$ 2 describes the photometric and spectroscopic observations. In $\S$ 3 is 
presented the Expanding Photosphere Method and the individual EPM analysis of 12 SNe IIP. 
In $\S$ 4.1 are described external comparisons to other 
methods and previous EPM analysis, in $\S$ 4.2 are discussed the error analysis and the effect 
of reddening in the EPM distances. In \mbox{$\S$ 4.3} are shown 12 Hubble diagrams and the 
corresponding Hubble constants. In $\S$ 4.4 we propose an external calibration for the EPM.      
Finally, the conclusions are summarized in $\S$ 5.

\section{Observations}

In this work we use photometry and spectroscopy from four SN followup programs: 
the Cerro Tololo SN program (1986-1996), the Cal\'an/Tololo survey (CT; 1990-1993), the 
Supernovae Optical and Infrared Survey (SOIRS; 1999-2000) and the Carnegie Type II Supernova 
Program (CATS; 2002-2003). 
During these programs optical (and some IR) photometry and spectroscopy were obtained for
nearly 100 SNe, 51 of which belong to the Type II class. 
All of the optical data have already been reduced and they are in due course for 
publication \citep{Ham08}. We also complemented our dataset with some spectroscopic 
observations from other authors.                   

\subsection{Photometry}

\begin{figure}
\plotone{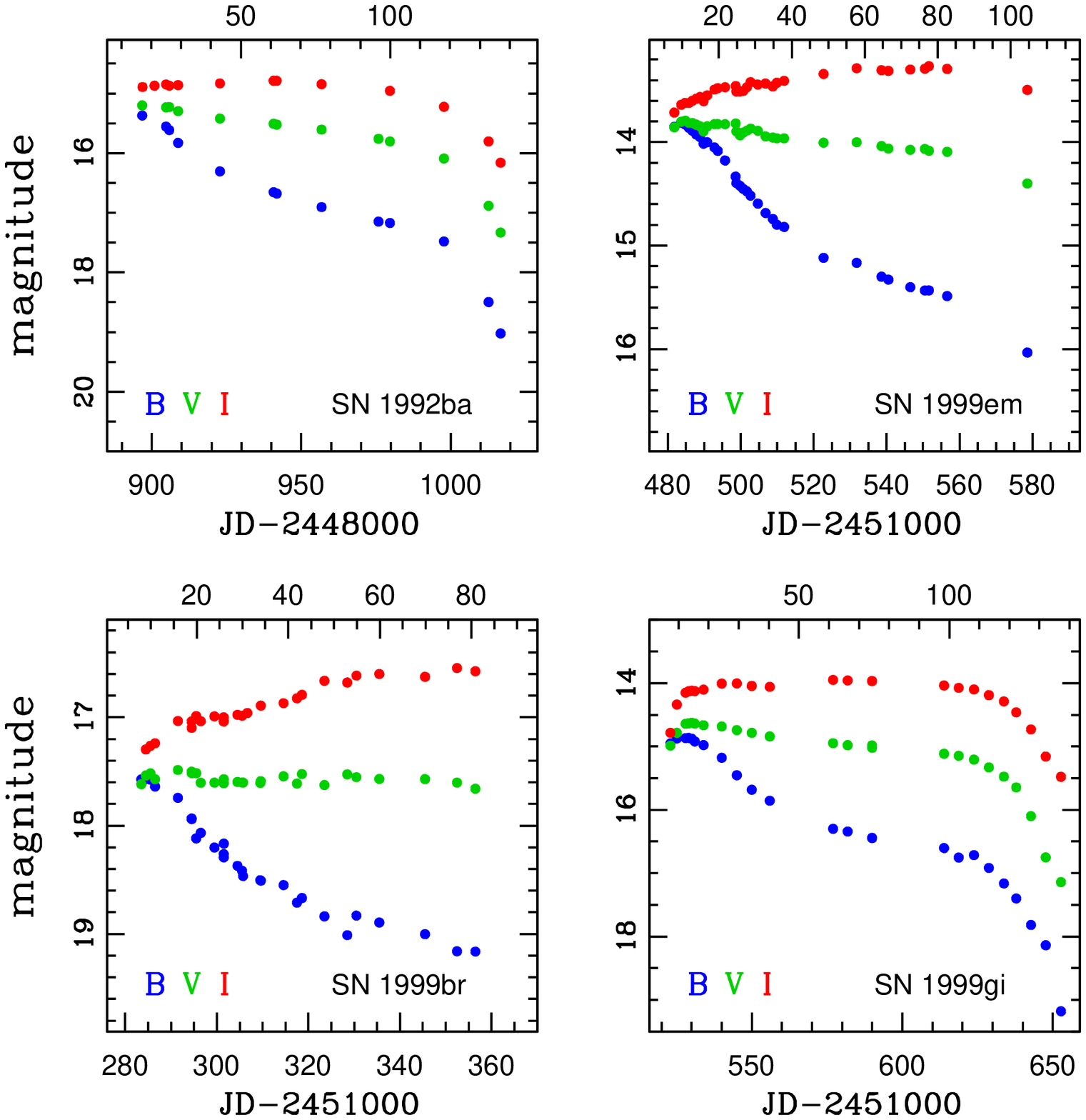}
\caption{Optical light curves of four SNe during the first $\sim$ 120 days of their evolution. In the
top of each panel is shown the date since the EPM explosion time derived using the {\it D05} models
(see Table \ref{tab_VI_DES.EPM}). ~\label{fig_lightcurve1}}
\end{figure}

\begin{figure}
\plotone{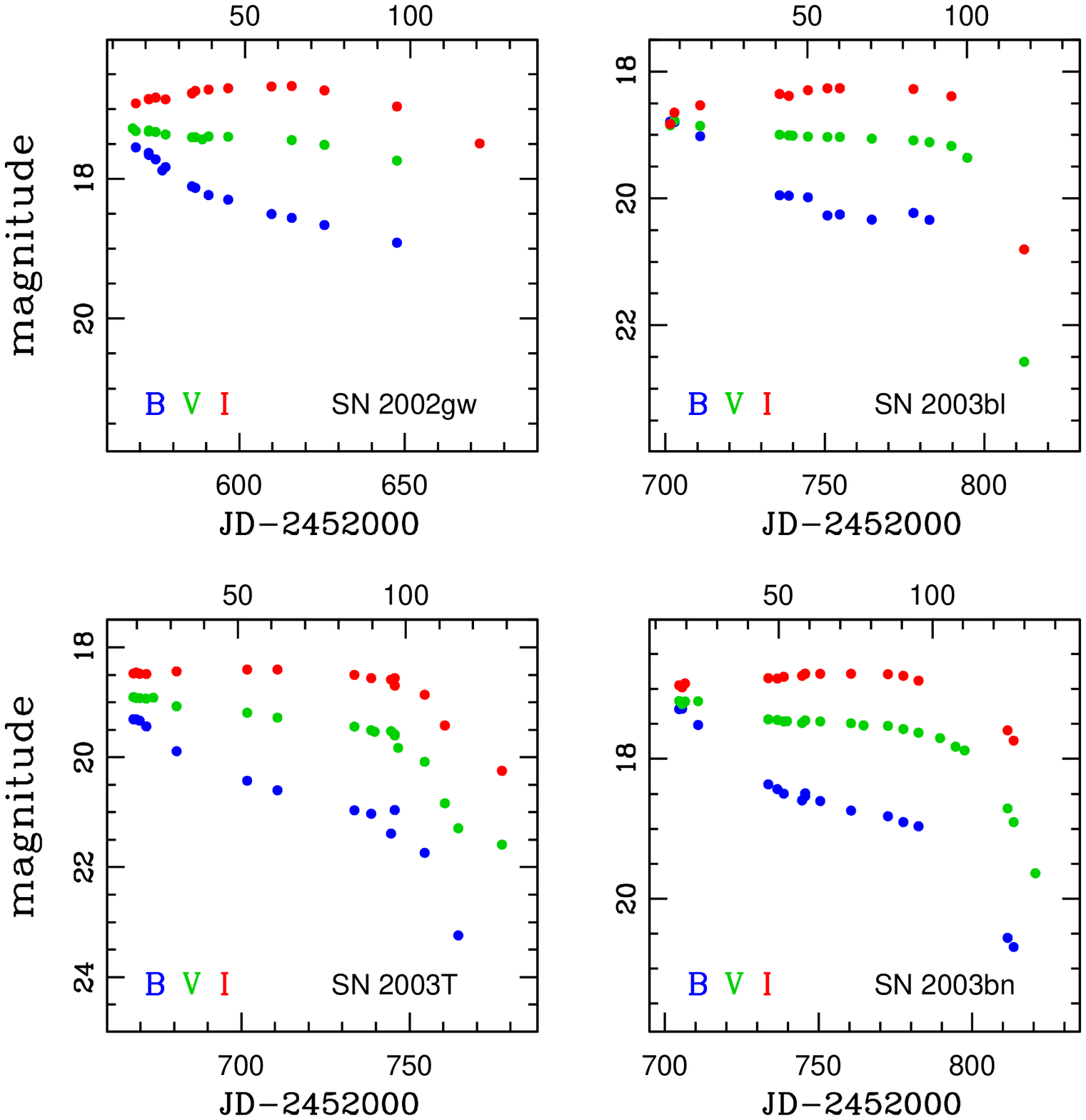}
\caption{Optical light curves of four SNe during the first $\sim$ 120 days of their evolution. In the
top of each panel is shown the date since the EPM explosion time derived using the {\it D05} models
(see Table \ref{tab_VI_DES.EPM}).
~\label{fig_lightcurve2}}
\end{figure}

\begin{figure}
\plotone{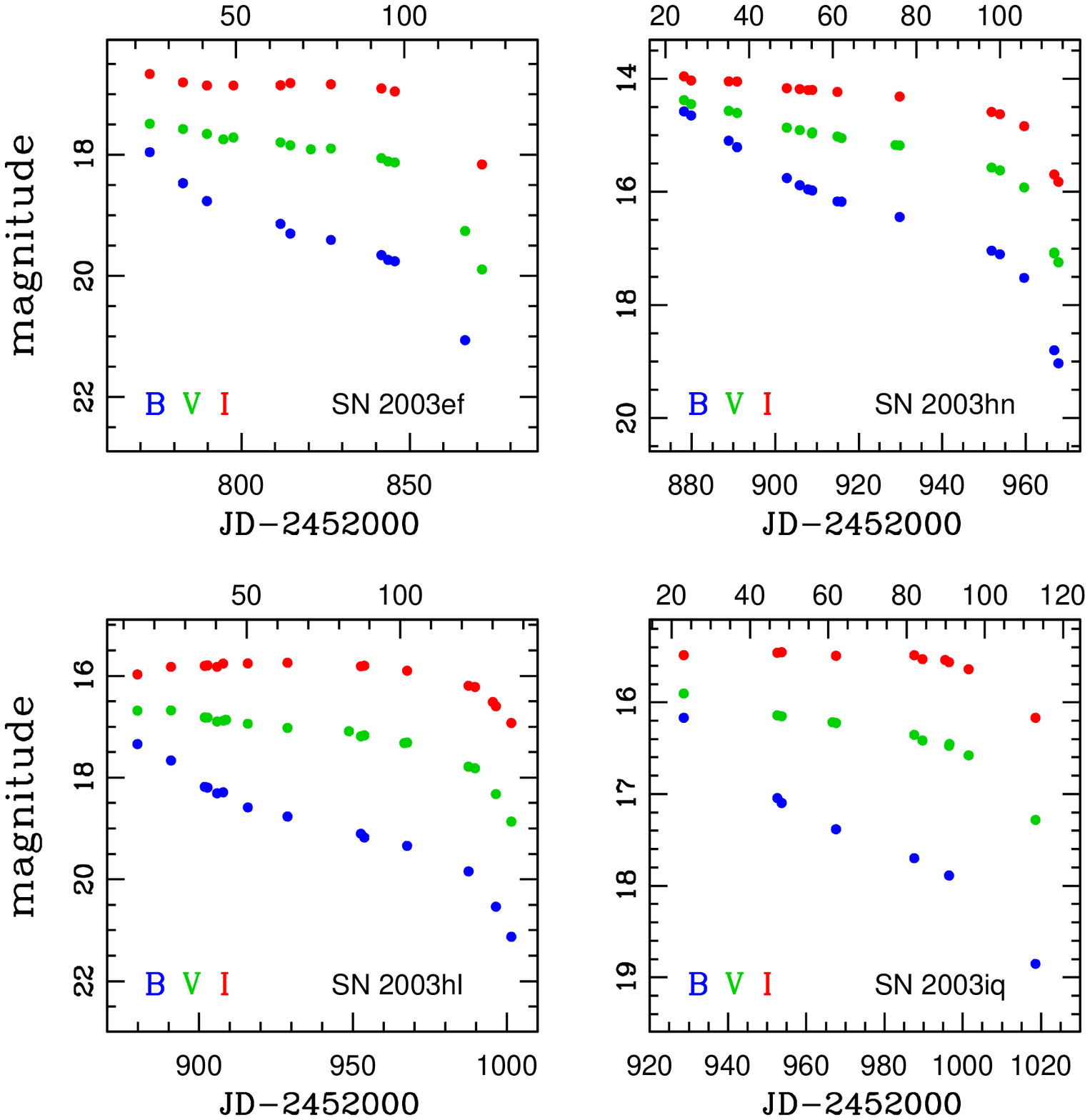}
\caption{Optical light curves of four SNe during the first $\sim$ 120 days of their evolution. In the
top of each panel is shown the date since the EPM explosion time derived using the {\it D05} models
(see Table \ref{tab_VI_DES.EPM}). ~\label{fig_lightcurve3}}
\end{figure}

The observations were made with telescopes from four different observatories: the 
{\it Cerro Tololo Inter-American Observatory} (CTIO), the {\it Las Campanas Observatory} (LCO), 
the {\it European Southern Observatory} (ESO) in La Silla and the {\it Steward Observatory} (S0).
Several telescopes and instruments were used to obtain the photometry as shown in Table 
\ref{tab_telescopes}.
In all cases CCD detectors and standard Johnson-Kron-Cousins {\it UBVRIZ} filters 
\citep{Jon66,Cou71} were employed. For a small subset of SNe observations in the {\it JHK} 
filters were also obtained. 
The data reductions were performed using IRAF \footnote{IRAF is distributed 
by the National Optical Astronomy Observatories, which are operated by the Association of 
Universities for Research in Astyronomy, Inc., under cooperative agreement with the National 
Science Foundation.} according to the procedure described in \citet{Ham08}. \newline \indent
The optical light curves of all the SNe used in this work are shown in Figures 
\ref{fig_lightcurve1}-\ref{fig_lightcurve3} which clearly reveal the plateau nature of
all these events.
            
\subsection{Spectroscopy}

Low resolution (R $\sim$ 1000) optical spectra (wavelength range $\sim$ 3200 - 10000 \AA)
were taken for each SN at various epochs using telescopes and instruments from four 
different observatories. Table \ref{tab_telescopes} lists all the telescopes and instruments 
used for the spectroscopy. Most of the time the spectra were taken orienting the slit along the 
paralactic angle. The wavelength calibration was performed using comparison lamp spectra 
taken at the same position of each SN. The flux calibration was done via observations of flux 
standards stars \citep{Ham92,Ham94}. For more details on the observational procedures see 
\citet{Ham08}. \newline \indent
The spectra were taken to the rest frame using the heliocentric redshifts given in \mbox{Table 
\ref{tab_SN_list}} in order to measure the SN ejecta velocities. In seven cases we were able to 
measure the redshifts from narrow emission lines of HII regions at the SN position (see 
Table \ref{tab_SN_list}). Also, in one case (SN 1999em) we adopted the value from 
\citet{Leo02} which corresponds to the redshift measured at the SN position. In four cases, 
we were unable to extract this information from our data and we had to rely on redshifts of 
the host galaxy centers. The latter does not take into account the rotation velocities of 
the host galaxies, which are typically $v \sim 200$ $km$ $s^{-1}$.  

\subsection{Sample of SNe used in this work}

51 SNe II were observed in the surveys described above. From this sample, only 11 SNe comply
with the EPM requirements, which are: 1) the optical SN light curve ($V$ and the $I$ bands)
must show a nearly constant luminosity during the photospheric phase, i.e, the SN must belong
to the SNe IIP class (see Figures \ref{fig_lightcurve1}-\ref{fig_lightcurve3});
2) the SN must to have early time photometry; 3) the SN must to have at least three early
spectroscopic observations. The need for all of these requirements is   
discussed in $\S$ 3.6.
To the sample of 11 SNe we added the SN IIP 1999gi, which has extensive photometry and
spectroscopy published by \citet{Leo02b}.

\section{The Expanding Photosphere Method}

\subsection{Basic ideas of the EPM}

The EPM is a geometrical technique that relates an angular size and a physical size of a SN, 
in order to derive its distance. 
Although the angular radius $\theta$ of a SN cannot be resolved spatially, it can be derived 
assuming a spherically symmetric expanding photosphere (reasonable assumption for  
SNe IIP, as discussed by \citet{L01}) that radiates as a black body ``diluted" by a factor 
$\zeta^2$, i.e,                      

\begin{equation}
\theta = \frac{R}{D} = \sqrt{\frac{(1+z)f_{\lambda}}{{\pi}{\zeta_{{\lambda}^{'}}^{2}}
B_{{\lambda}^{'} }(T)10^{-0.4[A(\lambda)+A^{'}({\lambda}^{'})]}}}
\end{equation} \newline
where $R$ is the photospheric radius, $D$ is the distance to the SN, $f_{\lambda}$ is the 
observed flux density, $\lambda$ is the observed wavelength, $B_{{\lambda}^{'}}$ is the Planck 
function in the SN rest frame, $T$ is the color temperature, ${\lambda}^{'} = {\lambda}/(1+z)$ 
is the corresponding wavelength in the SN rest frame,
$A(\lambda)$ is the foreground dust extinction and $A^{'}({\lambda}^{'})$ is the host 
galaxy extinction. 
The factor $\zeta_{\lambda^{'}}$ known as ``distance correction factor" or ``dilution
factor", accounts for the fact that a SN does not radiate as a perfect black body. 
There is flux dilution caused by grey electron scattering which makes the 
photosphere (defined as the region of total optical depth $\tau = 2/3$) to form in a 
layer above the thermalization surface. Also, the dilution factor accounts for line 
blanketing in the SN atmosphere. Since the electron scattering is the main source of continuum 
opacity, the total opacity is grey, and the photospheric angular radius is independent of 
wavelength in the optical and near IR \citep{E96}, which explains why $R$ and $\theta$ do not
have a wavelength subscript.
\newline \indent           
Because the gravitational binding energy (U$\sim10^{49}$ $erg$) of a SN progenitor is far less than the 
expansion kinetic energy (E$\sim10^{51}$ $erg$) of the ejecta, it is reasonable to assume free expansion. This assumption 
is supported by hydrodynamical models which show that the different layers of the ejecta reach 
$\sim$ 95\% of their terminal velocities $\sim$ 1 day after the explosion. During this brief period  
there is a transition from an acceleration phase due to the SN explosion, to homologous 
expansion \citep{Utr07,Ber08}. 
Due to the high expansion velocities ($\sim$ 10000 $km$ $s^{-1}$), the initial radius (typically 
$R_0 \sim 10^{13}$ $cm$ for a red supergiant) can be neglected after $\sim$ 1 day from explosion; 
hence after that period the physical radius of the SN can be approximated by

\begin{equation}
R \approx \frac{v(t-t_{0})}{1+z}
\end{equation} \newline 
where $v$ is the photospheric velocity (derived from spectral absorption lines) 
and $t_{0}$ the explosion date. Combining (1) and (2) we obtain 

\begin{equation}
\frac{\theta_{i}}{v_{i}} \approx \frac{(t_{i}-t_{0})}{(1+z)D}
\end{equation} \newline 
where $\theta_{i}$ and $v_{i}$ are the derived quantities measured at time $t_{i}$, which are
estimated following the steps explained in the following sections. 
Equation 3 shows that the quantity $\theta/v$ increases linearly with time, so 
$D$ and $t_{0}$ can be derived with two or more spectroscopic and photometric observations.  
More observations allow us to check the internal consistency of the method.

\subsection{Dilution factors}

\begin{figure}
%%\epsscale{.80}
\plotone{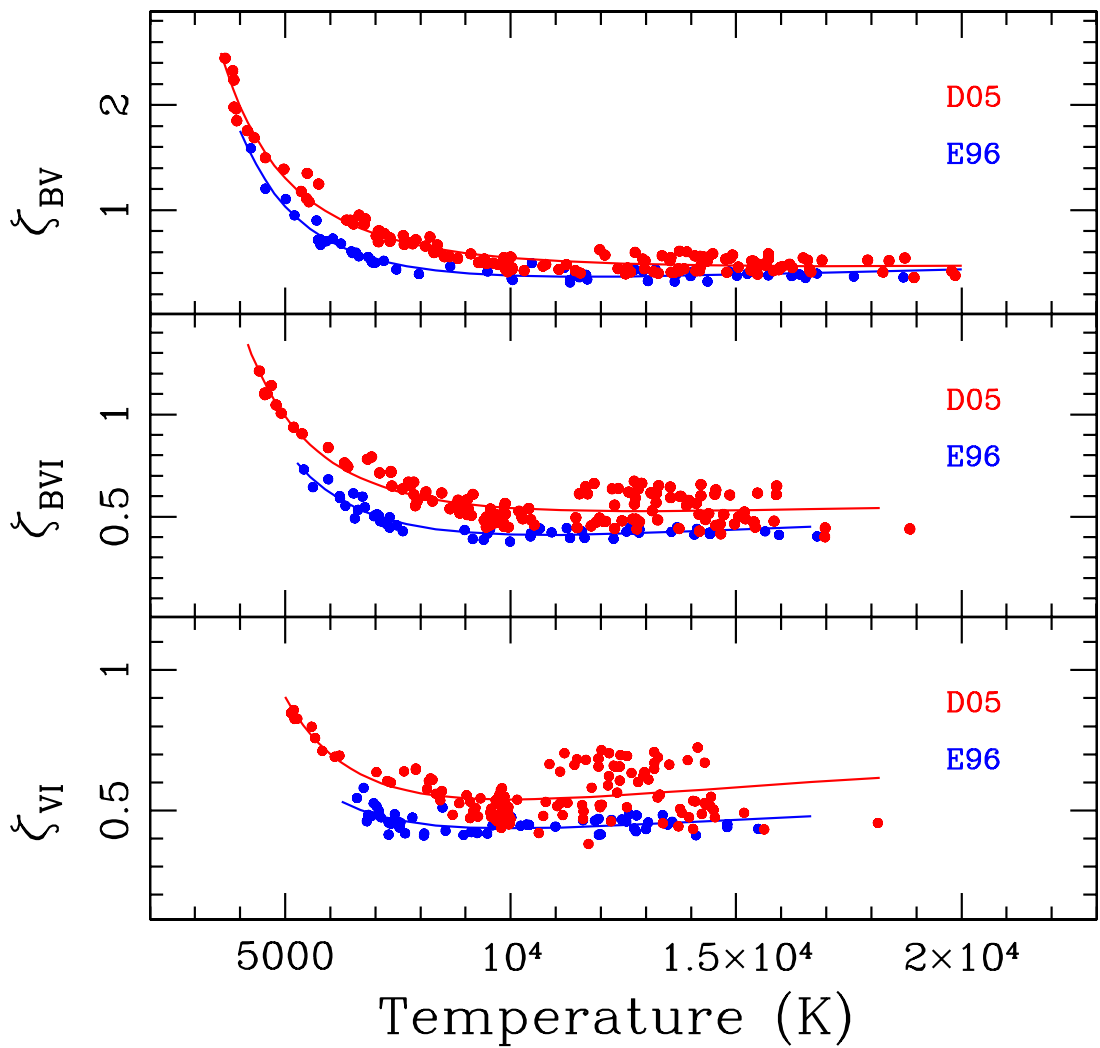}
\caption{Dilution factors $\zeta$ as a function of the color temperature,
computed at $z=0$ from the {\it E96} (blue dots) and {\it D05} (red dots) atmosphere models for three 
different filter subsets ($\{BV\},\{BVI\},\{VI\}$). 
The blue (red) line correspond to the polinomial fit
performed to the {\it E96} ({\it D05}) dilution factors. ~\label{fig_zeta}}
\end{figure}

The dilution factors correspond to the ratio of the luminosity of a SN atmosphere model 
($L_{\lambda^{'}}$) and the corresponding black body luminosity, i.e.,

\begin{equation}
\zeta_{\lambda^{'}}^2 = \frac{L_{\lambda^{'}}}{{\pi}B_{\lambda^{'}}(T)4{\pi}R^2}
\end{equation} \newline
In practice, the dilution factors must be determined for the same filter subsets employed to 
determine the color temperature ($T$) of a SN. In this work we focussed on three 
different optical filter subset, $\{BV\}$, $\{BVI\}$ and $\{VI\}$, and we used two 
SN atmosphere models, namely, those by\citet{E96} ({\it E96} hereafter), and \citet{D05b} ({\it D05} 
hereafter) to compute the dilution factors. See also \citet{D05a} for more details of the 
imput parameters of the {\it D05} models. Because the color temperature of the SNe were determined from 
colors measured in the observer's rest frame, both the atmosphere models and the black body 
function must be redshifted, therefore the dilution factors must be computed for the 
specific redshift of each SN. \newline \indent
We computed {\it B,V,I} synthetic magnitudes using 58 spectra from {\it E96} atmosphere 
models and 138 spectra from {\it D05} atmosphere models. For each filter subset S 
($S = \{BV\}, \{BVI\}, \{VI\}$) we fit black body functions in the SN rest frame 
$B_{\lambda^{'}}(T_s)$, and solved for $T_s$ and $\zeta_{S,z}$ by minimizing the quantity

\begin{equation}
\epsilon=\sum_{{\overline{\lambda}} \in S}{[M_{\overline{\lambda}}} + 5\log(\frac{R}{10pc}) + 
5\log(\zeta_{S,z}) - b_{\overline{\lambda}}(T_s,z)]^2
\end{equation} \newline
Here $R$ is the photospheric radius, $M_{\overline{\lambda}}$ is the redshifted synthetic 
absolute magnitude of the atmosphere model for a band with central wavelength 
${\overline{\lambda}}$, and $b_{\overline{\lambda}}(T_s,z)$ is the synthetic magnitude of 
${\pi}B_{\lambda^{'}} (T_s)10^{-0.4[A(\lambda)+A^{'}(\lambda^{'})]}/(1+z)$, given by

\begin{equation}
b_{\overline{\lambda}} = -2.5\log_{10}\int\frac{{{\pi{\lambda}B_{\lambda^{'}}(T_s)
10^{-0.4[A(\lambda)+A^{'}({\lambda}^{'})]}}}}{hc(1+z)}S(\lambda)d\lambda + ZP
\end{equation} \newline
where $S(\lambda)$ is the filter transmision function and $ZP$ the zero point of the
photometric system \citep{Ham01}. The constant $h$ and $c$ are the Planck constant and 
the speed of light, respectively. Clearly the dilution factors depend on the specific 
redshift of the SN and on the filter subset used to obtained the color temperature of 
the models. Figure \ref{fig_zeta} shows the resulting dilution factors versus temperature
at $z=0$. We performed polynomial fits to $\zeta(T_s)$ of the form  

\begin{equation}
\zeta(T_s) = \sum_{j=0}^{2} b_{s,j}\left( \frac{10^{4}K}{T_s} \right )^{j} 
\end{equation} \newline
Table \ref{tab_zeta_coef} lists the $b_{s,j}$ coefficients at $z=0$ for three filter subsets 
and both atmosphere models, {\it E96} and {\it D05}. 
The corresponding polinomial fits are shown as solid lines in \mbox{Figure \ref{fig_zeta}.} 
\newline \indent   
The {\it D05} dilution factors are quite insensitive to the color temperature above $\sim$ 
$9000$ $K$, and lie around 0.5, while at lower temperature they increase sharply with decreasing 
temperature, reaching a value over unity below $\sim$ $5000$ $K$. The {\it E96} dilution factors 
present the same pattern, but they are systematically lower than the {\it D05} dilution factors 
by \mbox{$\sim$ $15\%$.} The origin of these differences is unclear. \citet{D05a} discuss that the
discrepancy might be related to the different approach used to handle the relativistic 
terms. 
Also, {\it D05} solved the non-Local Thermodynamic Equilibrium (non-LTE) for all the species, 
and employed a very complex atom model. {\it E96}, on the other hand, solved the non-LTE problem 
for a few species and for the rest of the metals the excitation and ionization were assumed 
to be given by the Saha-Boltzmann equation, and the opacity was taken as pure scattering.
Another important difference between the {\it E96} and {\it D05} dilution factors is the dependence 
on the parameters involved in the atmosphere modelling. While the {\it E96} dilution factors 
show little sensivity to a broad range of phyical parameters other than temperature, the 
{\it D05} models show a larger dispersion at a given color temperature.
On average, the {\it E96} models lead to a dispersion of $\sigma\sim 0.03$ in $\zeta$, 
while the {\it D05} models yield to $\sigma\sim 0.07$. 
%This is far the biggest contribution to the error in the EPM solutions when the {\it D05} dilution 
%factors are employed. In the other hand, when {\it E96} is employed, the uncertainty introduces 
%by the dispersion in the dilution factors is comparable to that produces by the 
%photospheric velocity conversion.

%It can be noted that all three set of dilution factors have the same pattern. 
%Above $\sim$ $9000K$ the dilution factors are quite insensitive to the color temperature, 
%but below that value it increases over small variations of the color temperature, reaching
%over 2 at $\sim$ $4000K$ employing the $\{BV\}$ filter subset. In the $\{VI\}$ case, the 
%dilution factors remains below unity at low color temperatures. Also, there are systematic 
%differences between the dilution factors calculated from {\it E96} and 
%{\it D05} atmosphere models, 
%been the latters $\sim$ $15\%$ higher than those computed from $E96$ atmosphere
%models, which leads to higher EPM distances.         

\subsection{Angular radii}

An apparent angular radius ($\theta\zeta_{s}$) and a color temperature ($T_s$) of the SN can 
be obtained by fitting a Planck function $B_{{\lambda^{'}}}(T_{s})$ to the observed broad 
band magnitudes (see eq. 1). Here S is the filter subset combination, i.e., $S = \{BV\}, \{BVI\}, 
\{VI\}$.  Since we have two unknowns ($\theta\zeta_{s}$,$T_{s}$), the subsets must contain 
at least two filters. In order to derive these parameters, we used a least-squares technique 
at each spectroscopic observation epoch (see $\S$ 3.6), by minimizing the quantity

\begin{equation}
\chi^{2} = \sum_{s}\frac{[m_{\overline{\lambda}} + 5log(\theta\zeta_{s,z})-
b_{\overline{\lambda}}(T_s,z)]^2}{\sigma_m^2}
\end{equation} \newline
Here $m_{\overline{\lambda}}$ is the apparent magnitude in the filter with central wavelength 
$\overline{\lambda}$, i.e., $m_{\overline{\lambda}}\in\{B,V,I\}$, $\sigma_{m}$ is the 
photometric error in the magnitude $m_{\overline {\lambda}}$ and $b_{\overline{\lambda}}$ is 
defined in eq. 6.
Because $\zeta_s$ is mainly a function of the color temperature (Figure 
\ref{fig_zeta}), it is possible to use $T_s$ to solve for $\zeta_{s}$ and determine the true 
angular radius $\theta$, from $\theta\zeta_{s}$.

\subsection{Physical radii}

\begin{figure}
\plotone{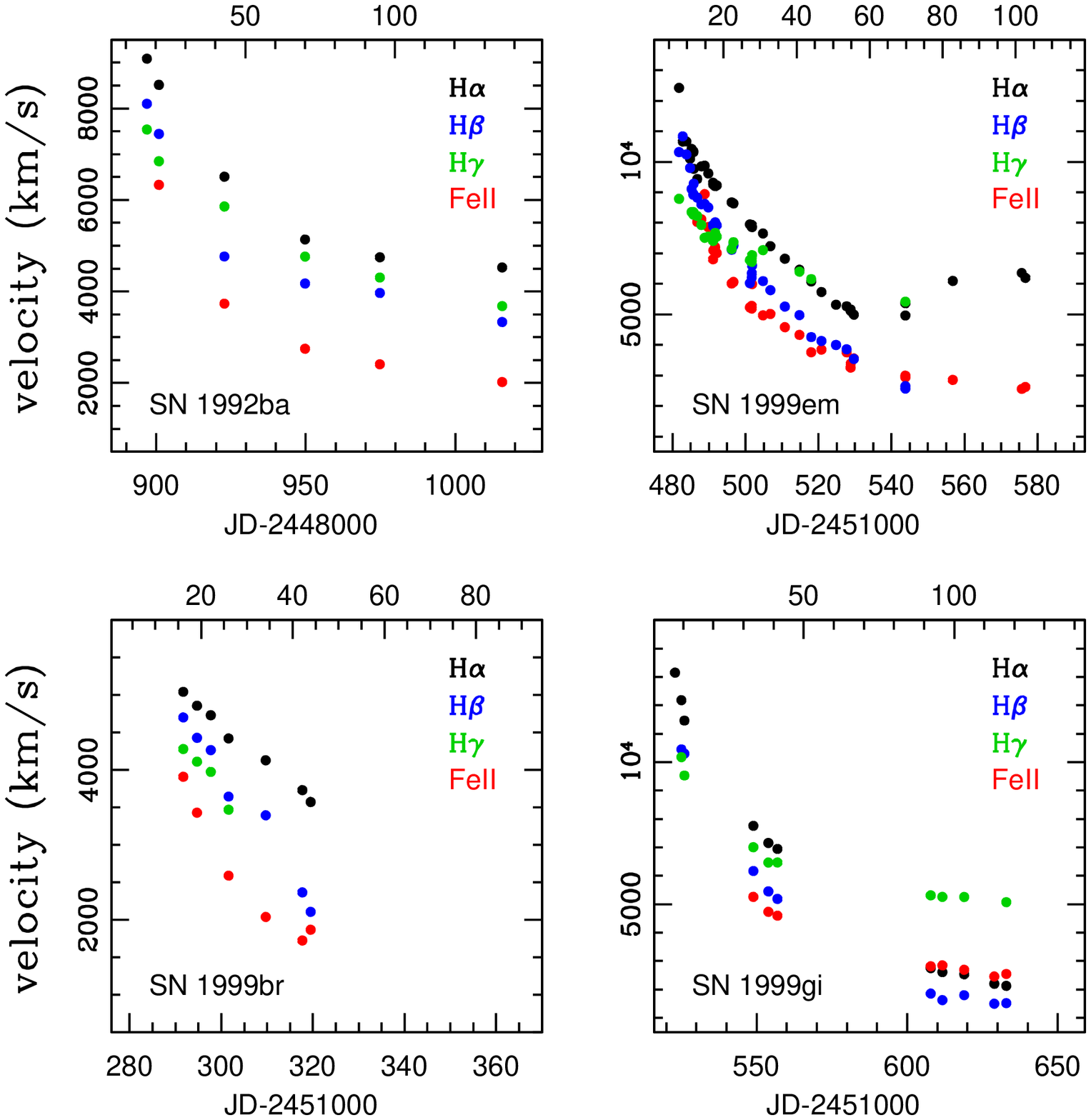}
\caption{Line velocity evolution determined from the P Cygni absorption minima of four different
features during $\sim$ 100 days after discovery. In the
top of each panel is shown the date since the EPM explosion time derived using the {\it D05} models
(see Table \ref{tab_VI_DES.EPM}).
~\label{fig_SNe_velocities1}}
\end{figure}

\begin{figure}
\plotone{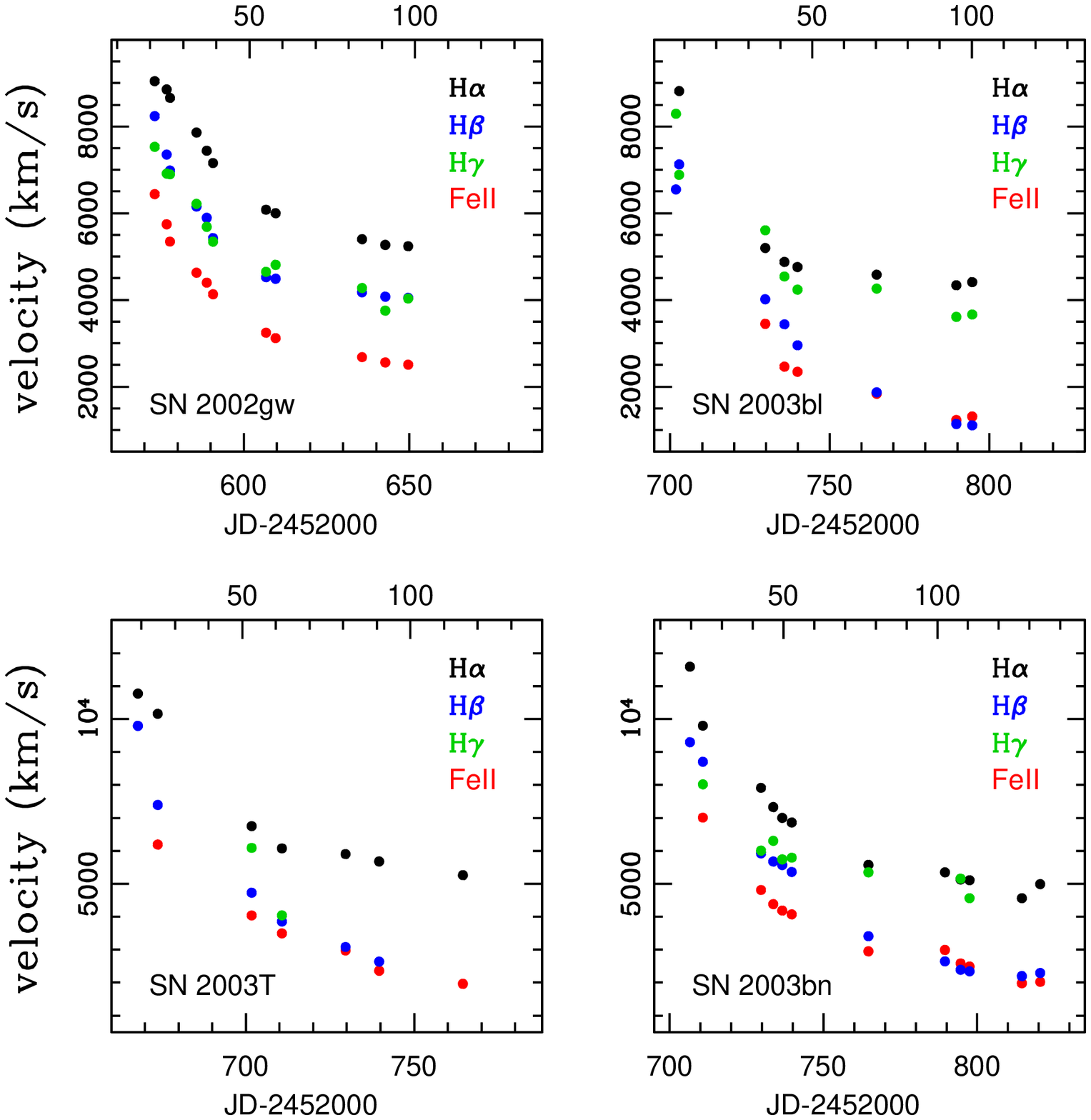}
\caption{Line velocity evolution determined from the P Cygni absorption minima of four different
features during $\sim$ 100 days after discovery. In the
top of each panel is shown the date since the EPM explosion time derived using the {\it D05} models
(see Table \ref{tab_VI_DES.EPM}).
~\label{fig_SNe_velocities2}}
\end{figure}

\begin{figure}
\plotone{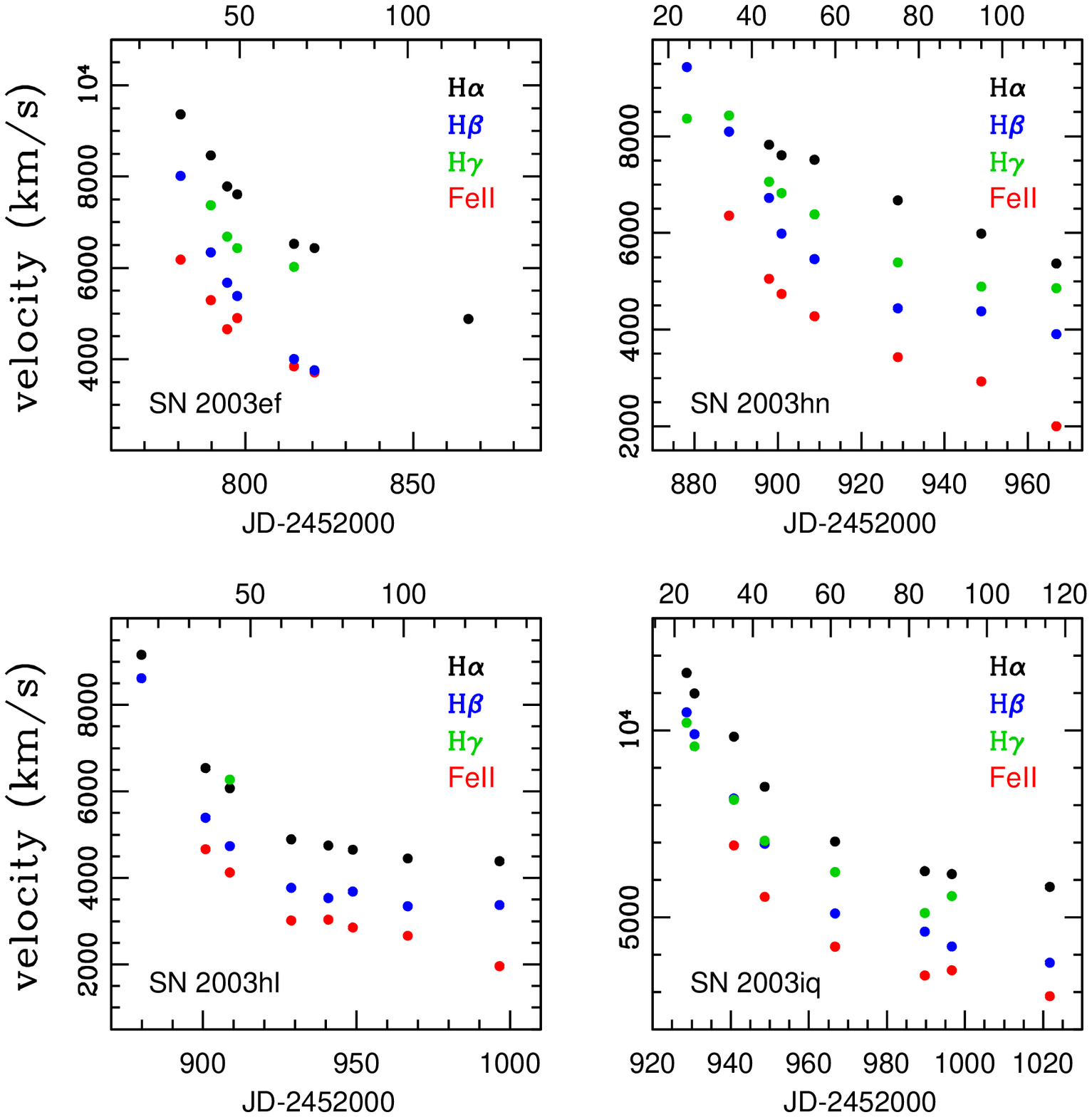}
\caption{Line velocity evolution determined from the P Cygni absorption minima of four different
features during $\sim$ 100 days after discovery. In the
top of each panel is shown the date since the EPM explosion time derived using the {\it D05} models
(see Table \ref{tab_VI_DES.EPM}).
~\label{fig_SNe_velocities3}}
\end{figure}

\begin{figure}
\plotone{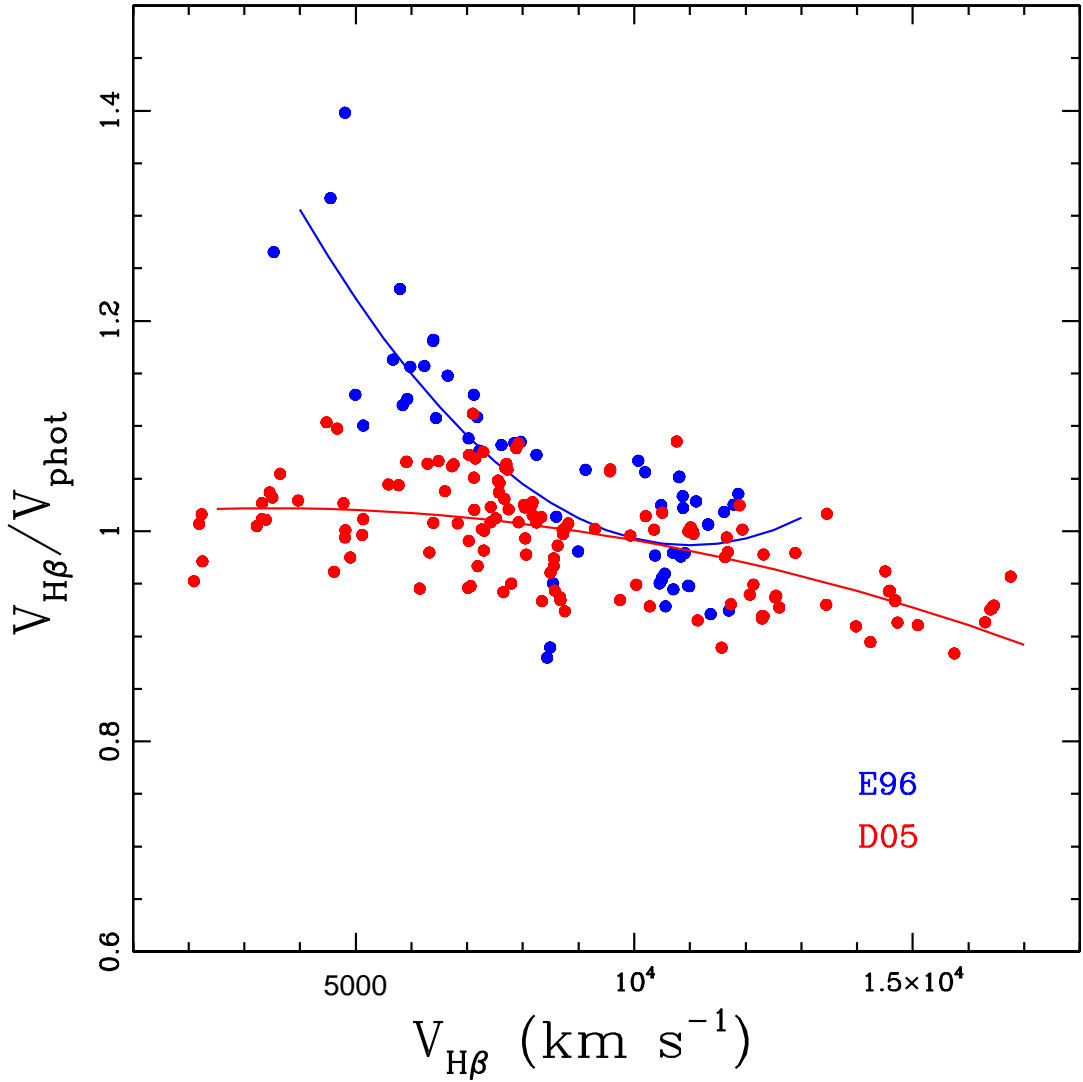}
\caption{Ratio between the H$\beta$ and the photospheric velocity versus the
H$\beta$ velocity of the individual SN models. The blue dots correspond to {\it E96} models and 
the red dots to {\it D05} models. The blue (red) line corresponds to the polinomial fit 
performed to the {\it E96} ({\it D05}) photospheric velocity conversion. 
~\label{fig_models_VHbeta_Vphot}}
\end{figure}

\begin{figure}
\plotone{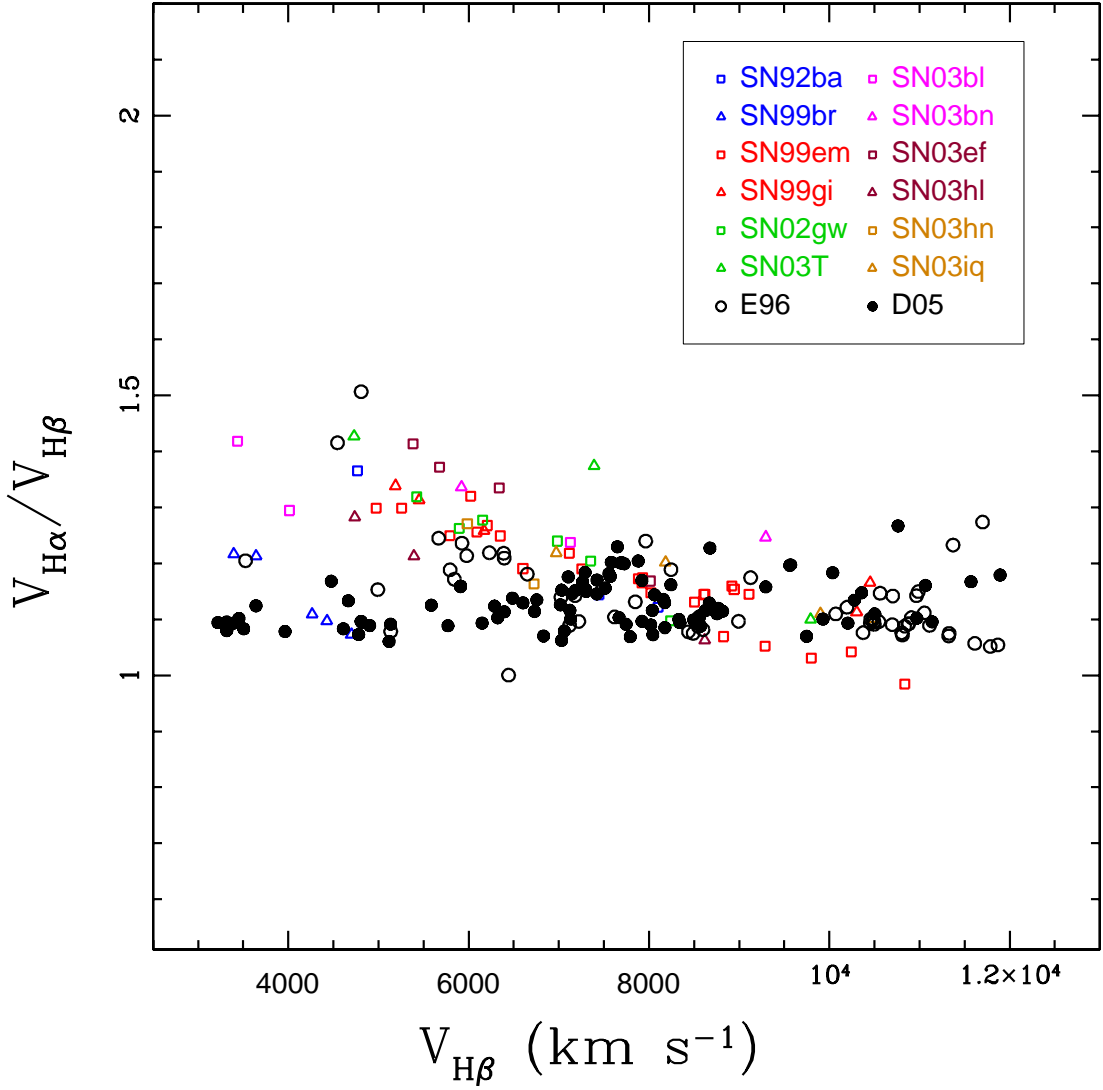}
\caption{Ratio between the H$\alpha$ and H$\beta$ velocity as a function of the 
H$\beta$ velocity. The triangles and the squares represent velocities measured from the 
spectra of our SN sample. The open and filled black circles correspond to the  
velocity ratio measured from the synthetic spectra of {\it E96} and {\it D05} respectively . 
~\label{fig_obs_vel.ratio}}
\end{figure}

Once $\theta$ is determined, the next step is to measure the photospheric velocity          
(see eq. 3). The photospheric velocity of the SN at a given epoch 
can be obtained from the absorption lines in the spectra. Table \ref{tab_spec_vel} 
lists the spectroscopic velocities measured from the minima of H$\alpha$, H$\beta$, H$\gamma$
and Fe {\sc ii} $\lambda5169$ lines, for all 12 SNe. Figures 
\ref{fig_SNe_velocities1}-\ref{fig_SNe_velocities3} show the temporal evolution of the spectral 
line velocities. 
\newline \indent
To date the photospheric velocities have been estimated using weak spectral absorption 
features such as Fe {\sc ii} lines $\lambda 5169$, $\lambda 5018$, $\lambda 4924$ and 
Sc {\sc ii} $\lambda 4670$ \citep{SKE92,Leo02}. 
The physical assumption is that these lines are optically thin and are formed near the 
photosphere of the SN. However, there are two problems with this approach: 1) at early
times the spectra are dominated by Balmer lines and the weak lines are absent and 2) the
synthetic spectra show that even the weak lines do not necessarily yield true 
photospheric velocities \citep{D06}. One way to circumvent these problems is to use the
Balmer lines which are present in the spectra over most of the evolution of the SN.   
Although the Balmer lines are much more optically thick than the \mbox{Fe {\sc ii}} lines, 
\citet{D06} argued that, contrary to what is usually believed, optically thick lines do not 
necessarilly overestimate the photospheric velocity, and the offset from the photospheric
velocity can be measured from the synthetic spectra.
In this work we decided to use the minimum of the H$\beta$ absorption line to derive the
photospheric velocity because this line is present during all the plateau phase, it can be 
easily identified, and it does not present any blend, at least in the first $\sim$ 50 days
after explosion. \newline \indent
To convert from observed H$\beta$ spectroscopic velocities 
to true photospheric velocities we used the synthetic spectra from {\it E96} and {\it D05}.
Figure \ref{fig_models_VHbeta_Vphot} shows (in red) the ratio of H$\beta$ velocity and the 
photospheric velocity, as a function of H$\beta$ velocity for all the {\it D05} models. 
Note that the {\it D05} models predict that the H$\beta$ line forms quite close to the 
photosphere at all epochs (for all values of $v_{H\beta}$). 
Also plotted in Figure \ref{fig_models_VHbeta_Vphot} (in blue) 
are the {\it E96} models which confirms that the H$\beta$ forms close to the photosphere at early
epochs, when the $v_{H\beta}$ is high. 
However, at later epochs (lower $v_{H\beta}$) {\it E96} predict that H$\beta$ forms in outer 
layers (higher velocities) than {\it D05}. 
Also it is important to note that the {\it E96} models cover 
a shorter range in velocity ($\sim$ $4500-12000$ $km$ $s^{-1}$) than the {\it D05} models 
($\sim$ $2000-17000$ $km$ $s^{-1}$), which restricts the EPM analysis using the {\it E96} models.        
%This effect at later epochs produces a departure from linearity in 
%the EPM solutions, therefore we ruled out the {\it E96} photospheric velocity conversion. 
\newline \indent
To derive the ratio between the H$\beta$ and the photospheric velocity we used a polinomial 
fit (as plotted in Figure \ref{fig_models_VHbeta_Vphot}) of the form 

\begin{equation}
\frac{v_{H\beta}}{v_{phot}} = \sum_{j=0}^{2} a_{j}(v_{H\beta})^{j}
\end{equation} \newline
The $a_{j}$ coefficients are listed in Table \ref{tab_ratio_coef}. The {\it E96} models lead to 
a dispersion of $\sigma = 0.06$ and the {\it D05} models to $\sigma = 0.04$. The photospheric 
velocity $v_{i}$ can be obtained from a measurement of $v_{H\beta}$    
 
\begin{equation}
v_{i} = \frac{v_{H\beta}}{\displaystyle{\sum_{j=0}^{2} a_{j}(v_{H\beta})^{j}}}
\end{equation} \newline
In order to examine which of the adopted photospheric velocity conversion was closer to 
reality, we compared the ratio between the H$\alpha$ and H$\beta$ velocities 
measured from the observed spectra of our sample of SNe and from the synthetic spectra 
of the {\it E96} and {\it D05} models. Figure \ref{fig_obs_vel.ratio} shows 
the $H\alpha / H\beta$ velocity ratio as a function of the H$\beta$ velocity. It can be seen 
that, while there is good agreement between theory and observations at high H$\beta$ 
velocities ($\sim$ 6500 - 10500 $km$ $s^{-1}$), the {\it D05} models underestimate the H$\alpha$ 
velocities (or overestimate the H$\beta$ velocities) at lower expansion velocities, while 
the $H\alpha / H\beta$ velocity ratio predicted by the {\it E96} models is in good agreement 
with the observations at all H$\beta$ velocities.                               
This suggests that {\it E96} predict more realistic line profiles in the SN ejecta 
than {\it D05} and therefore should provide a better photospheric velocity conversion. 
%This is because the SN photosphere lies very close to the H recombination front, therefore if 
%the models predict the true H line forming regions they should predict the true photospheric
%forming region and the corresponding photospheric velocity.

%Also we did the same analysis, but this time using the Fe{\sc ii} 5169 velocity instead of
%the H$\alpha$ velocity. Figure \ref{fig_obs_Fe_vel.ratio} shows the Fe{\sc ii} / H$\beta$ 
%velocity ratio as a function of the H$\beta$ velocity. As can be seen the {\it D05} models 
%overestimate the Fe{\sc ii} velocities (or underestimate the $H\beta$ velocities) at all
%range of H$\beta$ velocity (early and late epochs), while the {\it E96} models agree with the 
%observation at $V_{H\beta}$ greater than $\sim$ 6500 $km$ $s^{-1}$, but at lower $V_{H\beta}$
%values, {\it E96} underestimate the $V_{FeII}$ velocities.

\subsection{Extinction}      

\begin{figure}
\plotone{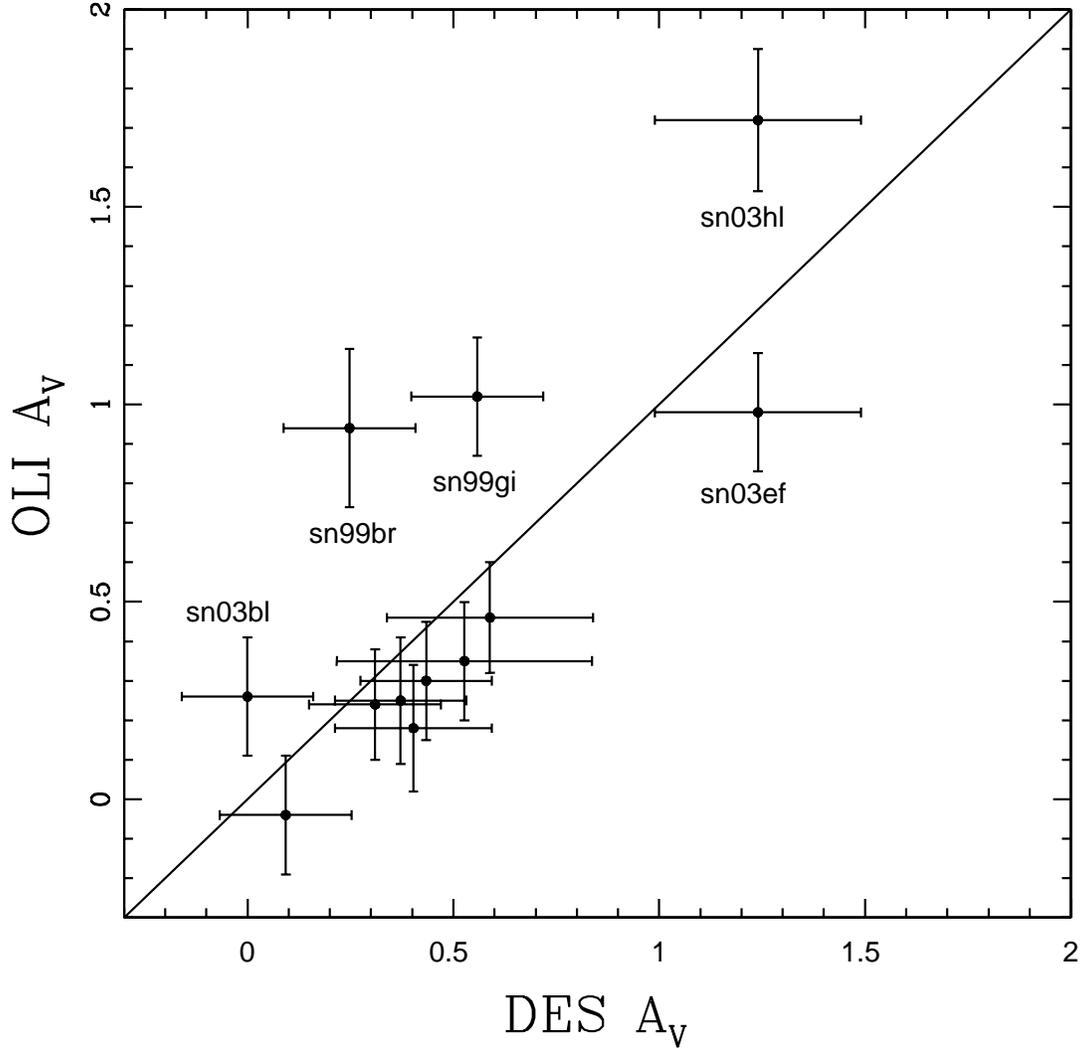}
\caption{Comparison between the {\it DES} and {\it OLI} reddening methods for the 12 SNe.
The straight line has a slope of one. The more deviant SNe are explicitly
marked. ~\label{fig_DES_OLI_reddening}}
\end{figure}

To estimate the amount of Galactic foreground extinction we used the IR dust maps of 
\citet{Sch98}. Table \ref{tab_reddening} summarizes the foreground extinction adopted. 
In this work we adopted two different methods for host galaxy reddenings of our SN sample, 
a spectroscopic method ({\it DES} hereafter), and a method based on the color evolution of the 
SNe ({\it OLI} hereafter). 
The former was developed by \citet{DES08} and consists in fitting different model spectra to
the early time spectra of a SN. The two fitting parameters are the amount of reddening and 
the photospheric temperature.
The color-based technique was developed by \citet{OLI08} and is based on the assumption 
that the color at the end of the plateau phase is the same for all SNe IIP. 
%For more details see \citet{DES08} and \citet{OLI08}, respectively. 
In both cases they adopted the \citet{Car89} extinction law (with $R_{V}=3.1$). 
Table \ref{tab_reddening} lists the host galaxy visual extinction values $A_V$ obtained from both 
methods. Also, in Figure \ref{fig_DES_OLI_reddening} are plotted the {\it OLI} versus {\it DES} 
visual extinctions. As can be seen, there are no systematic differences between both 
models. However, there are individual differences, specially in five SNe, in which cases 
their names are explicitly marked in the plot.

\subsection{Implementation of EPM}

\begin{figure}
\plotone{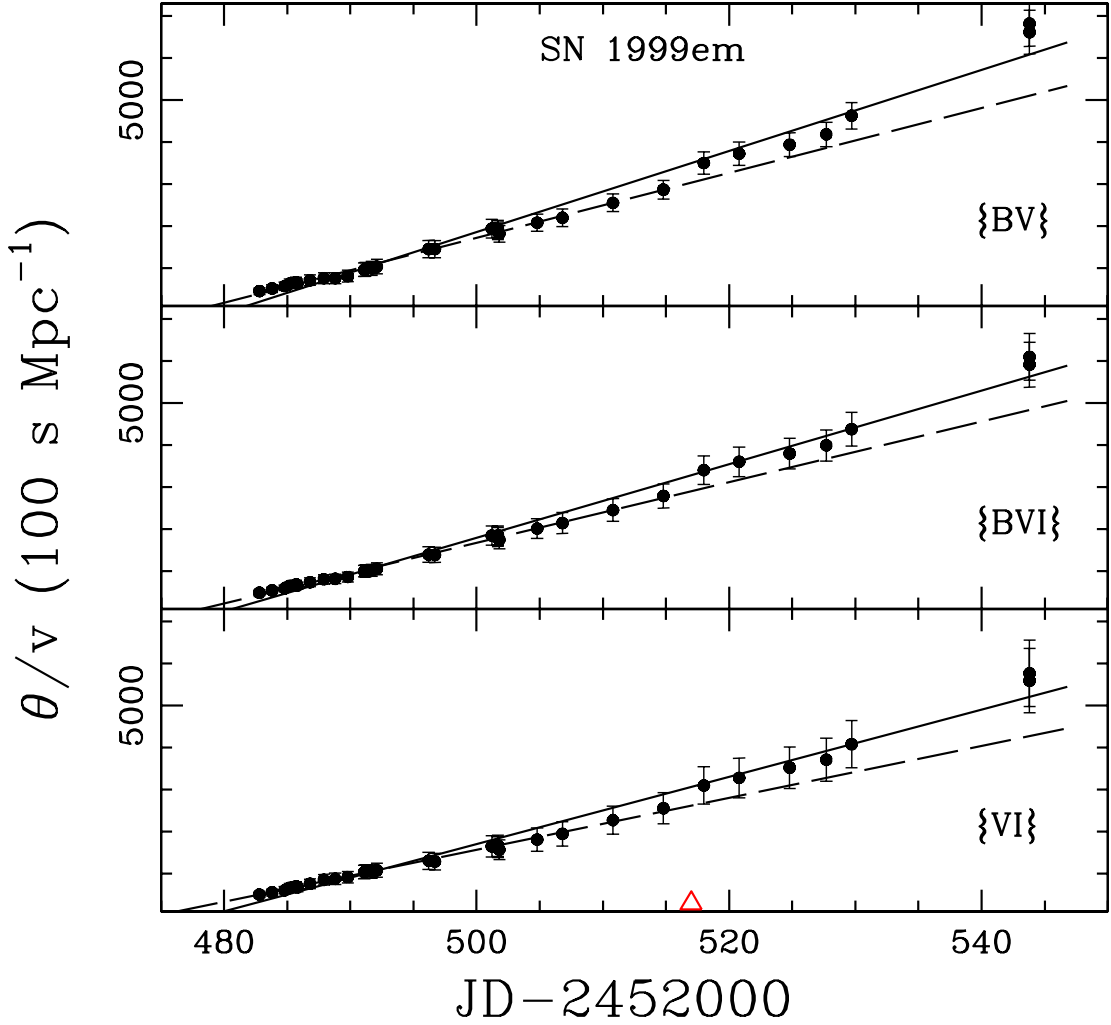}
\caption{$\theta/v$ as a function of time for SN 1999em using the $\{BV\}$, $\{BVI\}$ and $\{VI\}$
filter subsets and the {\it D05} models. The solid (dashed) lines correspond to unweighted 
least-squares fits to the derived EPM quantities using $\sim$ 70 (40) days after explosion. 
The red triangle in the bottom panel shows the day $\sim$ 40 after explosion. 
~\label{fig_full_SN99em.EPM}}
\end{figure}

The EPM method is only valid in the optically thick phase of a H-rich expanding atmosphere.
Observationally this period corresponds to the plateau phase of Type II SNe and thus
justifies our first selection criterion in $\S$ 2.3. \newline \indent
The EPM requires at least two simultaneaus photometric and spectroscopic observations 
(see eq. 3), but we recommend the use of at least three points in order to obtain an
internal check.  
The photometry is used to determine the angular size of the SN and the 
spectroscopy is used to measure the expansion velocities of the SN.       
The requirement of simultaneous photometric and spectroscopic observations 
is not always accomplished because most of the time 
the photometry and the spectroscopy of a SN are taken at different epochs. To overcome
this problem, it is necessary to interpolate the photometry or the velocities measured
from the spectroscopy. 
In this work we decided to interpolate the photometry for two reasons: 1) the number of
photometric observations in our sample of SNe is far greater than the number of spectroscopic 
observations and 2) the optical apparent magnitude of the Type II-P SNe is nearly constant 
during the plateau phase, which makes the interpolation more reliable than the velocity 
interpolation, which has a steeper dependence with time. 
To interpolate a magnitude at the epoch of a given spectroscopic observation we use a
quadratic polynomial fit, using four neighboring points, i.e., four photometric observations
around the spectroscopic date. \newline \indent
In this study, we restricted the EPM analysis to the first $\sim45-50$ days after explosion
because there is a clear departure from linearity in the $\theta/v$ versus $t$ plots after
this date. In Figure \ref{fig_full_SN99em.EPM} are plotted the EPM solutions for SN 1999em
(because it has extensive photometric and spectroscopic observations during the plateau phase)
using the $\{BV\}, \{BVI\}$ and $\{VI\}$ filter subsets and the {\it D05} models. 
The solid line corresponds to the 
least-squares fit to the derived EPM quantities using the first $\sim$ 70 days after explosion, 
while the dashed line correspond to the least-squares fit using only the first \mbox{$\sim$ 40} 
days after explosion. 
As can be noted, after \mbox{$\sim$ 40} days from explosion (marked with a red
triangle) there is departure from the linear $\theta/v$ versus $t$ relation in all three cases.
This justifies our second and third selection criteria in \mbox{$\S$ 2.3}.
This restriction severely lowers the number of SNe of our sample to which we can
apply the EPM. Out of the initial 51 SNe of the \citet{Ham08} sample, only 11 objects 
fulfill the requirement of having a plateau behavior  and having 
early time photometry and spectroscopy for the EPM analysis. 
%The EPM analisys should be applied at any epoch on the plateau phase if photometric and 
%spectroscopic observations were available. However, the steady-state atmosphere models 
%fails to reproduce the observed spectra at late epochs on the plateau (from $\sim$ 40 - 50 
%days after the explosion) which restricts the EPM only to be applied if there is available 
%photometry and spectroscopy of the early epochs on the SN evolution.           

\subsubsection{EPM analysis to individual SNe}

\begin{figure}
\plotone{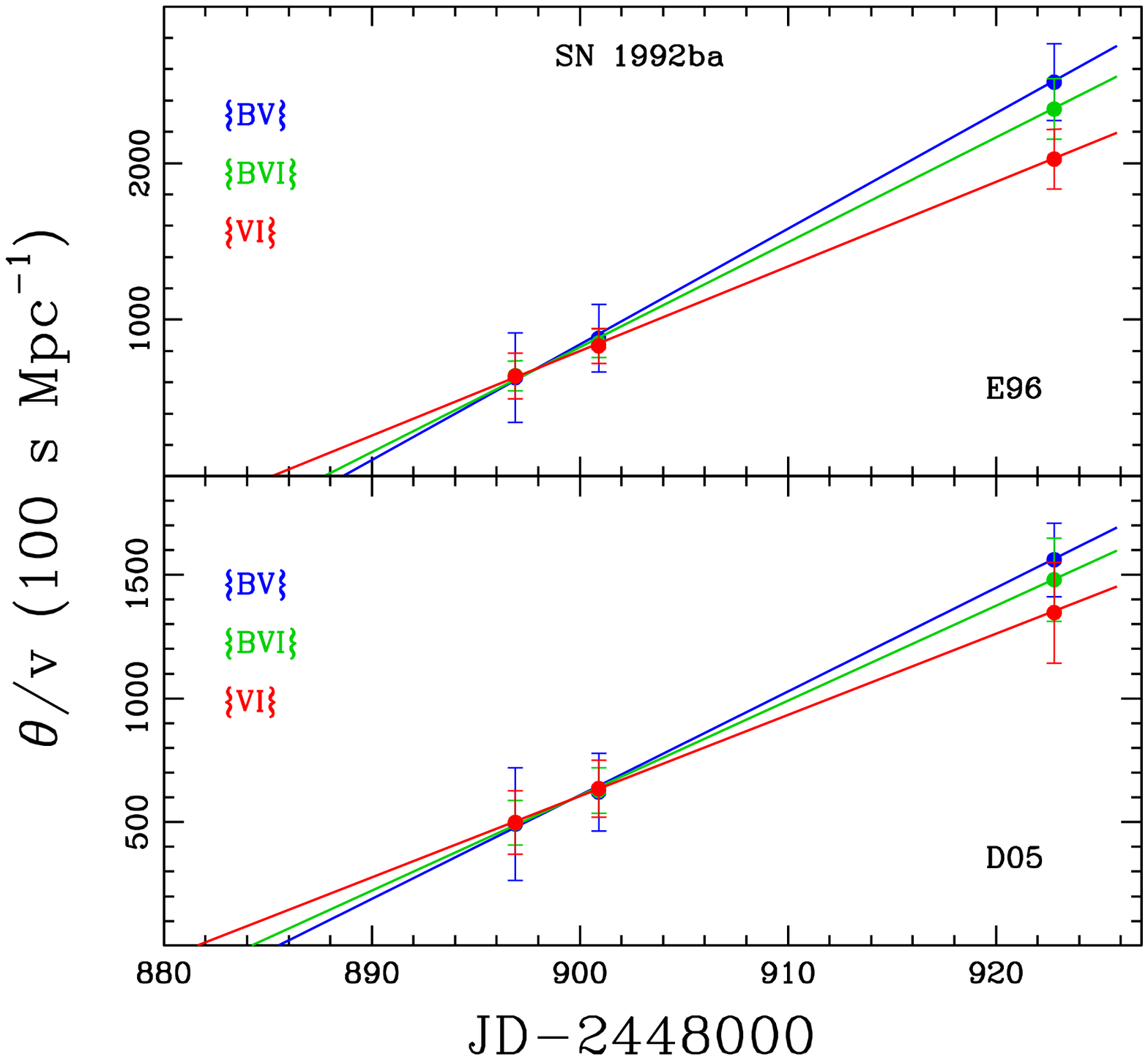}
\caption{$\theta/v$ as a function of time for SN 1992ba using the $\{BV\}$, $\{BVI\}$ and $\{VI\}$ 
filter subsets. The ridge lines correspond to unweighted least-squares fits to the
derived EPM quantities. The upper and lower panel shows the results using {\it E96} and {\it D05} 
dilution factors respectively. In all cases we employ the {\it DES} reddening.
~\label{fig_SN92ba_EPM}}
\end{figure}

\begin{figure}
\plotone{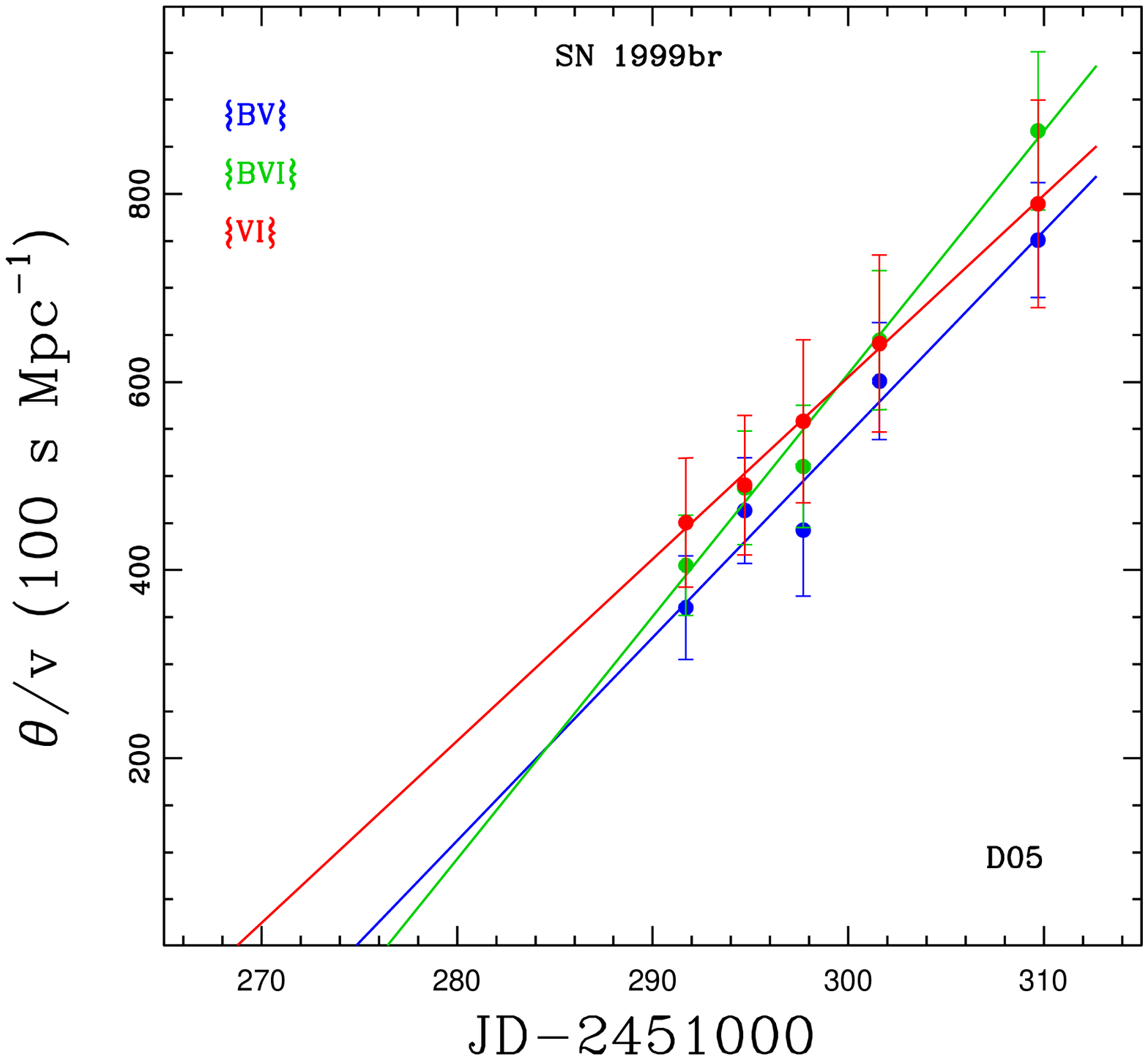}
\caption{$\theta/v$ as a function of time for SN 1999br using the $\{BV\}$, $\{BVI\}$ and $\{VI\}$
filter subsets and the {\it D05} models. The ridge lines correspond to unweighted least-squares 
fits to the derived EPM quantities. In all cases we employ the {\it DES} reddening.
~\label{fig_SN99br_EPM}}
\end{figure}

\begin{figure}
\plotone{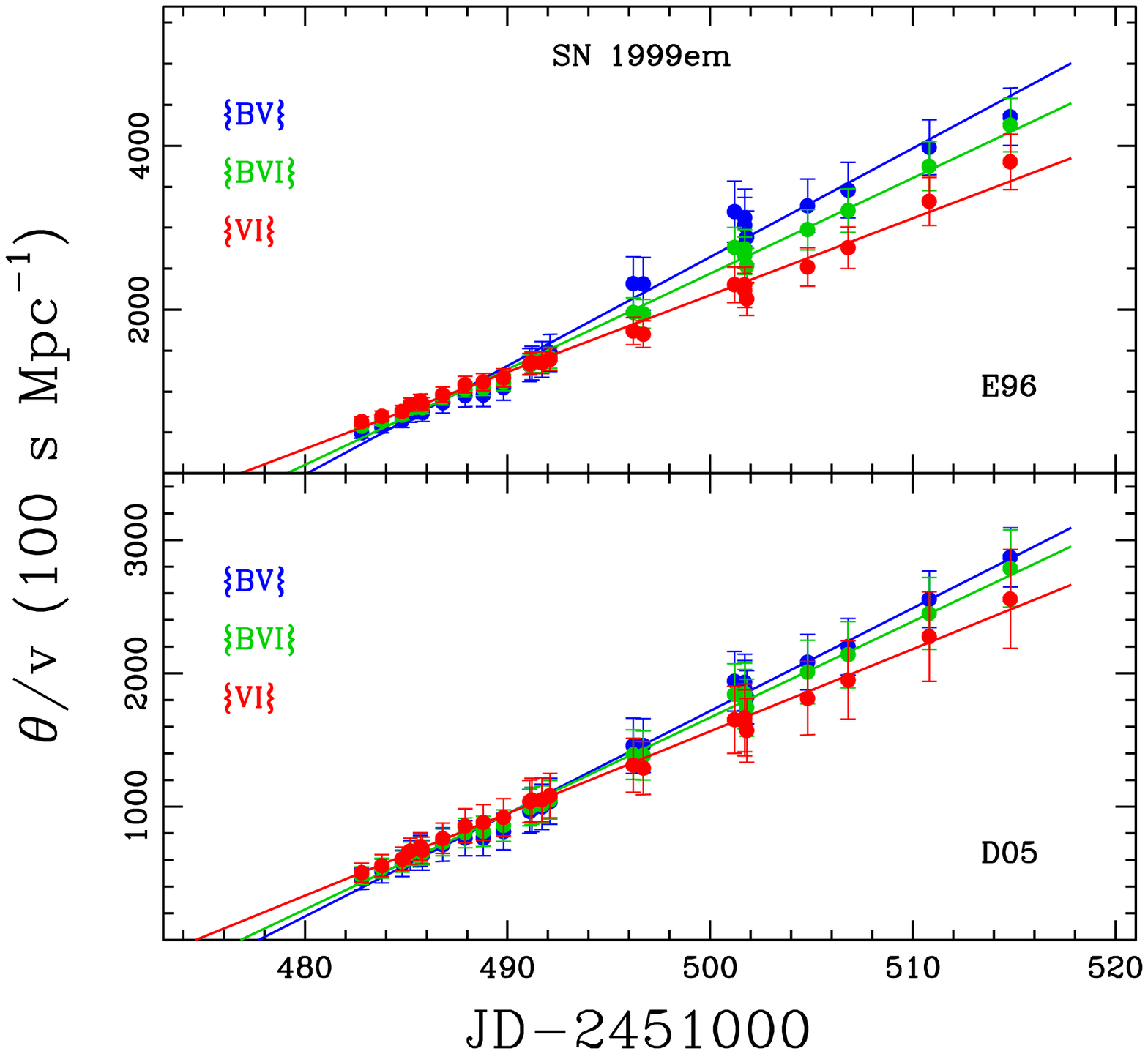}
\caption{$\theta/v$ as a function of time for SN 1999em using the $\{BV\}$, $\{BVI\}$ and $\{VI\}$
filter subsets. The ridge lines correspond to unweighted least-squares fits to the
derived EPM quantities. The upper and lower panel shows the results using {\it E96} and {\it D05} 
dilution factors respectively. In all cases we employ the {\it DES} reddening.
~\label{fig_SN99em_EPM}}
\end{figure}

\begin{figure}
\plotone{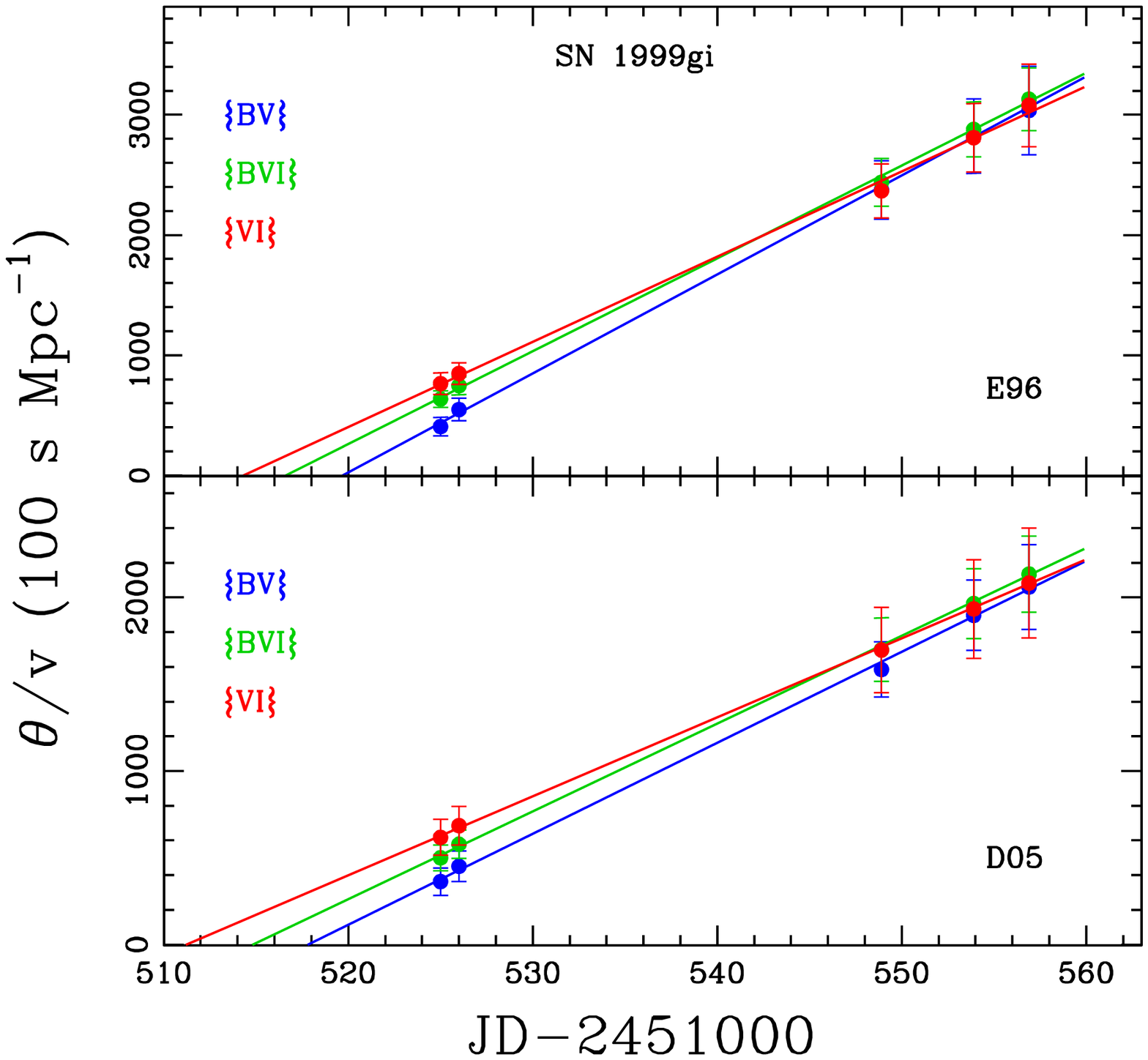}
\caption{$\theta/v$ as a function of time for SN 1999gi using the $\{BV\}$, $\{BVI\}$ and $\{VI\}$
filter subsets. The ridge lines correspond to unweighted least-squares fits to the
derived EPM quantities. The upper and lower panel shows the results using {\it E96} and {\it D05} dilution factors respectively. In all cases we employ the {\it DES} reddening.
~\label{fig_SN99gi_EPM}}
\end{figure}

\begin{figure}
\plotone{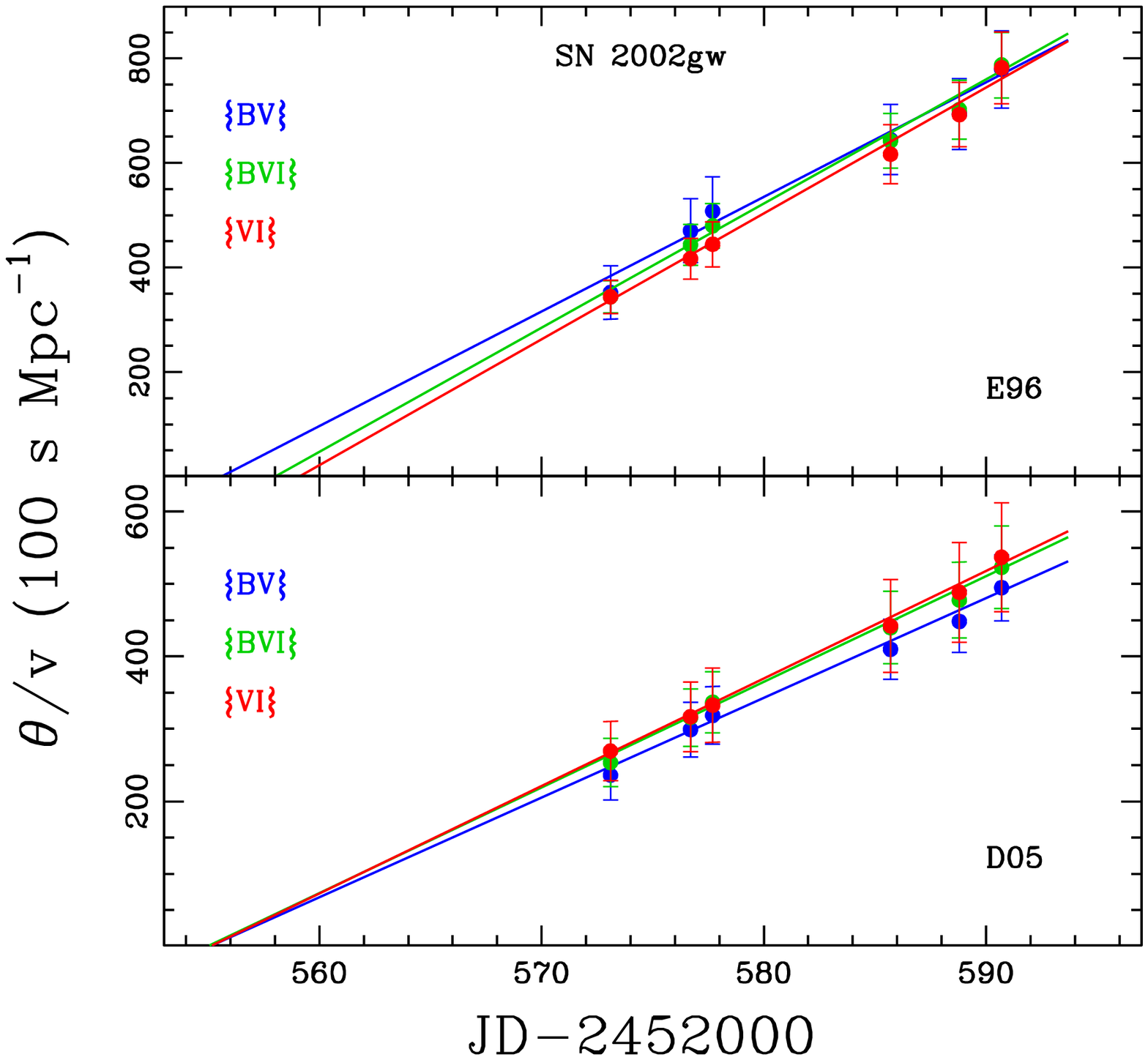}
\caption{$\theta/v$ as a function of time for SN 2002gw using the $\{BV\}$, $\{BVI\}$ and $\{VI\}$
filter subsets. The ridge lines correspond to unweighted least-squares fits to the
derived EPM quantities. The upper and lower panel shows the results using {\it E96} and {\it D05} dilution factors respectively. In all cases we employ the {\it DES} reddening.
~\label{fig_SN02gw_EPM}}
\end{figure}

\begin{figure}
\plotone{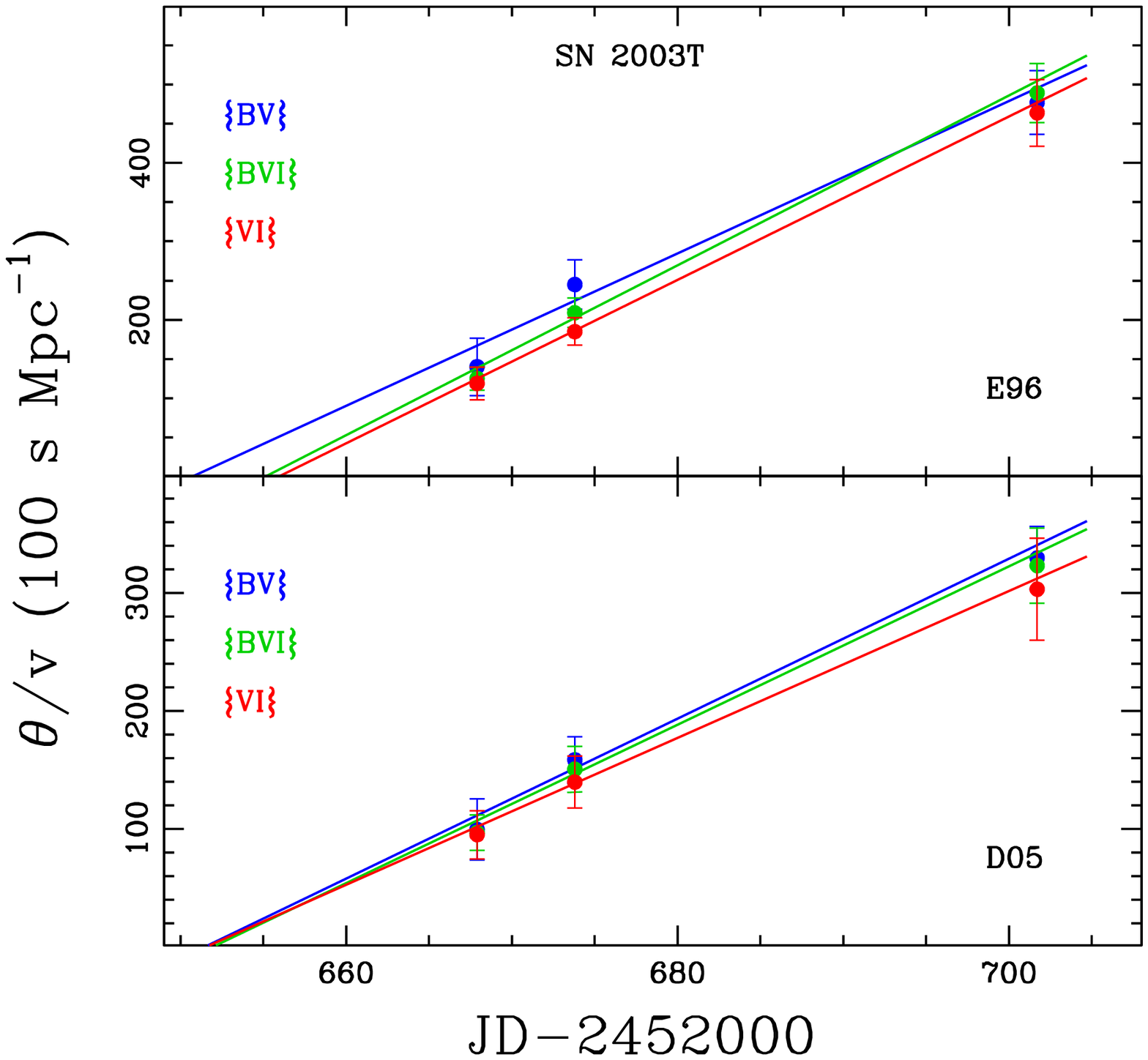}
\caption{$\theta/v$ as a function of time for SN 2003T using the $\{BV\}$, $\{BVI\}$ and $\{VI\}$
filter subsets. The ridge lines correspond to unweighted least-squares fits to the
derived EPM quantities. The upper and lower panel shows the results using {\it E96} and {\it D05} dilution factors respectively. In all cases we employ the {\it DES} reddening.
~\label{fig_SN03T_EPM}}
\end{figure}

\begin{figure}
\plotone{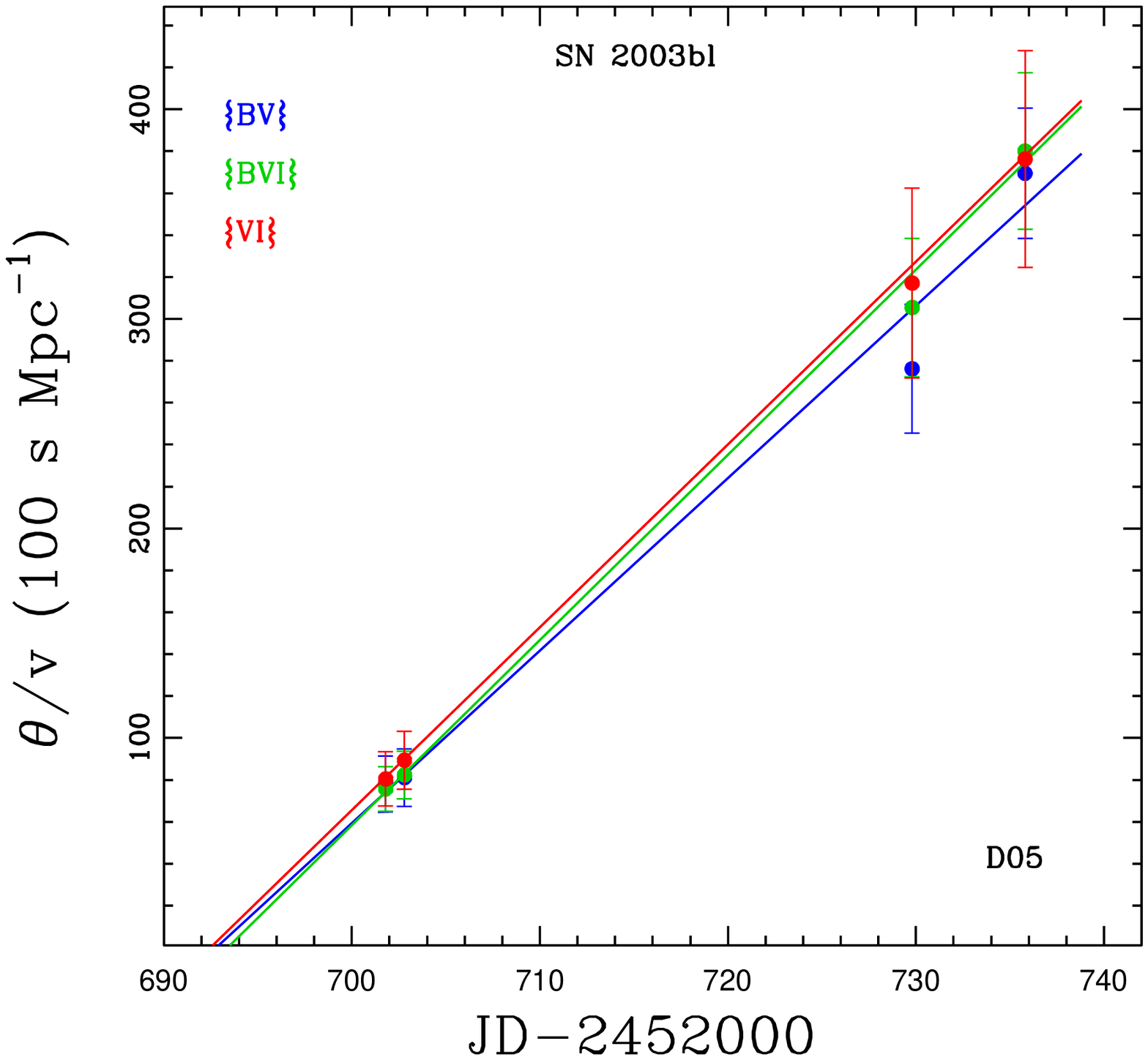}
\caption{$\theta/v$ as a function of time for SN 2003bl using the $\{BV\}$, $\{BVI\}$ and $\{VI\}$
filter subsets and the {\it D05} models. The ridge lines correspond to unweighted least-squares
fits to the derived EPM quantities. In all cases we employ the {\it DES} reddening.
~\label{fig_SN03bl_EPM}}
\end{figure}

\begin{figure}
\plotone{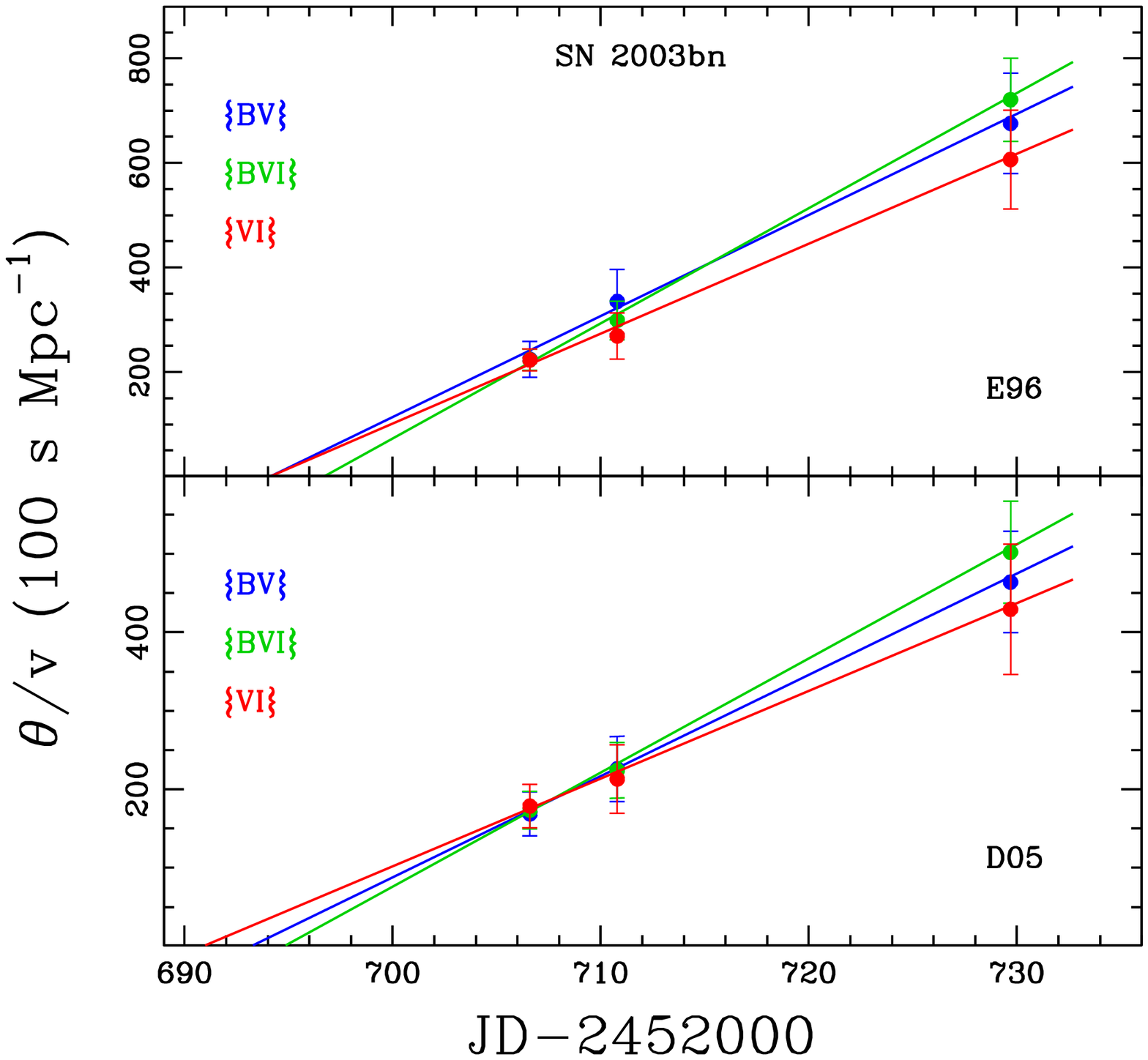}
\caption{$\theta/v$ as a function of time for SN 2003bn using the $\{BV\}$, $\{BVI\}$ and $\{VI\}$
filter subsets. The ridge lines correspond to unweighted least-squares fits to the
derived EPM quantities. The upper and lower panel shows the results using {\it E96} and {\it D05} dilution factors respectively. In all cases we employ the {\it DES} reddening.
~\label{fig_SN03bn_EPM}}
\end{figure}

\begin{figure}
\plotone{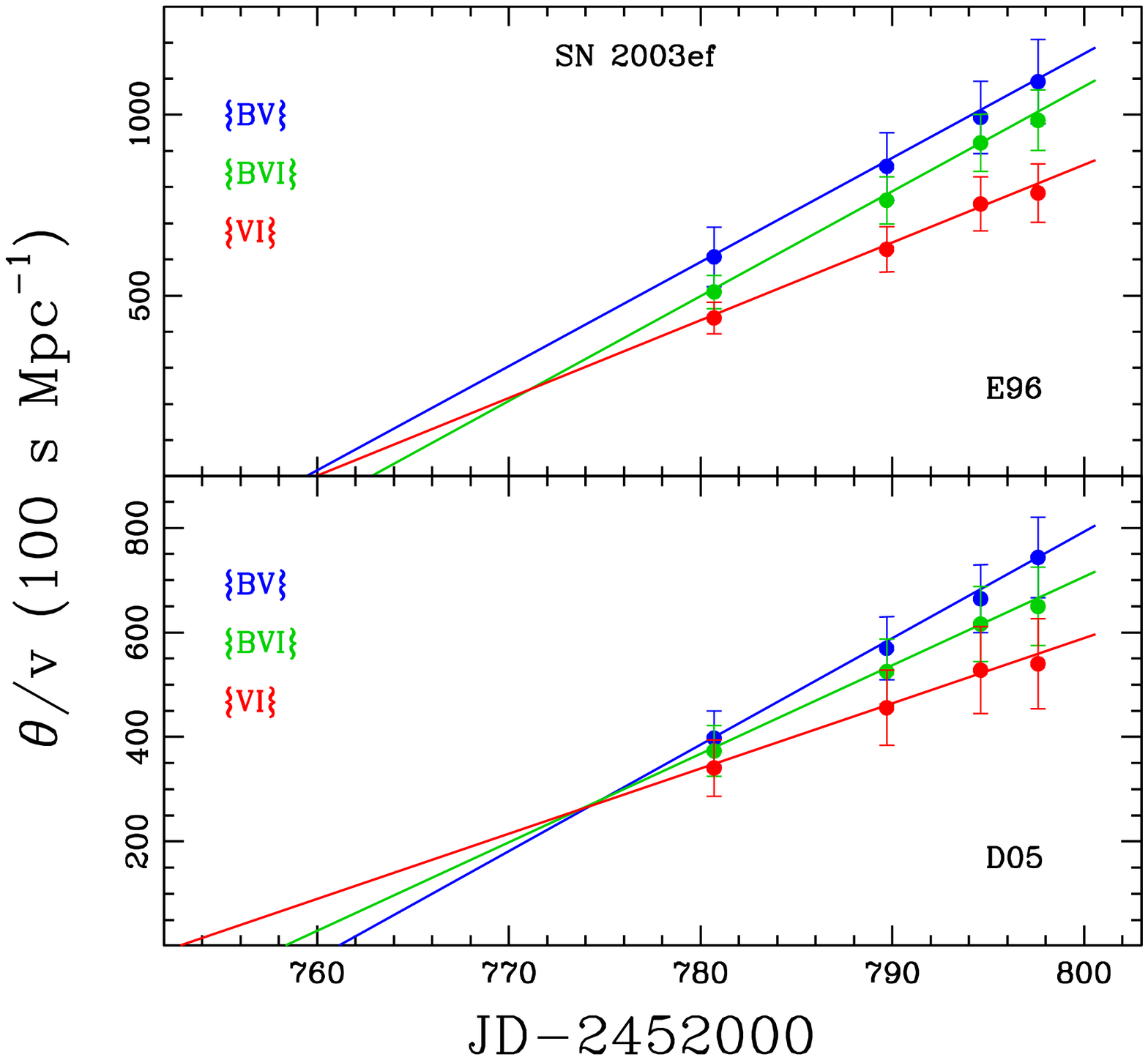}
\caption{$\theta/v$ as a function of time for SN 2003ef using the $\{BV\}$, $\{BVI\}$ and $\{VI\}$
filter subsets. The ridge lines correspond to unweighted least-squares fits to the
derived EPM quantities. The upper and lower panel shows the results using {\it E96} and {\it D05} dilution factors respectively. In all cases we employ the {\it DES} reddening.
~\label{fig_SN03ef_EPM}}
\end{figure}

\begin{figure}
\plotone{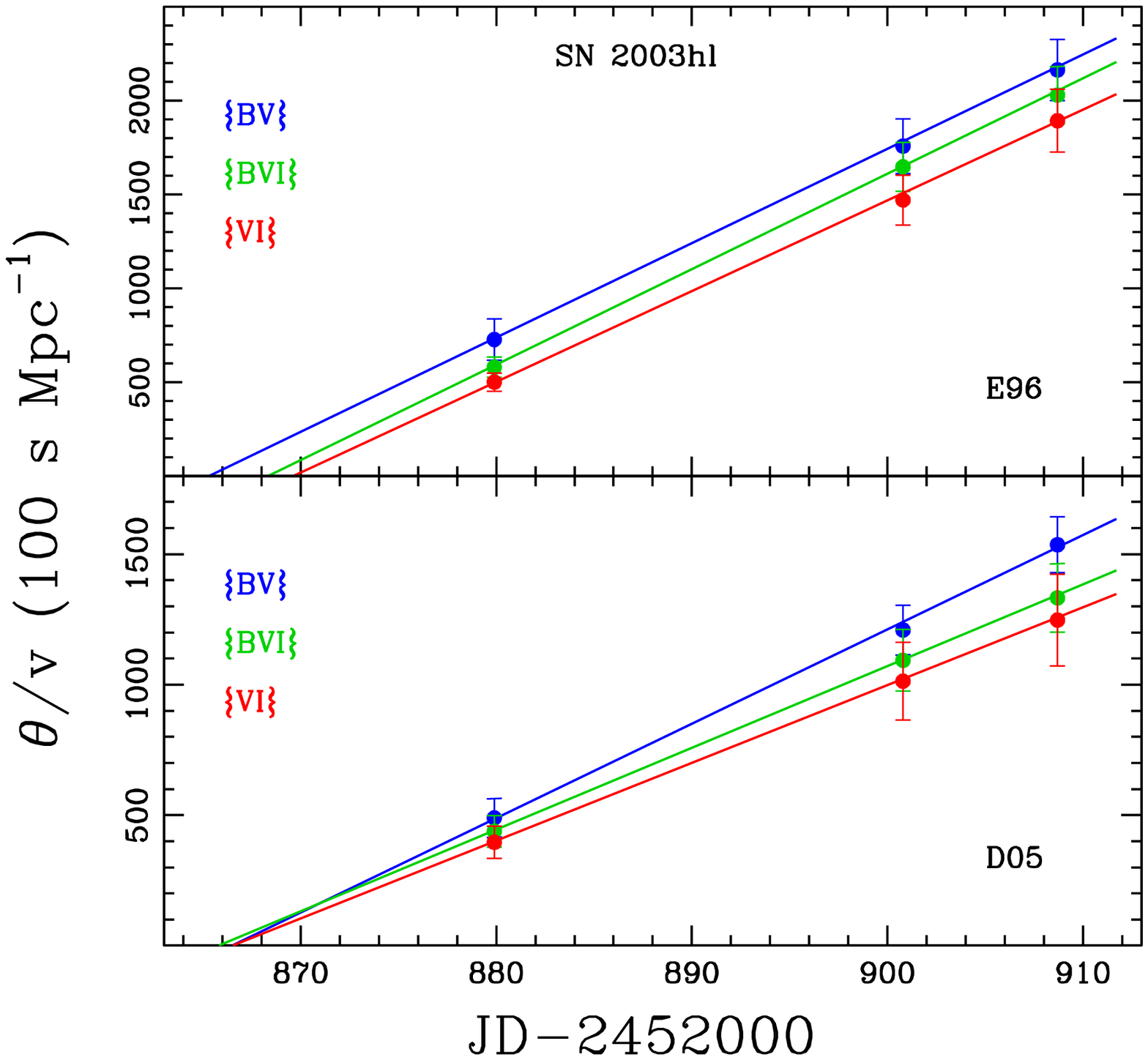}
\caption{$\theta/v$ as a function of time for SN 2003hl using the $\{BV\}$, $\{BVI\}$ and $\{VI\}$
filter subsets. The ridge lines correspond to unweighted least-squares fits to the
derived EPM quantities. The upper and lower panel shows the results using {\it E96} and {\it D05} dilution factors respectively. In all cases we employ the {\it DES} reddening.
~\label{fig_SN03hl_EPM}}
\end{figure}

\begin{figure}
\plotone{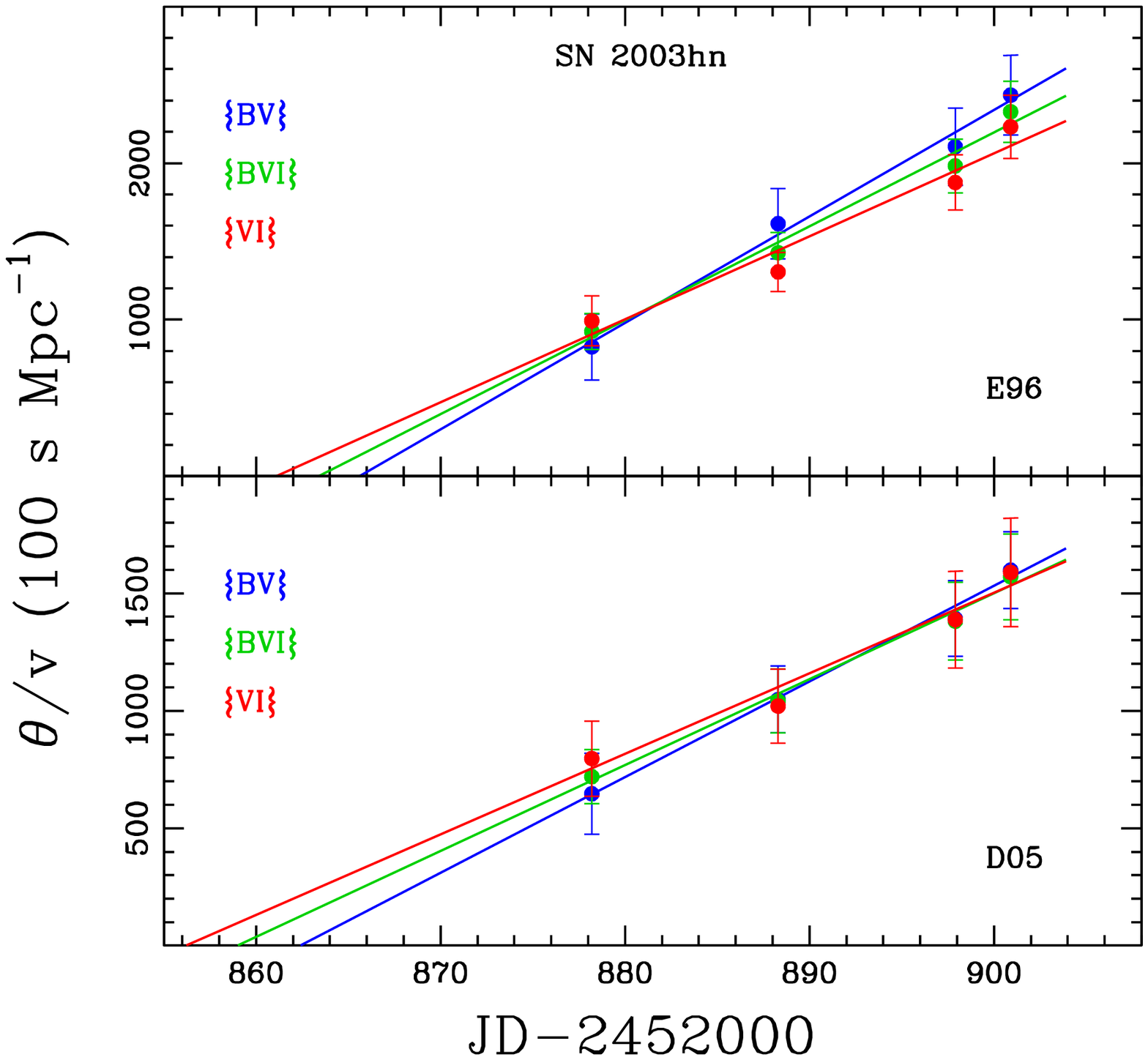}
\caption{$\theta/v$ as a function of time for SN 2003hn using the $\{BV\}$, $\{BVI\}$ and $\{VI\}$
filter subsets. The ridge lines correspond to unweighted least-squares fits to the
derived EPM quantities. The upper and lower panel shows the results using {\it E96} and {\it D05} dilution factors respectively. In all cases we employ the {\it DES} reddening.
~\label{fig_SN03hn_EPM}}
\end{figure}

\begin{figure}
\plotone{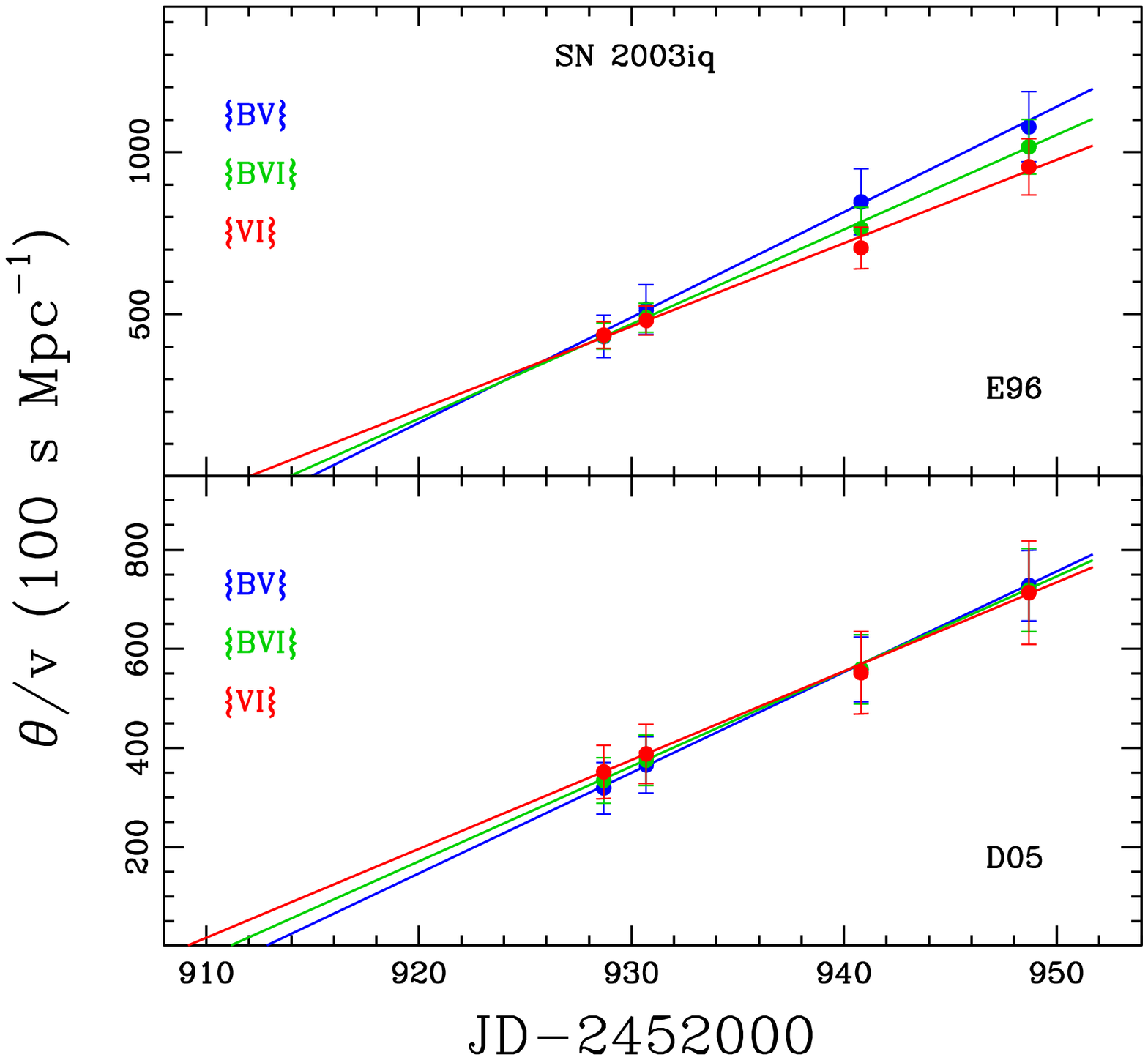}
\caption{$\theta/v$ as a function of time for SN 2003iq using the $\{BV\}$, $\{BVI\}$ and $\{VI\}$
filter subsets. The ridge lines correspond to unweighted least-squares fits to the
derived EPM quantities. The upper and lower panel shows the results using {\it E96} and {\it D05} dilution factors respectively. In all cases we employ the {\it DES} reddening.
~\label{fig_SN03iq_EPM}}
\end{figure}

In this section we present the EPM analysis for 12 SNe IIP (11 from our database and one 
from the literature) with early spectroscopic and photometric observations. 
We carried out the analysis using three different filter subsets
(\{BV\}, \{BVI\}, \{VI\}), two sets of host galaxy extinctions ({\it OLI}, {\it DES}) and 
two atmosphere models ({\it E96}, {\it D05}), which yields a total of 12 solutions for each SN. 
The tables that summarize the EPM 
quantities are available in electronic format for all 144 cases.
%As we shall see in $\S$ 3.7, the lowest dispersions in the Hubble diagrams are obtained
%with the {\it DES} reddening, therefore for brevity in the remainder of this section we will use 
%only this reddening. Figures \ref{fig_SN92ba_EPM}-\ref{fig_SN03iq_EPM} show the EPM analysis 
%for the six {\it DES} cases. \newline \indent
In the remainder of this section we restrict the presentation to the 6 solutions that use 
the {\it DES} extinction because they give the lowest dispersion in the Hubble diagrams. Figures 
\ref{fig_SN92ba_EPM}-\ref{fig_SN03iq_EPM} show these 6 solutions for each of the 12 SNe.
In the following, we provide the EPM distance $D$ and the explosion date $t_0$ and 
their uncertainties, using {\it DES} and the \{VI\} filter subset, and we compare the time of 
explosion to the range restricted by pre-SN images of the host galaxies. These results are 
also summarized in Table \ref{tab_VI_DES.EPM}. In order to obtain a more realistic 
estimation of the error in the distance and the explosion date, we computed 100 Monte
Carlo simulations for each SN, in which we varied all the parameters involved in the EPM
(see Table \ref{tab_errors}), and we averaged the 100 distances and explosion dates to 
derive the EPM $D$ and $t_0$. This produces small differences between the results 
computed from the initial single EPM solution and that obtained from the 100 Monte Carlo 
simulations, but the latter provides a much more realistic estimate of the uncertainties.  
Finally, in Tables \ref{tab_SN92ba_EPM}-\ref{tab_SN03iq_EPM} we reproduce the results 
(computing the 100 Monte Carlo simulations) for each SN using the specific $\{VI\}$, {\it DES} 
and {\it D05} combination, which leads to the lowest dispersion in the Hubble diagrams among all 
12 possible combinations. \newline
 
%As we shall see in section 3.7, the lowest dispersions in the Hubble diagrams are obtained
%with the {\it DES} reddening. For brevity in the remainder of this section we show the explicit
%analysis for the six DES cases. In Tables \ref{tab_SN92ba_EPM}-\ref{tab_SN03iq_EPM} 
%we reproduce the results for each SN using the $\{VI\}$, {\it DES} and {\it D05} combination. 
%\newline \indent
%For each SN we provide the distance $d$ and the explosion date $t_0$ derived from the EPM, 
%using the specific case of the $\{VI\}$ filter subset, and we compare the latter to the range 
%restricted by pre-SNe images of the host galaxies.
%We also include the error in the distance and in $t_0$. In order to obtain a realistic 
%estimation of the error in the distance and in the explosion date, we computed 100 Monte Carlo 
%simulations for each SN, in which we varied all the parameters involved in the EPM 
%(see Table \ref{tab_errors}), assuming a gaussian distribution of the errors.
%The resulting distances, explosion times and their errors are tabulated in Table 
%\ref{VI_DES.EPM} for the 12 SNe.

\subsubsection*{SN 1992ba}
Figure \ref{fig_SN92ba_EPM} shows $\theta/v$ versus time for SN 1992ba using the $\{BV\},
\{BVI\}$ and $\{VI\}$ filter subsets and the {\it E96} and {\it D05}. 
Table \ref{tab_SN92ba_EPM} summarizes the EPM quantities derived from the $\{VI\}$ 
filter subset and the {\it D05} models. 
We used 3 epochs (JD 2448896.9 - 2448922.8) to compute the distance to this SN. 
In order to use the velocities measured on JD 2448896.9 and 24448900.9 we had to 
extrapolate the $I$ band photometry until JD 2448896.9. \newline \indent
SN 1992ba was discovered by \citet{Eva92} on JD 2448896.3. \citet{McN92}  
reported that the SN was not present on a plate taken on JD 2448883.2 with 
limiting magnitude 19. 
The EPM solution yields $t_{0}$ = 2448883.9 $\pm$ 3.0 using the {\it E96} models and 
$t_{0}$ = 2448879.8 $\pm$ 5.6 with {\it D05}. 
These results  agree (within one $\sigma$) with the explosion date constrained by the pre and 
post explosion observations. The distances derived to SN 1992ba are $D$ = 16.4 $\pm$ 2.5 Mpc and 
$D$ = 27.2 $\pm$ 6.5 Mpc using the {\it E96} and the {\it D05} dilution factors, respectively.

\subsubsection*{SN 1999br}

Figure \ref{fig_SN99br_EPM} shows $\theta/v$ versus time for SN 1999br using the $\{BV\},
\{BVI\}$ and $\{VI\}$ filter subsets and the {\it D05} models. 
Table \ref{tab_SN99br_EPM} summarizes the EPM quantities from the $\{VI\}$ filter subset 
and the {\it D05} models. 
We used 5 epochs (JD 2451291.7 - 2451309.7) to compute the distance to this SN. The 
EPM solution shows some departure from linearity using the \{BV\} and \{BVI\} filter subsets. 
SN 1999br presents very low expantion velocities, therefore we were unable to obtain its 
distance using the {\it E96} models. This is because the photospheric velocity conversion
factor $V_{H\beta}/V_{phot}$ is not defined at low expansion velocities (see $\S$ 3.4 
and Figure \ref{fig_models_VHbeta_Vphot}). 
The EPM solution yields $t_{0}$ = 2451275.6 $\pm$ 7.7 using the {\it D05} models. 
This  result compare very well with the observations, because SN 1999br was discovered by 
the Lick Observatory Supernova Search (LOSS) on JD 2451280.9 \citep{Kin99}. 
An image taken on JD 2451264.9 showed nothing at the SN position at a limiting magnitudes 
of 18.5 \citep{Li99a}. The EPM distance to SN 1999br is $D$ = 39.5 $\pm$ 13.5 Mpc using the 
{\it D05} dilution factors.                 

\subsubsection*{SN 1999em}

SN 1999em is the best ever observed SN IIP. Many photometric and spectroscopic observations 
were made by different observers during the plateau phase. 
Figure \ref{fig_SN99em_EPM} shows $\theta/v$ versus time for the SN 1999em using the $\{BV\},
\{BVI\}$ and $\{VI\}$ filter subsets and the {\it E96} and {\it D05} models.  
Table \ref{tab_SN99em_EPM} summarizes the EPM quantities derived from the $\{VI\}$ filter 
subset. We used 25 epochs (JD 2451482.8 - 2451514.8) to derive the distance to SN 1999em. 
Four spectra were taken from \citet{Ham01} and the other 21 from \citet{Leo02}.
In some cases there were two spectra taken at the same epoch from both sources.
In those cases we used them individuallly in the EPM solution instead of averaging the 
measured velocities from each spectrum. We removed the first spectrum (JD 2451481.8) from the 
EPM solution because it shows a clear departure from the linear $\theta/v$ versus $t$ 
relation. 
The EPM solutions using {\it E96} and {\it D05} are quite linear and show great detail in 
the evolution of $\theta/v$ due to the high quality spectroscopic and photometric coverage.
However, the {\it E96} solution shows a small departure from linearity in the last two 
spectroscopic epochs. This effect is probably due to the high rise in the $V_{H\beta}
/V_{phot}$ ratio at low velocities in the {\it E96} models. \newline \indent
SN 1999em was discovered on JD 2451480.9 by the LOSS program \citep{Li99b}. An image taken
at the position of the SN on JD 2451472.0 showed nothing at a limiting magnitude of 19.0.
The EPM yields $t_{0}$ = 2451476.3 $\pm$ 1.1 and $t_{0}$ = 2451474.0 $\pm$ 2.0 using the 
{\it E96} and {\it D05} models.
These explosions dates are between the pre-discovery and the discovery
date. The distances derived to SN 1999em are $D$ = 9.3 $\pm$ 0.5 Mpc from {\it E96} and 
$D$ = 13.9 $\pm$ 1.4 Mpc from {\it D05}. 

%The $D05$ solution agrees within the error bars with the cepheid distance derived 
%to SN 1999em host galaxy (NGC 1637) of 11.7 $\pm$ 1.0 Mpc \citep{Leo03}. The {\it E96} solution
%subestimates the Cepheids distance by about $\sim 20\%$.
%Our distance estimates can be compare with the EPM distance derived from
%other authors \citep{Ham01,Leo02,D06}. Also, \citet{Bar04}, has applied a similar
%technique (SEAM) to derive the distance to SN 1999em. \newline \indent

\subsubsection*{SN 1999gi}

Figure \ref{fig_SN99gi_EPM} shows $\theta/v$ versus time for SN 1999gi using the $\{BV\},
\{BVI\}$ and $\{VI\}$ filter subsets and the {\it E96} and {\it D05} models. 
Table \ref{tab_SN99gi_EPM} summarizes the EPM quantities derived from the $\{VI\}$ filter 
subset.
We used 5 epochs (JD 2451525.0 - 2451556.9) to apply the EPM method. All the spectra and
the photometry were taken from \citet{Leo02b}. The first spcetrum (JD 2451522.9) was remove
from the EPM solutions because it yields an H$\beta$ velocity of $\sim 26.000$ $km$ $s^{-1}$, 
well above the range of the photospheric velocity conversion (see $\S$ 3.4 and Figure 
\ref{fig_models_VHbeta_Vphot}). 
The explosion dates of SN 1999gi obtained using the EPM are $t_{0}$ = 2451517.0 $\pm$ 1.2 
using {\it E96} models and $t_{0}$ = 2451515.6 $\pm$ 2.4 with {\it D05}. 
These results agreed with the observations because a pre-discovery image taken on JD 2451515.7 
\citep{Tr99} showed nothing at the SN position (limiting unfiltered magnitude of 18.5).
SN 1999gi was discovered on JD 2451522.3 \citep{Na99} on unfiltered CCD frames, so the 
explosion date can be constrained in a range of only 6.6 days.
We derive a distance of $D$ = 11.7 $\pm$ 0.8 and $D$ = 17.4 $\pm$ 2.3 Mpc using the {\it E96} 
and {\it D05} models, respectively.
%Our distance and explosion date can be compared with the EPM results derived by 
%\citet{Leo02b}.

\subsubsection*{SN 2002gw}

Figure \ref{fig_SN02gw_EPM} shows $\theta/v$ versus time for SN 2002gw using the $\{BV\},
\{BVI\}$ and $\{VI\}$ filter subsets and the {\it E96} and {\it D05} models.                     
Table \ref{tab_SN02gw_EPM} summarizes the EPM quantities from the $\{VI\}$ filter subset.
The EPM solutions were obtained using 6 epochs (JD 2452573.1 - 2452590.7). 
The EPM yields explosion times of $t_{0}$ = 2452557.9 $\pm$ 2.7 and $t_{0}$ = 2452551.7 $\pm$ 
7.6 (using {\it E96} and {\it D05} dilution factors, respectively). SN 2002gw was discovered on 
JD 2452560.8 \citep{Mon02}. An image taken on JD 2452529.6 shows nothing at the SN position 
at a limiting magnitude of 18.5. Also, an unfiltered CCD image taken on JD 2452559.1 shows 
the SN at magnitude 18.3 \citep{Ita02}. The EPM explosion dates are in agreement with the
SN explosion date constrained by the observations. The EPM distances are $D$ = 37.4 $\pm$ 4.9 
Mpc and $D$ = 63.9 $\pm$ 17.0 Mpc using {\it E96} and {\it D05}, respectively.

\subsubsection*{SN 2003T}

Figure \ref{fig_SN03T_EPM} shows $\theta/v$ versus time for SN 2003T using the $\{BV\},
\{BVI\}$ and $\{VI\}$ filter subsets and the {\it E96} and {\it D05} models.                     
Table \ref{tab_SN03T_EPM} summarizes the EPM quantities from the $\{VI\}$ filter subset.
The EPM explosion dates are $t_{0}$ = 2452654.2 $\pm$ using {\it E96} models and $t_{0}$ = 
2452648.9 $\pm$ 3.4 with {\it D05}. In both cases the third epoch used to derive the distance is 
beyond $\sim$ 45 ays after the EPM $t_{0}$, but it proves neccesary to include it to compute 
the EPM analysis. This SN was discovered by LOTOSS on JD 2452664.9 \citep{Sch03}. An image
taken on JD 2452644.9 shows nothing at a limiting magnitude of 19.0, in good agreement 
with the EPM analysis. The EPM distances are $D$ = 87.8 $\pm$ 13.5 Mpc using {\it E96} and 
$D$ = 147.3 $\pm$ 35.7 Mpc with {\it D05}.

\subsubsection*{SN 2003bl}

Figure \ref{fig_SN03bl_EPM} shows $\theta/v$ versus time for SN 2003bl using the $\{BV\},
\{BVI\}$ and $\{VI\}$ filter subsets and {\it D05} models.
Table \ref{tab_SN03bl_EPM} summarizes the EPM quantities derived for SN 2003bl from 
the $\{VI\}$ filter subset.
The EPM solutions were obtained using 4 epochs (JD 2452701.8 -2452735.8). 
As with the SN 1999br, we were unable to apply the EPM using {\it E96} because we only had two 
spectra with velocities higher than $4500$ $km$ $s^{-1}$, and so the photospheric velocity 
correction could not be applied (see $\S$ 3.4 and Figure \ref{fig_models_VHbeta_Vphot}).          
SN 2003bl was discovered by LOTOSS on JD 2452701.0 \citep{Swi03}. A pre-discovery image
taken on JD 2452438.8 shows nothing at the SN position at a limiting magnitud of 19.0.
The EPM yields $t_{0}$ = 2452692.6 $\pm$ 2.8, consistent with the SN discovery date . 
The EPM distance is $D$ = 92.4 $\pm$ 14.2 Mpc.

\subsubsection*{SN 2003bn}

Figure \ref{fig_SN03bn_EPM} shows $\theta/v$ versus time for SN 2003bn using the $\{BV\},
\{BVI\}$ and $\{VI\}$ filter subsets and the {\it E96} and {\it D05} models.                      
Table \ref{tab_SN03bn_EPM} summarizes the EPM quantities from the $\{VI\}$ filter subset.
We computed the EPM analisys using 3 epochs (JD 2452706.6 - 2452729.7). 
The EPM yields explosions dates of $t_{0}$ = 2452693.4 $\pm$ 2.7 and $t_{0}$ = 2452687.0
$\pm$ 9.0 from {\it E96} and {\it D05}, respectively. SN 2003bn was discovered on JD 2452698.0 
\citep{Wood03}. Two pre-discovery NEAT images shows nothing at the SN position on JD 
2452691.5 (limiting magnitude of 21.0) and the SN at a magnitude of 20.2 on JD 2452692.8, 
which restricted the explosion date in a range of only 1.3 days. This value for $t_0$ is in
agreement within one $\sigma$ with the EPM $t_0$ derived using {\it E96} and {\it D05}. 
The EPM distances from {\it E96} and {\it D05} are $D$ = 50.2 $\pm$ 7.0 Mpc and $D$ = 87.2 
$\pm$ 28.0 Mpc, respectively.

\subsubsection*{SN 2003ef}

Figure \ref{fig_SN03ef_EPM} shows $\theta/v$ versus time for SN 2003ef using the $\{BV\},
\{BVI\}$ and $\{VI\}$ filter subsets and the {\it E96} and {\it D05} models.                     
Table \ref{tab_SN03ef_EPM} summarizes the EPM quantities from the $\{VI\}$ filter subset.
We computed the EPM analysis using 4 epochs (JD 2452780.7 -2452797.6). 
The explosion date derived are $t_{0}$ = 2452759.8 $\pm$ 4.7 and $t_{0}$ = 2452748.4 $\pm$ 
15.6 with {\it E96} and {\it D05}, respectively. SN 2003ef was discovery by the LOTOSS on 
JD 2452770.8 (mag. about 16.3) \citep{Wei03}, consistent with the EPM $t_{0}$. A KAIT image 
taken on JD 2452720.8 showed nothing at the SN position at a limiting magnitude of 18.5. 
The EPM distances are $D$ = 38.7 $\pm$ 6.53 Mpc with {\it E96} and $D$ = 74.4 $\pm$ 30.3 Mpc with {\it D05}.
 
\subsubsection*{SN 2003hl}

Figure \ref{fig_SN03hl_EPM} shows $\theta/v$ versus time for SN 2003hl using the $\{BV\},
\{BVI\}$ and $\{VI\}$ filter subsets and the {\it E96} and {\it D05} models.                       
Table \ref{tab_SN03hl_EPM} summarizes the EPM quantities derived from the $\{VI\}$ filter 
subset. 
The EPM solutions were obtained using 3 epochs (JD 2452879.9 -2452908.7).
We estimated the explosion dates on $t_{0}$ = 2452872.3 $\pm$ 1.7 and $t_{0}$ = 2452865.4
$\pm$ 5.9 using {\it E96} and {\it D05}, respectively. 
SN 2003hl was discovered on JD 2452872.0 during the LOTOSS program at a magnitude of 
16.5 \citep{Moo03}. A pre-discovery KAIT image taken 
on JD 2452863.0 shows nothing at the SN position at a limiting magnitude of 19.0. This image
restricts the explosion date in a range of 9 days. The EPM explosion dates are in agreement with 
the observations (within one $\sigma$). We derived EPM distances of $D$ = 17.7 $\pm$ 2.1 Mpc with 
{\it E96} and $D$ = 30.3 $\pm$ 6.3 Mpc with {\it D05}. 
%We derived EPM distances of D = 17.7 $\pm$ 2.1 with {\it E96} and D = 30.3 $\pm$ 6.3 with D05.
%The derived distances disagrees with the EPM distance derived for SN 2003iq, which explodes
%in the same galaxy. We think that this SN could be intrinsically red, which leads to 
%an overestimation of the reddening, and therefore a subestimation in its distance. 
%This effect is still presents when we used the OLI reddening. In that case, the discrepancy 
%is bigger.                          
%We estimated the explosion dates on $t_{0}$ = 2452872.3 $\pm$ 1.7 and $t_{0}$ = 2452865.4 
%$\pm$ 5.9. \newline \indent
%SN 2003hl was discovered on JD 2452872.0 during the LOTOSS program at a magnitude of 
%16.5 \citep{Moo03}.

\subsubsection*{SN 2003hn}

Figure \ref{fig_SN03hn_EPM} shows $\theta/v$ versus time for SN 2003hn using the $\{BV\},
\{BVI\}$ and $\{VI\}$ filter subsets and the {\it E96} and {\it D05} models. 
Table \ref{tab_SN03hn_EPM} summarizes the EPM quantities from the $\{VI\}$ filter subset. 
The EPM solutions were obtained using 4 epochs (JD 2452878.2 - 2452900.9).
The EPM explosion dates derived are $t_{0}$ = 2452859.5 $\pm$ 3.8 and $t_{0}$ = 2452853.8 
$\pm$ 9.3 using the {\it E96} and {\it D05} dilution factors, respectively.
This SN was discovered on JD 2452877.2 at mag. 14.1 by \citet{Eva03}. Evans also reported
that the SN was not visible at mag. 15.5 on JD 2452856.5. This date agrees with the explosion 
date derived from {\it E96} and is less than one $\sigma$ lower than that derived from {\it D05}.
The EPM solutions leads to \mbox{$D$ = 16.9 $\pm$ 2.2} Mpc and $D$ = 26.3 $\pm$ 7.1 Mpc using {\it E96} 
and {\it D05}, respectively. 

\subsubsection*{SN 2003iq}

Figure \ref{fig_SN03iq_EPM} shows $\theta/v$ versus time for SN 2003iq using the $\{BV\},
\{BVI\}$ and $\{VI\}$ filter subsets and the {\it E96} and {\it D05} models.
Table \ref{tab_SN03iq_EPM} summarizes the EPM quantities from the $\{VI\}$ filter subset. 
The EPM solutions were obtained using 4 epochs (JD 2452928.7 - 2452948.7).
This SN was discovered by \citet{LLa03} on JD 2452921.5, while monitoring SN 2003hl in
the same host galaxy. A pre-discovery image taken on 2452918.5 shows nothing at the SN
position. These reports constrain the explosion date to a range of only three days.
The EPM yields $t_{0}$ = 2452909.6 $\pm$ 4.3 using {\it E96} and $t_{0}$ = 2452905.6
$\pm$ 9.5 using {\it D05}. In both cases the explosion date is far earlier than expected
because the SN was not present on JD 2452918.5. This implies that the EPM solutions
to this SN are not satisfactory. 
We derived EPM distances of $D$ = 36.0 $\pm$ 5.6 Mpc with $E96$ and D = 53.3 $\pm$ 17.1 
Mpc with {\it D05}.

%We derived EPM distances of D = 36.0 $\pm$ 5.6 with $E96$ and D = 53.3 $\pm$ 17.1 with $D05$.
%Both distances derived are far larger than the distances derived to the host galaxy using
%SN 2003hl (see subsection SN 2003hl). \newline \indent

\section{Discussion}

%\subsubsection{The distance to NGC 0772}

%NGC 0772 is of particular interest because hosted two SNe studied in this work, SN 200hl 
%and SN 2003iq. This case offers a great oportunity to prove if the method is internally
%consistent. We would expected that both SNe lead to a similar distance, however this is not
%the case. As can be seen in the 
         
\subsection{External comparison}

\begin{itemize}
\item {\bf Previous EPM distances.} 

The EPM method has been already applied to SN 1999em by other authors. \citet{Ham01} employed
the {\it E96} dilution factors and eight different filter subsets to perform the EPM analysis to 
this SN. They used a cross-correlation technique to estimate the photospheric velocity and
adopted a host galaxy extinction of $A_V=0.18$.
They derived a distance of 6.9 $\pm$ 0.1, 7.4 $\pm$ 0.1 and 7.3 $\pm$ 0.1 Mpc from the
$\{BV\}$, $\{BVI\}$ and $\{VI\}$ filter subsets, respectively. These values are in agreement
with our estimates of 6.9 $\pm$ 0.6, 7.5 $\pm$ 0.6 and 9.3 $\pm$ 0.5 Mpc (from the 
$\{BV\}$, $\{BVI\}$ and $\{VI\}$ filter subsets, respectively), except in the $\{VI\}$ case. 
Also \citet{Leo02} employed the {\it E96} models to derive the distance to SN 1999em. They used
four weak unblended spectral features (Fe {\sc ii} 4629, 5276, 5318 and Sc {\sc ii} 
$\lambda$4670) as the photospheric velocity indicators.                                  
They adopted a host galaxy reddening of $A_V=0.31$, the same value predicted by {\it DES}.        
They derived a distance of 7.7 $\pm$ 0.2, 8.3 $\pm$ 0.2 and 8.8 $\pm$ 0.3 Mpc from the 
$\{BV\}$, $\{BVI\}$ and $\{VI\}$ filter subsets, respectively. These results are in agreement
with our {\it E96} distances.
Finally, \citet{D06} applied the EPM method to SN 1999em using {\it E96} and {\it D05}. 
They adopted the SN 1999em {\it DES} reddening value of $A_V=0.31$.
Using the {\it E96} models, they derived a distance of 8.6 $\pm$ 0.8, 9.7 $\pm$ 1.0 and 11.7 
$\pm$ 1.5 Mpc from the $\{BV\}$, $\{BVI\}$ and $\{VI\}$ filter subsets, respectively, which
are somewhat greater than our distances. 
Using the {\it D05} models they derived a distance of 13.5 $\pm$ 1.5, 12.5 $\pm$ 1.6 and 14.6 $\pm$ 
1.9 Mpc from the $\{BV\}$, $\{BVI\}$ and $\{VI\}$ filter subsets, respectively, which are 
significantly larger than our values of 11.2 $\pm$ 0.2, 12.0 $\pm$ 0.2 and 14.0 $\pm$ 0.2 Mpc, 
respectively. This is probably due to a different implementation of the EPM. 

\item {\bf SEAM distance}   
The Spectral-fitting Expanding Atmosphere Method (SEAM) is a similar technique to the EPM, but 
it avoids the use of dilution factors and includes the synthetic spectral fitting to the observed 
spectra of the SN. \citet{Bar04} applied this method to SN 1999em. They derived a distance of 
$D = 12.5 \pm 2.3$ Mpc, in good agreement with our distances derived using the {\it D05} models
(11.2 $\pm$ 0.2, 12.0 $\pm$ 0.2 and 14.0 $\pm$ 0.2 Mpc from the $\{BV\}$, $\{BVI\}$ and $\{VI\}$ 
filter subsets, respectively), but significantly greater than the EPM distances derived using 
{\it E96} (6.9 $\pm$ 0.6, 7.5 $\pm$ 0.6 and 9.3 $\pm$ 0.5 Mpc from the $\{BV\}$, $\{BVI\}$ and 
$\{VI\}$ filter subsets, respectively).

\item {\bf Cepheid distance}

\citet{Leo03} identified 41 Cepheid variable stars in NGC 1637, the host galaxy of SN 1999em.
They derived a Cepheid distance to NGC 1637 of $D=11.7\pm1.0$ Mpc. As with the SEAM results, the
Cepheid distance is consistent with our EPM distances derived using the {\it D05} models (11.2 $\pm$ 
0.2, 12.0 $\pm$ 0.2 and 14.0 $\pm$ 0.2 Mpc from the $\{BV\}$, $\{BVI\}$ and $\{VI\}$ filter subsets, 
respectively). In all cases, the {\it E96} models lead to significantly lower distances (6.9 $\pm$ 
0.6, 7.5 $\pm$ 0.6 and 9.3 $\pm$ 0.5 Mpc from the $\{BV\}$, $\{BVI\}$ and $\{VI\}$ filter subsets, 
respectively).
\end{itemize}

\subsection{Error analysis}

\subsubsection{Effects of reddening}

\begin{figure}
\plotone{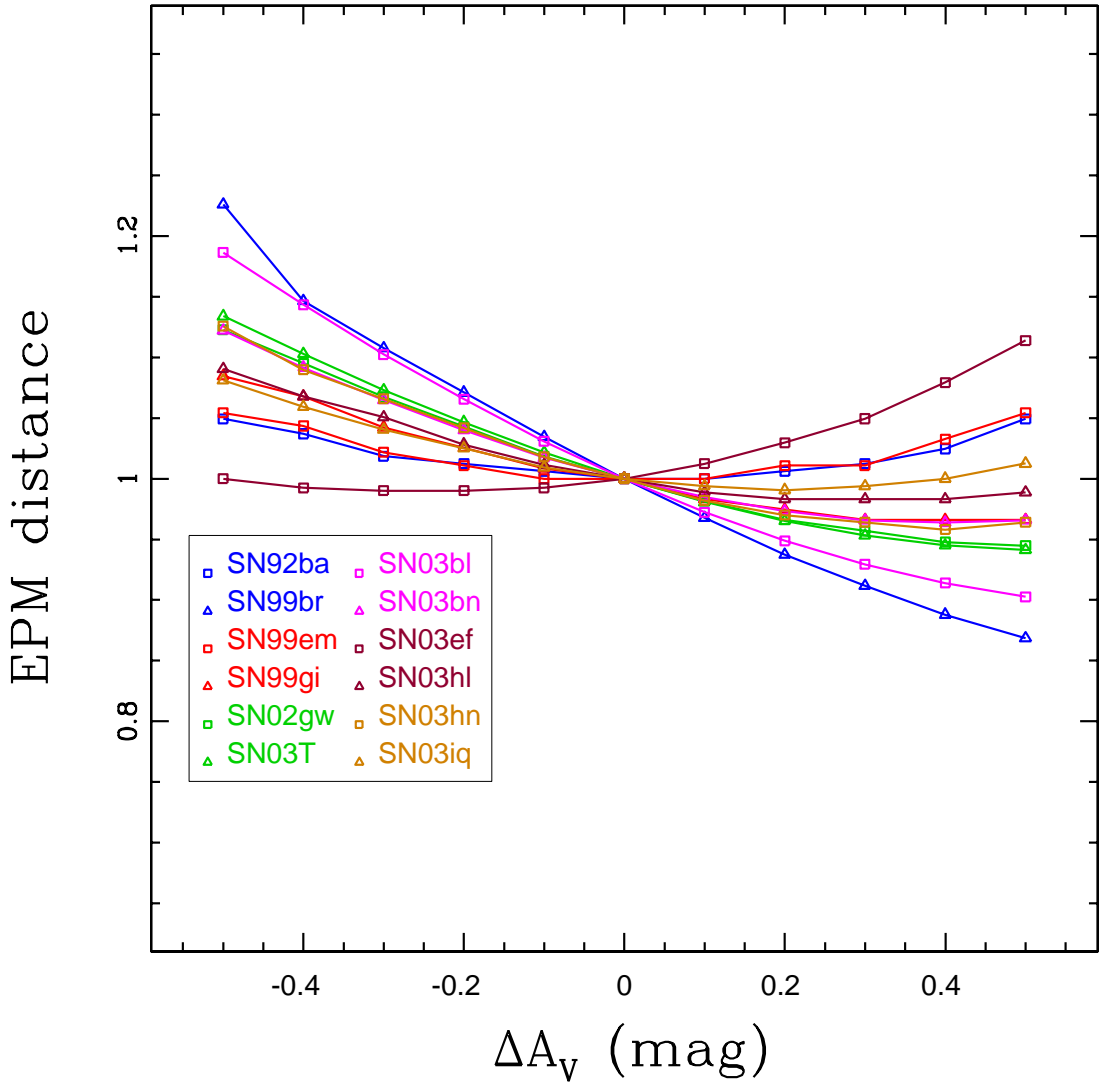}
\caption{Normalized EPM distances as a function of the host galaxy visual extinction relative
to the {\it DES} value ($\Delta A_V = 0$).
 ~\label{fig_delta_reddening}}
\end{figure}

While the \citet{Sch98} IR maps provide a precise estimate of the amount of Galactic 
foreground extinction, the determination of host galaxy extinction is a more challenging 
task. This is a potential problem because the distances derived using EPM depend on the 
adopted host galaxy extinction. In order to investigate the sensitivity of the distances 
to dust extinction, we performed the EPM analysis to all the SNe in our sample using the $\{VI\}$ 
filter subset by varying the amount of host galaxy visual extinction $A_V$ in steps of $\Delta A_V=0.1$ 
{\it mag}. Figure \ref{fig_delta_reddening} shows the normalized EPM distances as a function 
of host galaxy visual extinction $A_V$ relative to the {\it DES} value ($\Delta A_V=0$). As can be
seen, the EPM is quite insensitive to the amount of host galaxy extinction adopted. On average, the 
distances change by less than $\sim 10 \%$ from $\Delta A_V$ = 0.0 to $\Delta A_V$ = 0.5 and by less 
than $\sim 20 \%$ going from $\Delta A_V$ = 0.0 to $\Delta A_V = -0.5$. 
%However, this effect is 
%because in seven cases the resulting $A_V$ (the {\it DES} value minus 0.5) were negative, which 
%leads to an overestimation in the distance of $\sim$ $20\%$, as in the SN 1999br and SN 2003bl 
%cases (see Table \ref{tab_reddening} for the {\it DES} values).  
Therefore, even a systematic error of 0.5 in $A_V$, produces a small error in the EPM distance.

\subsubsection{Other sources of error}

Table \ref{tab_errors} lists all the error sources in EPM and their typical 
values. In order to investigate which source contributes the most to the uncertainty in                
the EPM distance, we performed the EPM analysis to SN 1999gi (whose photometry and 
spectroscopy coverage is representative of our sample) and we changed the error
of a single source (listed in Table \ref{tab_errors}) leaving all others unchanged. 
We found two main sources of errors.
In the {\it E96} case, the errors in the photospheric velocity conversion and the dilution 
factors have the largest effect in the distance uncertainty, each one contributing $\sim 30\%$ 
of the total error, while in the {\it D05} case the error in the dilution factors produces $\sim 70\%$
of the uncertainty in the distance, far greater than that due to the error in the photospheric 
velocity conversion ($\sim 10\%$ of the total error). All of the other errors 
have a secondary effect in the total error.

\subsection{Hubble Diagrams}

\begin{figure}
\plotone{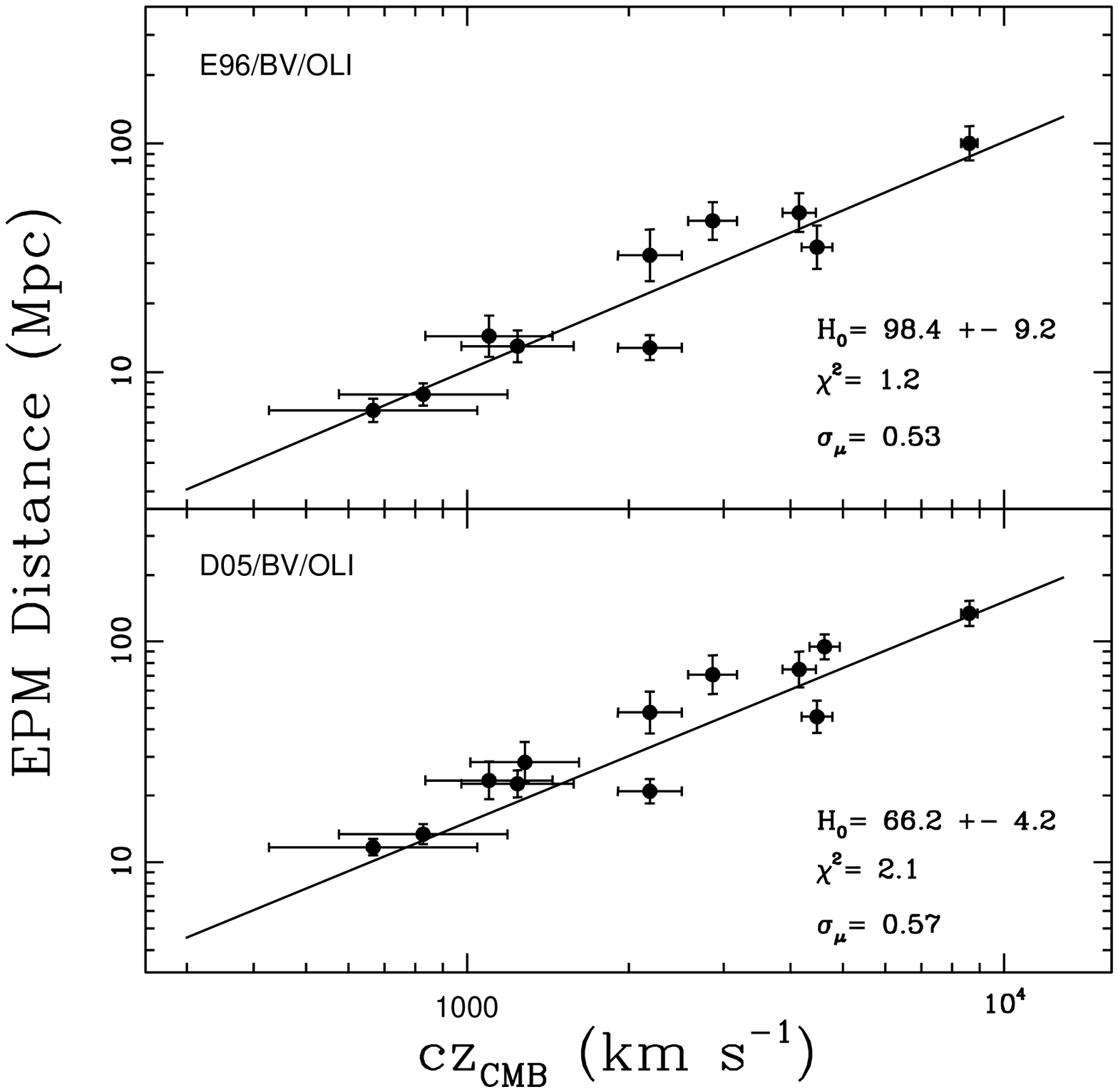}
\caption{Hubble diagram using the \{BV\} filter subset and {\it OLI} reddening. ~\label{fig_bv_OLI.HD}}
\end{figure}

\begin{figure}
\plotone{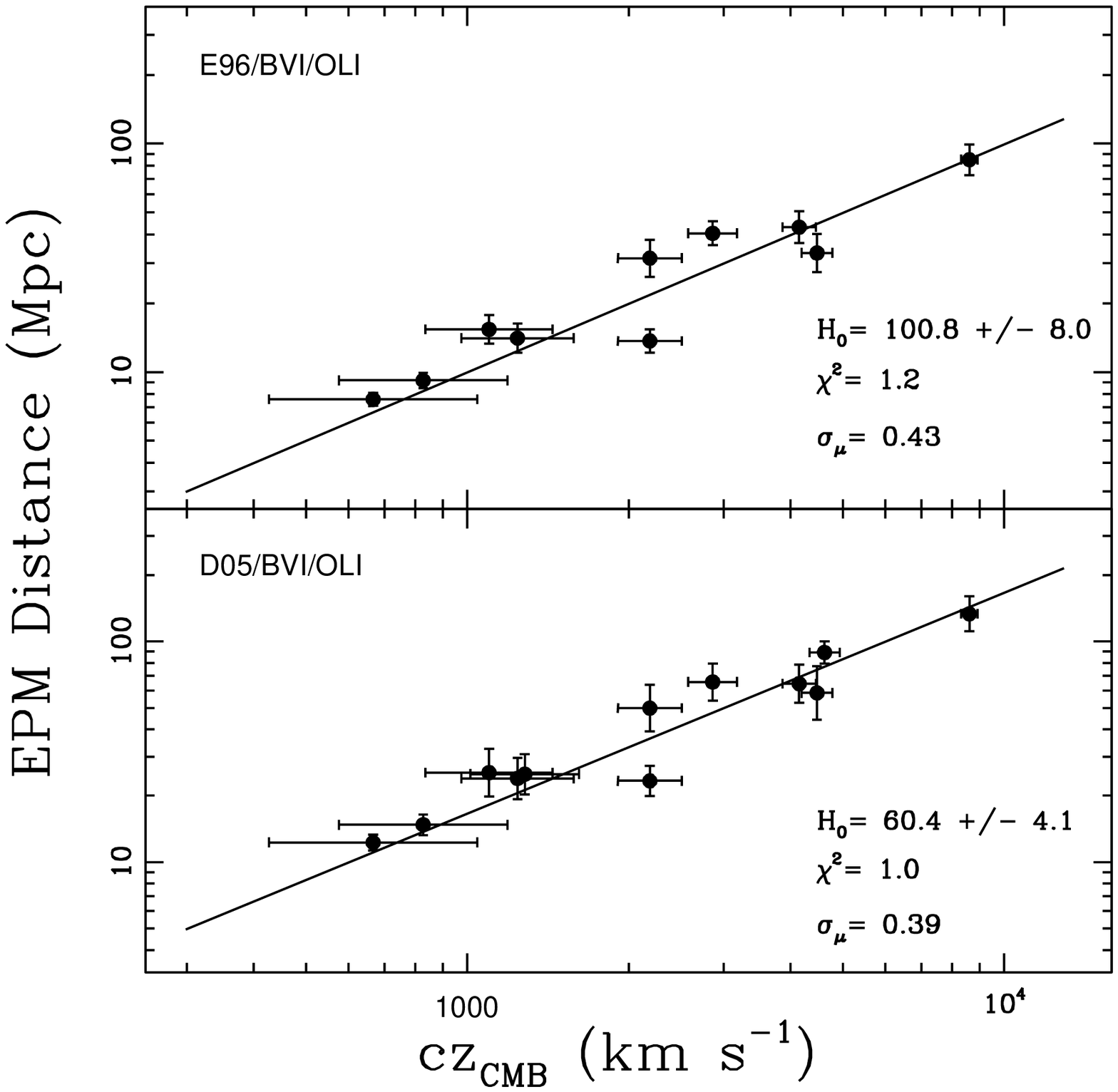}
\caption{Hubble diagram using the \{BVI\} filter subset and {\it OLI} reddening. ~\label{fig_bvi_OLI.HD}}
\end{figure}

\begin{figure}
\plotone{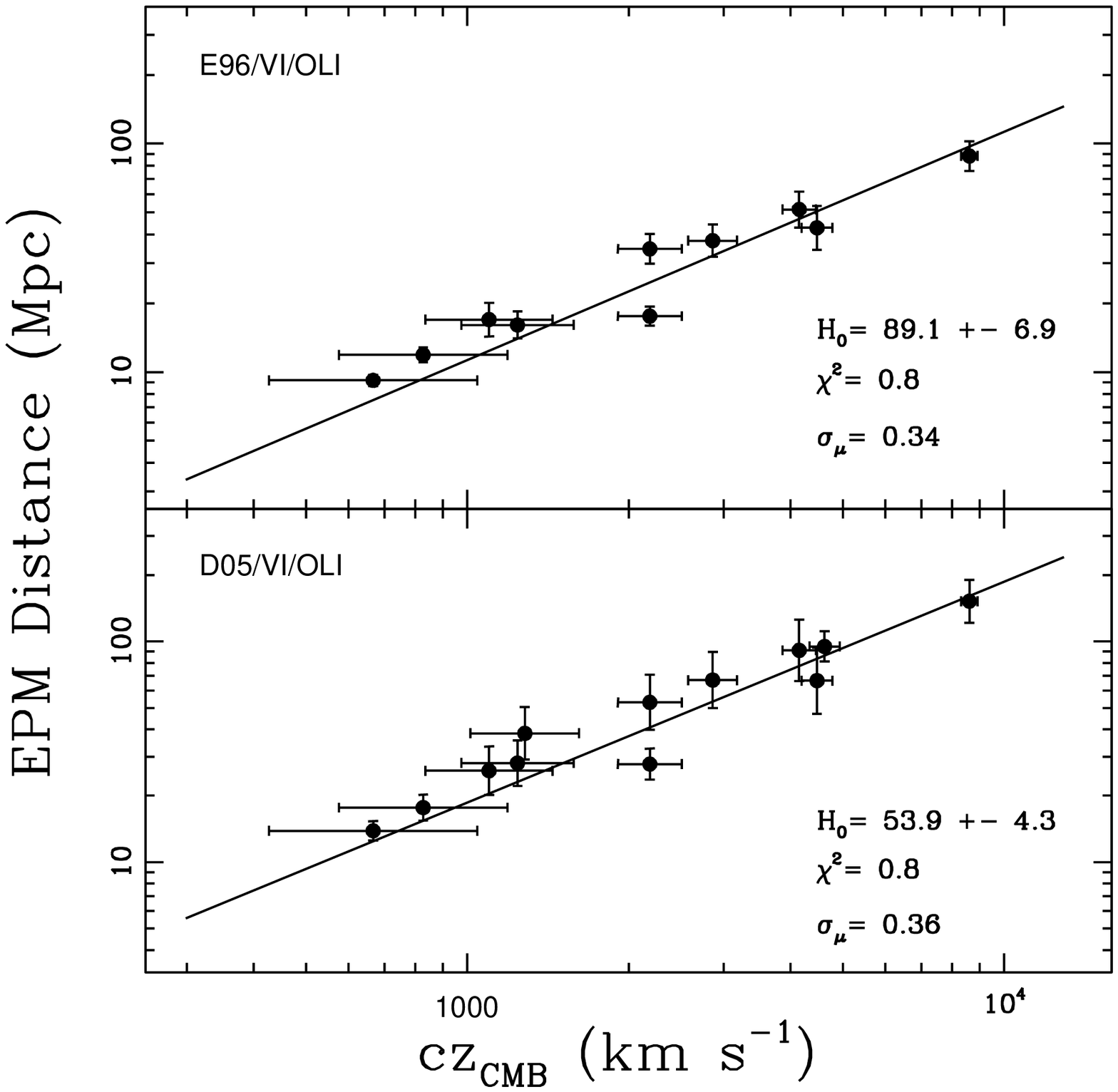}
\caption{Hubble diagram using the \{VI\} filter subset and {\it OLI} reddening. ~\label{fig_vi_OLI.HD}}
\end{figure}

\begin{figure}
\plotone{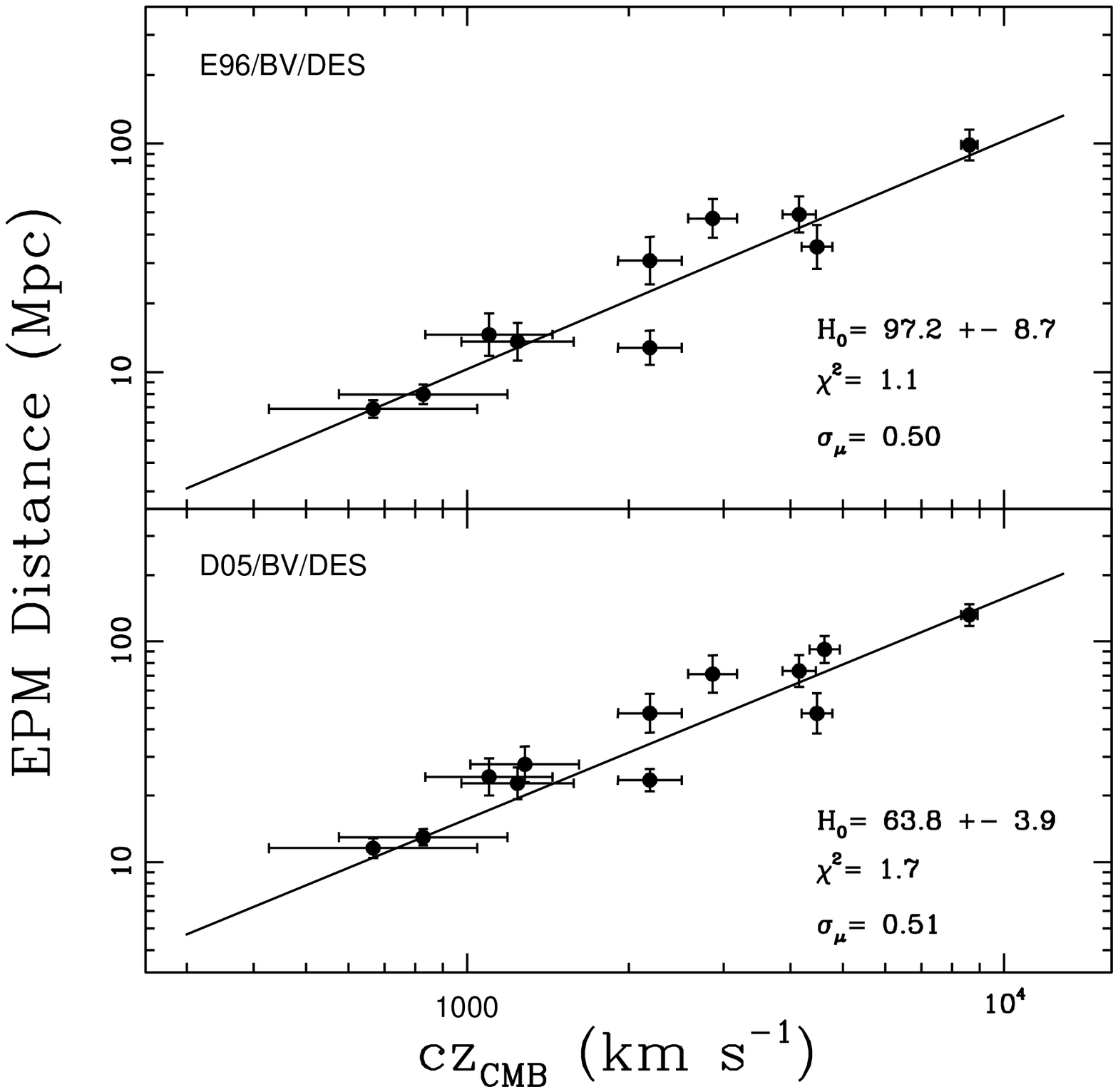}
\caption{Hubble diagram using the \{BV\} filter subset and {\it DES} reddening. ~\label{fig_bv_DES.HD}}
\end{figure}

\begin{figure}
\plotone{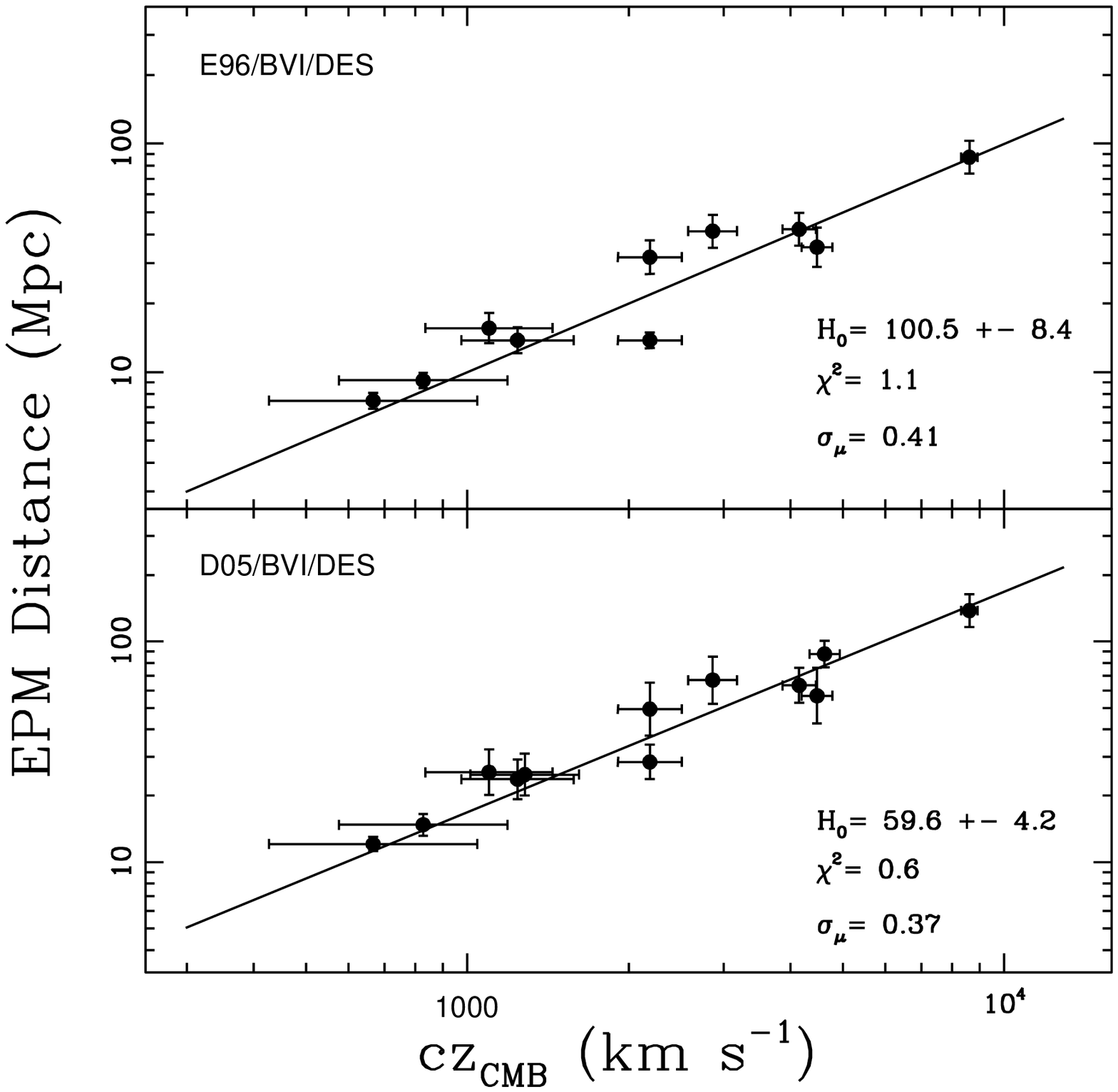}
\caption{Hubble diagram using the \{BVI\} filter subset and {\it DES} reddening. ~\label{fig_bvi_DES.HD}}
\end{figure}

\begin{figure}
\plotone{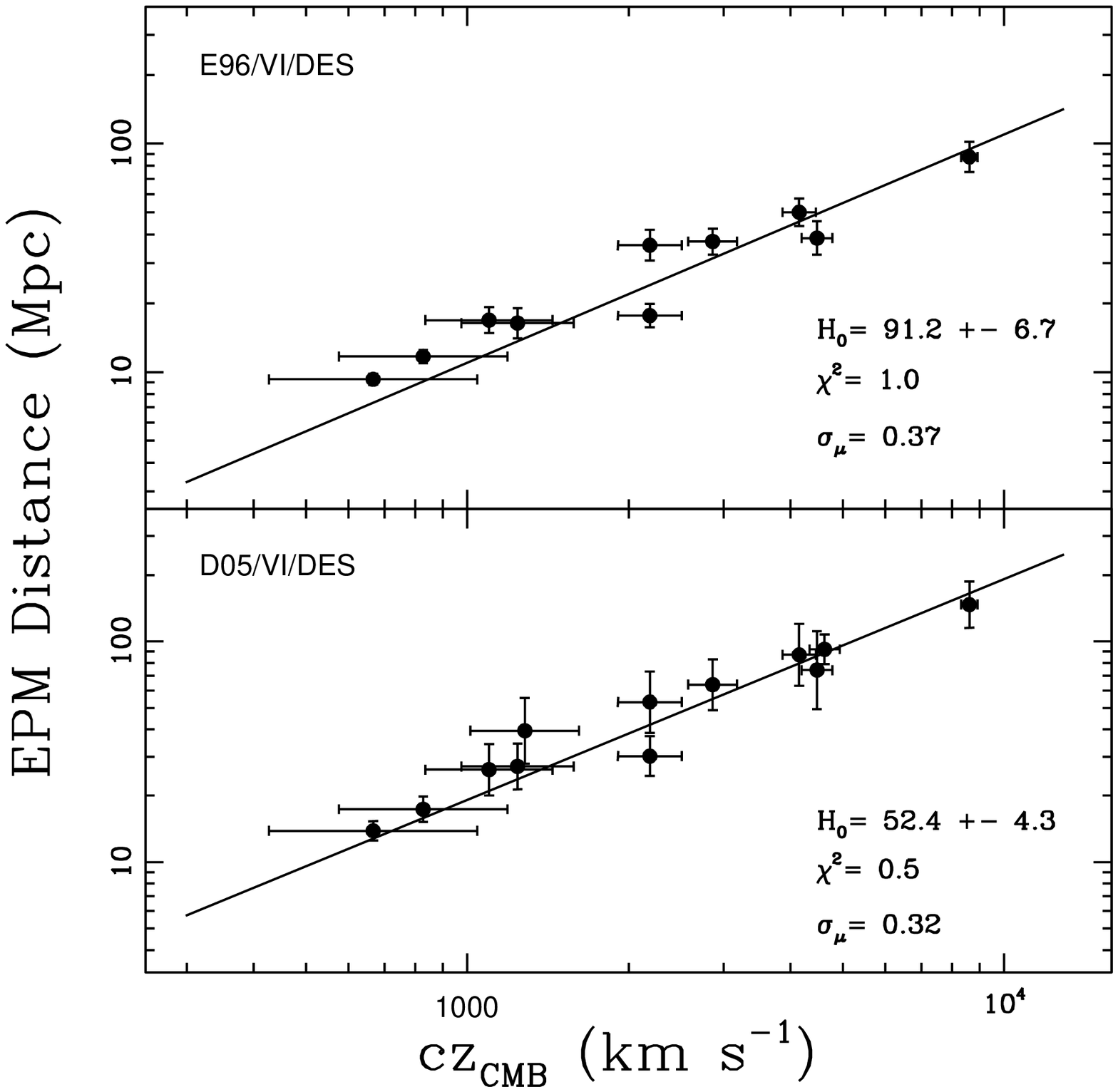}
\caption{Hubble diagram using the \{VI\} filter subset and {\it DES} reddening. ~\label{fig_vi_DES.HD}}
\end{figure}

Since the discovery of the expansion of the Universe \citep{Hub29}, the determination of the 
expansion rate, the Hubble constant ($H_{0}$), has become one of the most important 
challenges in astronomy and cosmology. Using the velocity-distance relation (Hubble diagram)
calibrated using the Cepheid period-luminosity relation, \citep{Hub31} obtained $H_{0}$ $\sim$ 
500 $km$ $s^{-1}$ $Mpc^{-1}$. During the second half of the 20th century, the Cepheid 
relation was significantly improved, and new Hubble diagrams were obtained, yielding  
Hubble constants in the range $\sim$ 50 - 100 $km$ $s^{-1}$ $Mpc^{-1}$. 
Today, the discrepancy is not over, but there is a convergence into a value of 
$H_{0}$ $\sim$ 60 - 75 $km$ $s^{-1}$ $Mpc^{-1}$ \citep{San06,Fre01}. \newline \indent
In this work we applied the EPM method to 12 SNe using two sets of dilution factors 
({\it E96}, {\it D05}), two extinction determination methods ({\it OLI}, {\it DES}) and three 
filter subsets (\{BV\}, \{BVI\} and \{VI\}) to derive their distances. 
In order to obtain the host galaxy redshifts relative to the {\it Cosmic Microwave Background}
(CMB), we corrected the heliocentric host galaxy redshifts for the peculiar velocity of the Sun 
relative to the CMB rest frame. For this purpose we added a velocity vector of 371 $km$ $s^{-1}$ 
in the direction $(l,b)=(264.14^{\circ},48.26^{\circ})$ \citep{Fix96} to the heliocentric 
redshifts. The resulting CMB redshift are given in Table \ref{tab_SN_list}. \newline \indent
Using the CMB host galaxy redshifts we constructed 12 different Hubble diagrams. 
Figures \ref{fig_bv_OLI.HD}-\ref{fig_vi_OLI.HD} show the Hubble 
diagrams obtained with {\it OLI} reddenings, from the \{BV\}, \{BVI\} and \{VI\} filter subsets, 
respectively. Figures \ref{fig_bv_DES.HD}-\ref{fig_vi_DES.HD} show the same diagrams but this time 
using {\it DES} extinctions. Each diagram is labeled with the derived Hubble constant, the reduced 
$\chi^{2}$ and the dispersion in distance modulus $\sigma_{\mu}$ from the linear fit. The resulting
$H_0$ values are summarized in Table \ref{tab_H0_values}.
\newline \indent
There is a systematic difference in the $H_{0}$ values obtained with the {\it E96} and {\it D05} 
models. Using {\it E96} we obtained $H_{0} = 89 - 101$ $km$ $s^{-1}$ $Mpc^{-1}$ while {\it D05}
yielded $H_{0}= 52 - 66$ $km$ $s^{-1}$ $Mpc^{-1}$. This difference arises mainly from      
the systematically higher {\it D05} dilution factors which lead to greater distances, and 
also from the distinct photospheric velocity conversion between both models. The former is currently the 
greatest source of systematic uncertainty in this method. \newline \indent
The use of different filter subsets leads to $H_{0}$ values consistent within 1$\sigma$ 
for a fixed atmosphere model. This is a very important result, because it shows the internal 
consistency of each set of atmosphere models.
However, the use of different filter subsets produces significant differences in 
dispersion, increasing from $\sigma_{\mu}\sim$ 0.3 (\{VI\}) to $\sigma_{\mu}\sim$ 0.4 
(\{BVI\}) and $\sigma_{\mu}\sim$ 0.5 (\{BV\}) (see Table \ref{tab_Dispersion_values}). 
The special case of {\it D05} with \{VI\} and {\it DES}, leads to $\sigma_{\mu}$ = 0.32, which 
corresponds to $\sim$ $15\%$ of error in distance. Clearly when the B band is employed, 
the dispersion in the Hubble diagram increases considerably. This effect could be explained 
by the presence of many absorption lines at those wavelengths, which makes the determination 
of the color temperature very sensitive to metallicity and to the opacity. 
However, both atmosphere models predict a modest effect of metallicity in the emergent 
flux at wavelength longer than $\sim$ 4000\AA, therefore the origin of the high dispersion 
when the B band is employed is not clear. \newline \indent
As expected, it can be noted that there are no significant differences in the $H_{0}$ values
and in the Hubble diagram dispersion between the {\it DES} and {\it OLI} reddening methods. 
This is because there is no systematic difference in the reddening between both methods 
(see \mbox{$\S$ 3.5).} However the {\it DES} method leads to somewhat lower dispersion in the Hubble 
diagrams than the {\it OLI} technique. \newline \indent
Finally, SN 2003hl and SN 2003iq are of particular interest because they both exploded in the 
same galaxy. To our disappointment all 12 posible 
combinations of filter subsets, reddening and atmosphere models lead to significant differences
in the EPM distance to the host galaxy. The most extreme case is the $\{BV\}$, {\it E96} and 
{\it OLI} combination, which leads to a distance of 32.5 $\pm$ 8.5 Mpc to SN 2003iq and 
12.8 $\pm$ 1.6 Mpc to SN 2003hl (a difference of 2.3 sigma). The smallest discrepancy occurs with
the $\{VI\}$, {\it DES} and {\it D05} combinations (30.3 $\pm$ 6.3 and 53.3 $\pm$ 17.1 Mpc for SN 
2003hl and SN 2003iq, respectively), which is also the combination that produces the lowest 
dispersion in the Hubble diagram. As discussed in \S 3, the EPM solutions  
to SN 2003iq yield an explosion time inconsistent with a pre-discovery image, therefore the 
EPM distance to SN 2003iq is quite suspicious.  
%Also, the DES and {\it OLI} methods predict a visual extinction greater than 1.5 {\it mag} for SN 
%2003hl. 

\subsection{External calibration and the internal precision of the EPM}

\begin{figure}
\plotone{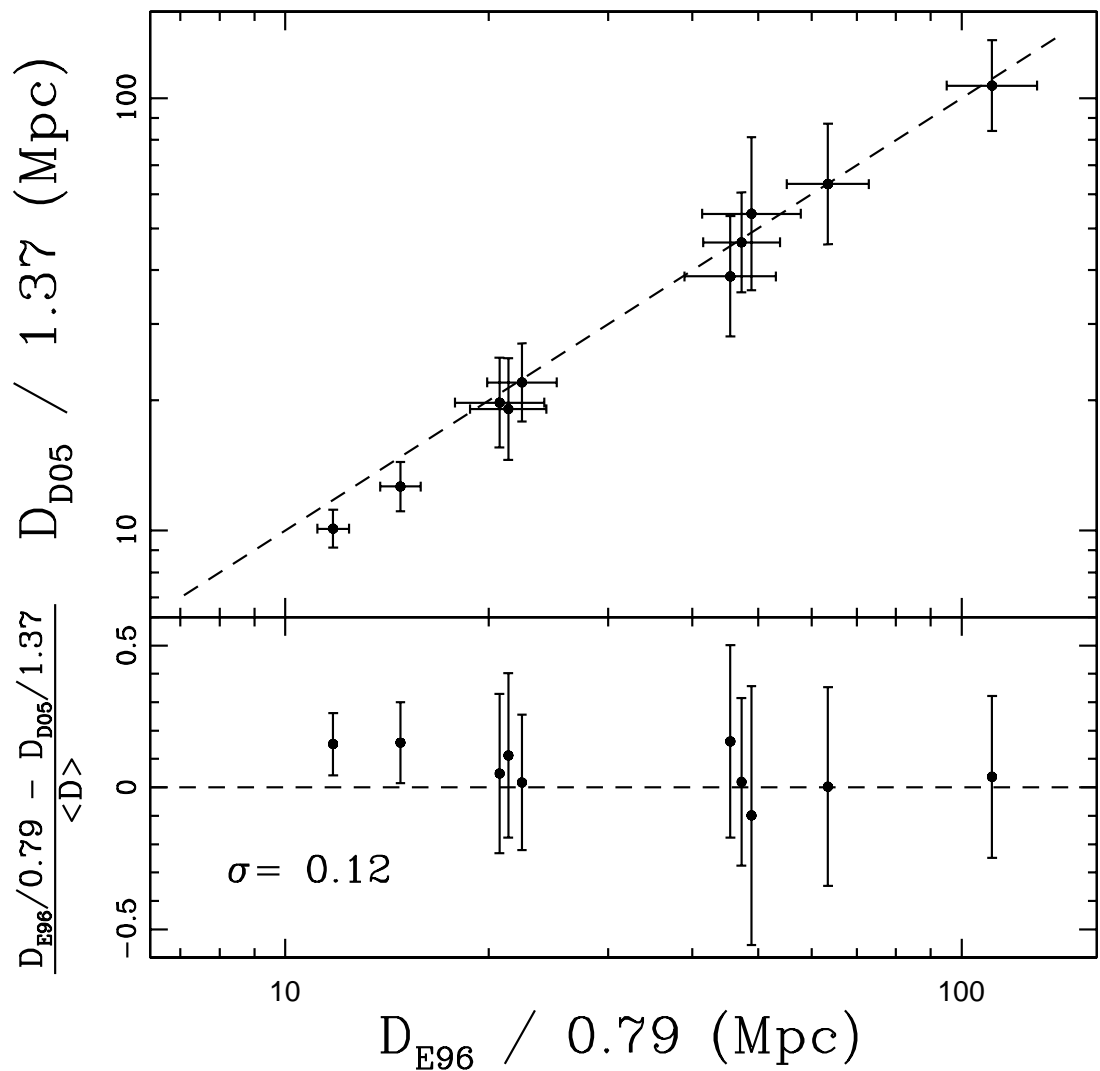}
\caption{Top panel: {\it D05} distances versus {\it E96} distances corrected to the 
{\it HST Key Project} Cepheid scale. The dashed line corresponds to the one to one relation. 
Bottom panel: differences between the corrected distances normalized to the average of the {\it E96} 
and {\it D05} corrected distances. The $12\%$ scatter reflects the internal precision of the EPM. 
~\label{fig_Cepheids_vi_DES.HD}}
\end{figure}

In the previous section we have shown that there is a systematic difference in the $H_0$ values
derived using the {\it E96} and the {\it D05} models. In order to remove this systematic effect we 
applied a calibration factor (given by the ratio between some external $H_0$ value and the EPM 
$H_0$ value) to the distances derived using {\it E96} and {\it D05}. 
%therefore the resulting distances between both models are in the same scale. 
For this purpose we used the value of $H_0=72$ $km$ $s^{-1}$ $Mpc^{-1}$ derived from the {\it HST Key 
Project} \citep{Fre01}. This external calibration allows us to bring the EPM distances 
to the Cepheids scale and allows us to remove the systematic difference in the EPM distances
between {\it E96} and {\it D05}. 
Figure \ref{fig_Cepheids_vi_DES.HD} shows (top panel) the {\it D05} distances versus the {\it E96}  
distances divided by a calibration factor of 1.37 and 0.79, respectively.
In both cases the EPM distances were derived using the $\{VI\}$ filter subset and the 
{\it DES} reddening. As can be seen, after applying this correction, the systematic differences disappear.
The dashed line in the top panel corresponds
to the one to one relation. Also, in Figure \ref{fig_Cepheids_vi_DES.HD} (bottom panel) are plotted 
the differences between the corrected {\it E96} and {\it D05} distances, normalized to the 
corresponding average between the corrected {\it E96} and {\it D05} distance. We found a standard 
deviation of $\sigma=0.12$. Since the dispersion arises from the combined errors in the {\it E96}
and the {\it D05} distances, the internal random errors in any of the EPM implementation must be less
than $12\%$. Note that this scatter is smaller than the $\sim 15\%$ dispersion seen in the Hubble
diagrams, which is affected by the peculiar motion of the host galaxies. The $12\%$ scatter is independent
of the redshift and must be an upper value of the internal precision of the EPM.

\section{Conclusions}

In this work we have applied the EPM method to 12 SNe IIP. We contructed 12 different Hubble
diagrams, using three different filter subsets ($\{BV\},\{BVI\},\{VI\}$), two atmosphere
models ({\it E96, D05}) and two methods to determine the amount of host galaxy extinction 
({\it DES, OLI}). Our main conclusions are the following:

\begin{itemize}

\item We found that the EPM must be restricted to the first $\sim 45-50$ days from explosion. 
After that epoch the method presents a departure from linearity in the $\theta/v$ versus time
relation and therefore an internal inconsistency.

\item We found that the results are less precise when the B band is used in the EPM analysis, 
regardless of the atmosphere models employed ({\it E96} or {\it D05}). The dispersion in the 
Hubble diagrams increases considerably from 0.3 to 0.5 {\it mag} when the B band is included 
and the V band is removed from the filter subset. Despite the loss in precision, there is no 
significant differences in the resulting distances when including or excluding the B filter.

\item We investigated the effect of host galaxy reddening in the EPM distances. For this 
purpose we computed many EPM solutions varying the amount of visual extinction, and we found 
that a difference of $\Delta A_V = 0.5$ {\it mag} leads on average to a difference of $\sim 5-10\%$ 
in distance. Therefore we conclude that the method is quite insensitive to the effect of dust.
 
\item We showed that systematic differences in the atmosphere models lead to $\sim50\%$ 
differences in the EPM distances and to values of $H_0$ between 52 and 101 $km$ $s^{-1}$ 
$Mpc^{-1}$. This effect is due to the systematic difference in the photospheric velocity
conversion provided by both models and the systematic differences in the dilution factors.
The latter is currently the greatest source of uncertainty in the EPM method.

\item The Hubble diagram with the lowest dispersion ($\sigma_{\mu}=0.32$ {\it mag}) was obtained
using the combination {\it D05}, $\{VI\}$, {\it DES}. Despite the systematic uncertainties in the EPM this
dispersion is quite low and corresponds to a precision of $\sim15\%$ in distance. This 
precision is similar to that of the SCM method for type II SNe \citep{Ham02,OLI08} and to the 
Tully-Fischer relation for spiral galaxies with a dispersion of $\sigma\sim0.30$ {\it mag} \citep{Sak00}. 
However, the EPM dispersion is considerably greater than that of the M/$\Delta m_{15}$ relation for 
Type Ia SNe, which has a dispersion of $\sigma\sim0.15-0.20$ {\it mag}, but we think that if the EPM 
is applied to a sample of SNe IIP in the Hubble Flow the dispersion in the Hubble diagram might decrease.
%However, this precision is far from that of the $\Delta m_{15}$ relation for Type Ia SNe 
%of $\sigma\sim0.15-0.20$. Finally, we think that applying the EPM to a bigger sample 
%of SN IIP in the Hubble Flow would lead to a lower dispersion in the Hubble diagram. 
 
\item Finally, despite the systematic differences in the $H_0$ value, we think that EPM
has great potential as an extragalactic distance indicator and that it can be applied to a 
sample of high redshift SNe IIP in order to check in an independent way the accelerating 
expansion of the universe.
\end{itemize}

%We derived $H_0$ in the range 52 $kms^{-1}Mpc^{-1}$ to 101 $kms^{-1}Mpc^{-1}$. 
%The Hubble diagram that produced the lowest dispersion ($\sigma_{\mu} = 0.32$) was obtained 
%using the combination $D05,\{VI\},DES$ and leads to $H_0 = 52.4 \pm 4.3$. 
%We found that the models are not accurate when the $B$ band is used in the EPM. Using both
%atmosphere models, the dispersion in the Hubble diagram increases considerably when the $B$ 
%band is include in the filter subset. Also, we have shown that the systematic differences 
%in the dilution factors computed from the E96 and D05 models, leads to differences in the 
%EPM distances and in the corresponding $H_0$ of $\sim50\%$. This effect is also due to the
%systematic difference in the photospheric velocity conversion provided by both models, 
%above all at low velocities. \newline \indent
%We also investigated how the reddening affect the EPM distances. We found that $H_0$ 
%essentially doesn't change when the reddening method was change. Also we computed many
%EPM solutions varying the amount of visual extinction, and we found that a difference of
%$\delta A_v = 1.0$ $mag$ on average leads to a difference of $\sim 10\%$ in the distance.
%Therefore we concluded that the method is quite insensitive to the effect of dust.           

\subsection*{Acknowledgments}

We thank Luc Dessart and Ronald Eastman for provide us their SN atmosphere models.
We are also very greatful to Brian Schmidt, Ryan Foley, Alexei Fillipenko, Robert Kirshner and
Thomas Matheson for share with us some spectra from a few SNe use in this work. We also acknowledge
to Doug Leonard for provide us the spectra taken to SN 1999gi. MJ acknowledges support from Centro
de Astrof\'isica FONDAP 15010003, support provided by Fondecyt through grant 1060808 and support
from the Millennium Center for Supernova Science through grant P06-045-F funded by ``Programa
Bicentenario de Ciencia y Tecnolog\'ia de CONICYT'' and ``Programa Iniciativa Cient\'ifica Milenio
de MIDEPLAN''.

%% Ahora estan las figuras

\clearpage

\clearpage

\clearpage

\clearpage

\clearpage

\clearpage

\clearpage

\clearpage

\clearpage

%\clearpage
%\begin{figure}
%\plotone{obs_Fe_vel.ratio.eps}
%\caption{Ratio between the Fe {\sc ii} 5169 and H$\beta$ velocity as a function of the
%H$\beta$ velocity. The triangles and the squares represent velocities measured from the
%spectra of our SN sample. The open and filled black circles correspond to the 
%velocity ratio measured from the synthetic spectra of {\it E96} and {\it D05} respectively .
%~\label{fig_obs_Fe_vel.ratio}}
%\end{figure}

\clearpage

\clearpage

\clearpage

\clearpage

\clearpage

\clearpage

\clearpage

\clearpage

\clearpage

\clearpage

\clearpage

\clearpage

\clearpage

\clearpage

\clearpage

\clearpage

\clearpage

\clearpage

\clearpage

\clearpage

\clearpage

\clearpage

%%%% Ahora las tablas

\clearpage
\begin{deluxetable}{ccc}
\tablewidth{0pt}
\tablecaption{Telescopes and instruments used in the photometric and spectroscopic  
observations \label{tab_telescopes}}
\tablehead{
\colhead{Telescope}  & \colhead{Instrument} & \colhead{Spec/Phot}}
\startdata
CTIO 0.9m    &  CCD      &  P   \\
YALO 1.0m    &  ANDICAM  &  P   \\
YALO 1.0m    &  2DF      &  S   \\
CTIO 1.5m    &  CCD      &  P   \\
CTIO 1.5m    &  CSPEC    &  S   \\
Blanco 4.0m  &  CSPEC    &  S   \\
Blanco 4.0m  &  2DF      &  S   \\
Blanco 4.0m  &  CCD      &  P   \\
Swope 1.0m   &  CCD      &  P   \\
du Pont 2.5m &  WFCCD    &  S/P \\ 
du Pont 2.5m &  MODSPEC  &  S   \\
du Pont 2.5m &  2DF      &  S   \\
du Pont 2.5m &  CCD      &  P   \\
Baade 6.5m   &  LDSS2    &  S/P \\
Baade 6.5m   &  B\&C      &  S   \\
Clay 6.5m    &  LDSS2    &  S/P \\
ESO 1.52m    &  IDS      &  S   \\
Danish 1.54m &  DFOSC    &  S/P \\
ESO 2.2m     &  EFOSC2   &  S   \\
NTT 3.58m    &  EMMI     &  S   \\
ESO 3.6m     &  EFOSC    &  S   \\
Kuiper 61"   &  CCD      &  P   \\
Bok 90"      &  B\&C      &  S   \\
\enddata
\end{deluxetable}

\clearpage
\begin{deluxetable}{ccccc}
\tablewidth{0pt}
\tablecaption{Heliocentric and CMB redshifts for the SNe used in this work. 
\label{tab_SN_list}}
\tablehead{
\colhead{SN} & \colhead{Host Galaxy} & \colhead{$cz_{helio}$} & 
\colhead{source\tablenotemark{a}} & \colhead{$cz_{CMB}$} \\
\colhead{}  & \colhead{} & \colhead{$(km$ $s^{-1})$} & 
\colhead{} &  \colhead{$(km$ $s^{-1})$} }
\startdata
1992ba &  NGC 2082                &  1092  & here & 1245 \\
1999br &  NGC 4900                &  960   & NED  & 1285 \\  
1999em &  NGC 1637                &  800   & L02  & 670  \\ 
1999gi &  NGC 3184                &  543   & here & 831  \\ 
2002gw &  NGC 0922                &  3117  & here & 2877 \\ 
2003T  &  UGC 04864               &  8373  & NED  & 8662 \\ 
2003bl &  NGC 5374                &  4385  & NED  & 4652 \\ 
2003bn &  2MASX J10023529-2110531 &  3832  & NED  & 4173 \\ 
2003ef &  NGC 4708                &  4440  & here & 4503 \\ 
2003hl &  NGC 0772                &  2265  & here & 2198 \\ 
2003hn &  NGC 1448                &  1347  & here & 1102 \\ 
2003iq &  NGC 0772                &  2364  & here & 2198 \\ 
\enddata
\tablenotetext{a}{\begin{small} The NED values correspond to the redshifts of the host 
galaxy center, while the values measured in this work (``here") were measured from narrow 
emission lines of HII regions at the SN position. Also L02 corresponds to the value adopted 
from \citet{Leo02}.\end{small}}
%\tablenotetext{b}{\begin{small} The CMB redshift correspond to that of the host galaxy 
%center. \end{small}}
\end{deluxetable}

\clearpage
\begin{deluxetable}{cccccccccccc}
\tablewidth{0pt}
\tablecaption{Dilution factors coefficients and dispersion for the \{BV\}, \{BVI\} and \{VI\}
filter subsets and the {\it E96} and {\it D05} models. \label{tab_zeta_coef}}
\tablehead{
\multicolumn{3}{c}{} & \multicolumn{2}{c}{{\it E96}} & \multicolumn{3}{c}{} & 
\multicolumn{2}{c}{{\it D05}} \\
%\colhead{} & \colhead{} & \colhead{} & \colhead{{\it E96}} & \colhead{} &
%\colhead{} & \colhead{} & \colhead{} & \colhead{{\it D05}} & \colhead{} \\
\cline{3-6} \cline{8-11}    
\colhead{Filter subset} & \colhead{} & \colhead{$b_0$} & \colhead{$b_1$} & \colhead{$b_2$} &  
\colhead{$\sigma$} & \colhead{} & \colhead{$b_0$} & \colhead{$b_1$} 
& \colhead{$b_2$} & \colhead{$\sigma$} }
\startdata
\{BV\}  & & 0.756 & -0.900 & 0.520 & 0.048 & & 0.593 & -0.450 & 0.403 & 0.075 \\
\{BVI\} & & 0.733 & -0.693 & 0.373 & 0.027 & & 0.711 & -0.476 & 0.308 & 0.068 \\
\{VI\}  & & 0.702 & -0.531 & 0.265 & 0.029 & & 0.915 & -0.747 & 0.371 & 0.077 \\
\enddata
\end{deluxetable}

\clearpage
\begin{deluxetable}{cccccc}
\tablewidth{0pt}
\tablecaption{Spectroscopic velocities for 12 SNe. \label{tab_spec_vel}} 
\tablehead{
%\multicolumn{1}{c}{} & \multicolumn{5}{c}{Velocities} \\
%\cline{3-6} \\
\colhead{SN}  & \colhead{JD-} & \colhead{H$\alpha$} &
\colhead{Fe {\sc ii} $\lambda5169$} & \colhead{H$\beta$}  & 
\colhead{H$\gamma$} \\
\colhead{}  &\colhead{2448000}  & \colhead{$(km$ $s^{-1})$} & \colhead{$(km$ $s^{-1})$} &
\colhead{$(km$ $s^{-1})$} & \colhead{$(km$ $s^{-1})$}} 
\startdata
%SN1992ba & & & & &  \\
 SN 1992ba & 896.9  &   9085.8  &   \nodata &   8101.1  &   7537.8  \\
          & 900.9  &   8513.6  &   6329.3  &   7442.9  &   6845.3  \\ 
          & 922.8  &   6508.0  &   3734.3  &   4765.5  &   5856.2  \\ 
          & 949.8  &   5136.2  &   2748.9  &   4173.4  &   4762.8  \\ 
          & 974.8  &   4747.3  &   2409.6  &   3966.5  &   4304.7  \\ 
          & 1015.7 &   4525.4  &   2022.1  &   3331.5  &   3680.7 \\ 
          & 1045.7 &   4139.6  &   980.8   &   2991.8  &   \nodata \\ 
          & 1067.6 &   3946.0  &   \nodata &   906     &   \nodata \\ 
\cline{1-6}
 SN 1999br & 3291.7  & 5043.2   & 3908.9   & 4701.0   & 4279.0  \\
          & 3294.7  & 4857.2   & 3429.0   & 4428.0   & 4110.3 \\ 
          & 3297.7  & 4729.3   & \nodata  & 4264.2   & 3974.1  \\ 
          & 3301.6  & 4420.2   & 2587.7   & 3643.7   & 3468.1  \\ 
          & 3309.7  & 4128.8   & 2037.2   & 3394.1   & \nodata \\ 
          & 3317.7  & 3730.2   & 1723.2   & 2364.4   & \nodata \\ 
          & 3319.5  & 3571.1   & 1866.4   & 2104.9   & \nodata \\ 
          & 3381.5  & 1130.1   & 1226.7   & 1044.2   & \nodata \\ 
\cline{1-6}
 SN 1999em & 3481.8  & 12422.8  & \nodata  & 10318.1  & 8784.7   \\     
          & 3482.8  & 10663.0  & \nodata  & 10835.5  & \nodata  \\
          & 3483.8  & 10667.2  & \nodata  & 10241.5  & \nodata  \\ 
          & 3484.8  & 10100.6  & \nodata  & 9799.5   & \nodata  \\ 
          & 3485.2  & 10423.9  & \nodata  & 9107.8   & 8351.9   \\ 
          & 3485.7  & 10318.3  & \nodata  & 8944.5   & 8271.5   \\ 
          & 3485.7  & 10342.4  & \nodata  & 8919.8   & 8353.5   \\         
          & 3485.8  & 9770.6   & \nodata  & 9288.6   & \nodata  \\ 
          & 3486.8  & 9438.7   & 8029.4   & 8827.7   & 8221.8   \\  
          & 3487.9  & 9851.0   & 8115.9   & 8608.6   & 7929.1   \\ 
          & 3488.8  & 9872.2   & 8940.6   & 8621.4   & 7509.6   \\ 
          & 3489.8  & 9619.9   & 7859.9   & 8506.5   & 7546.7   \\ 
          & 3491.1  & 9312.9   & 6806.4   & 7929.1   & 7407.9   \\ 
          & 3491.2  & 9246.6   & 7097.9   & 7884.5   & 7434.9   \\ 
          & 3491.7  & 9200.1   & 7203.2   & 8018.7   & 7671.8   \\        
          & 3492.1  & 9229.9   & 7010.4   & 7921.9   & 7547.2   \\ 
          & 3496.2  & 8671.2   & 6010.9   & 7117.0   & 7142.1   \\ 
          & 3496.7  & 8636.3   & 6062.1   & 7254.1   & 7364.4   \\        
          & 3501.2  & 7947.5   & 5228.2   & 6022.6   & 6775.9   \\ 
          & 3501.7  & 7868.3   & 5191.3   & 6207.1   & 6720.9   \\ 
          & 3501.7  & 7929.7   & 5280.3   & 6348.4   & 6788.6   \\       
          & 3501.8  & 7860.1   & 5992.8   & 6601.5   & 6939.5   \\ 
          & 3504.8  & 7650.8   & 4967.7   & 6091.8   & 7104.2   \\ 
          & 3506.8  & 7236.5   & 5015.8   & 5793.1   & \nodata  \\ 
          & 3510.8  & 6824.3   & 4580.0   & 5256.4   & \nodata  \\ 
          & 3514.8  & 6463.0   & 4327.2   & 4975.8   & 6399.7   \\ 
          & 3518.0  & 6078.2   & 3757.1   & 4255.9   & 6158.1   \\ 
          & 3520.8  & 5733.5   & 3845.3   & 4127.6   & \nodata  \\ 
          & 3524.8  & 5314.6   & \nodata  & 3997.9   & \nodata  \\ 
          & 3527.7  & 5265.1   & 3755.6   & 3855.6   & \nodata  \\ 
          & 3528.8  & 5116.4   & 3255.8   & \nodata  & \nodata  \\ 
          & 3528.8  & 5160.9   & 3386.3   & \nodata  & \nodata  \\ 
          & 3529.7  & 4992.4   & 3553.7   & 3537.5   & \nodata  \\ 
          & 3543.8  & 4963.3   & 2935.9   & 2569.4   & 5416.9   \\ 
          & 3543.8  & 5363.9   & 2990.2   & 2649.2   & \nodata  \\  
          & 3556.8  & 6095.4   & 2853.6   & \nodata  & \nodata  \\ 
          & 3575.7  & 6356.8   & 2555.0   & \nodata  & \nodata  \\ 
          & 3576.7  & 6196.6   & 2621.0   & \nodata  & \nodata  \\ 
          & 3604.7  & 4580.2   & 1916.4   & \nodata  & \nodata  \\ 
          & 3618.6  & 3829.3   & \nodata  & \nodata  & \nodata  \\ 
          & 3639.7  & 3681.8   & 1077.5   & 2381.7   & \nodata  \\ 
          & 3643.5  & 3663.6   & 1964.4   & 2284.7   & \nodata  \\    
          & 3794    & 3575.8   & \nodata  & \nodata  & \nodata  \\ 
          & 3814    & 3554.3   & 5112.2   & 2250.7   & \nodata  \\ 
%\cline{1-6}
 SN 1999gi & 3522.9  & 13151.3  & \nodata  & 26476.1  & 28329.4  \\
          & 3525.0  & 12184.3  & \nodata  & 10452.3  & 10178.0  \\
          & 3526.0  & 11465.1  & \nodata  & 10304.2  & 9530.3   \\
          & 3548.9  & 7762.4   & 5260.0   & 6169.9   & 7006.3   \\
          & 3553.9  & 7158.7   & 4733.5   & 5451.7   & 6467.8   \\
          & 3556.9  & 6946.8   & 4596.4   & 5189.7   & 6470.8   \\ 
          & 3607.8  & 2751.1   & 2814.2   & 1854.3   & 5312.5   \\  
          & 3611.7  & 2610.5   & 2848.0   & 1625.0   & 5260.5   \\ 
          & 3618.9  & 2534.2   & 2692.0   & 1798.5   & 5255.1   \\ 
          & 3628.9  & 2200.9   & 2457.8   & 1498.0   & \nodata  \\ 
          & 3632.9  & 2125.4   & 2544.9   & 1519.3   & 5071.4   \\ 
          & 3659.7  & 2222.4   & 1356.1   & 1681.0   & \nodata  \\ 
          & 3661.9  & 2385.5   & 2305.2   & 1887.1   & \nodata  \\ 
          & 3674.7  & 2564.9   & 1549.2   & 1986.7   & \nodata  \\ 
          & 3690.7  & 2969.3   & 916.5    & 1991.2   & \nodata  \\ 
\cline{1-6}
 SN 2002gw & 4573.1  & 9044.0   & 6438.8   & 8238.5   & 7530.2   \\
          & 4576.7  & 8852.7   & 5743.9   & 7350.8   & 6912.3   \\ 
          & 4577.7  & 8658.4   & 5343.8   & 6984.3   & 6896.9   \\ 
          & 4585.7  & 7860.0   & 4626.9   & 6154.3   & 6216.1   \\ 
          & 4588.8  & 7439.9   & 4397.3   & 5893.5   & 5685.8   \\ 
          & 4590.7  & 7157.7   & 4130.1   & 5424.8   & 5343.3   \\ 
          & 4606.7  & 6079.4   & 3243.8   & 4525.8   & 4648.3   \\ 
          & 4609.6  & 5998.4   & 3118.7   & 4489.2   & 4809.7   \\ 
          & 4635.7  & 5400.8   & 2680.1   & 4174.9   & 4275.0   \\ 
          & 4642.7  & 5267.8   & 2558.3   & 4075.5   & 3752.2   \\ 
          & 4649.6  & 5237.0   & 2506.4   & 4047.1   & 4031.8   \\ 
\cline{1-6}
 SN 2003T  & 4667.9  & 10771.8  & \nodata  & 9793.1   & \nodata  \\
          & 4673.8  & 10157.9  & 6190.2   & 7392.7   & \nodata  \\ 
          & 4701.7  & 6749.1   & 4038.8   & 4730.0   & 6088.6   \\ 
          & 4710.7  & 6073.9   & 3494.2   & 3854.3   & 4042.6   \\ 
          & 4729.6  & 5899.9   & 2974.7   & 3081.7   & \nodata  \\ 
          & 4739.6  & 5679.0   & 2361.3   & 2637.7   & \nodata  \\ 
          & 4764.5  & 5261.3   & 1963.6   & \nodata  & \nodata  \\ 
\cline{1-6}
 SN 2003bl & 4701.8  & \nodata  & \nodata  & 6548.6   & 8292.5   \\
          & 4702.8  & 8816.7   & \nodata  & 7125.8   & 6884.4   \\  
          & 4729.8  & 5197.1   & 3448.8   & 4015.2   & 5606.6   \\ 
          & 4735.8  & 4876.3   & 2460.9   & 3437.4   & 4539.5   \\ 
          & 4739.9  & 4758.5   & 2342.7   & 2953.4   & 4236.2   \\ 
          & 4764.7  & 4581.2   & 1834.8   & 1868.0   & 4260.7   \\ 
          & 4789.7  & 4337.1   & 1229.9   & 1139.5   & 3608.9   \\ 
          & 4794.7  & 4411.7   & 1310.1   & 1108.5   & 3662.5   \\ 
\cline{1-6}
 SN 2003bn & 4706.6  & 11587.2  & \nodata  & 9294.8   & \nodata  \\      
          & 4710.8  & 9795.8   & 7007.2   & 8703.0   & 8017.8   \\ 
          & 4729.7  & 7909.8   & 4811.9   & 5920.6   & 6006.6   \\ 
          & 4733.7  & 7327.3   & 4380.4   & 5675.9   & 6303.4   \\ 
          & 4736.6  & 6999.2   & 4186.8   & 5569.7   & 5741.0   \\ 
          & 4739.7  & 6859.7   & 4073.2   & 5359.7   & 5793.5   \\ 
          & 4764.6  & 5573.2   & 2950.5   & 3409.1   & 5346.8   \\ 
          & 4789.5  & 5347.3   & 2992.8   & 2646.3   & \nodata  \\ 
          & 4794.6  & 5131.5   & 2584.4   & 2390.0   & 5156.7   \\ 
          & 4797.5  & 5108.8   & 2490.2   & 2344.1   & 4563.6   \\ 
          & 4814.5  & 4561.5   & 1981.7   & 2198.2   & \nodata  \\ 
          & 4820.5  & 4988.8   & 2022.0   & 2292.3   & \nodata  \\ 
\cline{1-6}
 SN 2003ef & 4780.7  & 9363.4   & 6182.0   & 8014.1   & \nodata \\
          & 4789.7  & 8462.8   & 5292.7   & 6339.8   & 7373.2  \\  
          & 4794.6  & 7784.3   & 4657.6   & 5675.4   & 6684.9  \\ 
          & 4797.6  & 7611.3   & 4900.8   & 5384.9   & 6435.3  \\ 
          & 4814.5  & 6526.2   & 3842.4   & 4002.8   & 6023.1  \\ 
          & 4820.6  & 6434.6   & 3709.1   & 3757.9   & \nodata \\ 
          & 4866.5  & 4881.2   & \nodata  & \nodata  & \nodata \\ 
%\cline{1-6}
 SN 2003hl & 4879.9  & 9161.0   & \nodata  & 8617.5   & \nodata  \\
          & 4900.8  & 6541.3   & 4665.6   & 5392.6   & \nodata  \\ 
          & 4908.7  & 6074.7   & 4126.5   & 4737.0   & 6274.0   \\ 
          & 4928.7  & 4892.1   & 3016.6   & 3771.5   & \nodata  \\ 
          & 4940.8  & 4750.0   & 3036.0   & 3535.5   & \nodata  \\ 
          & 4948.8  & 4653.4   & 2854.9   & 3684.9   & \nodata  \\ 
          & 4966.7  & 4450.6   & 2662.8   & 3346.8   & \nodata  \\ 
          & 4996.6  & 4387.3   & 1960.4   & 3372.6   & \nodata  \\ 
          & 5021.7  & 4184.7   & \nodata  & \nodata  & \nodata  \\ 
\cline{1-6}
 SN 2003hn & 4878.2  & \nodata  & \nodata  & 9432.0   & 8367.2   \\
          & 4888.3  & \nodata  & 6356.9   & 8096.0   & 8430.9   \\ 
          & 4897.9  & 7826.1   & 5051.7   & 6726.1   & 7061.4   \\ 
          & 4900.9  & 7606.9   & 4738.6   & 5985.2   & 6824.4   \\ 
          & 4908.8  & 7514.7   & 4276.3   & 5459.5   & 6384.6   \\ 
          & 4928.8  & 6674.2   & 3431.2   & 4440.9   & 5391.8   \\ 
          & 4948.8  & 5985.6   & 2928.8   & 4380.5   & 4890.5   \\ 
          & 4966.8  & 5367.5   & 2002.2   & 3905.8   & 4858.5   \\ 
          & 4989.7  & 4778.7   & 827.7    & 3331.3   & \nodata  \\ 
          & 4996.7  & 4680.7   & 988.3    & 3275.7   & 3120.2   \\ 
          & 5040.7  & 4331.9   & \nodata  & \nodata  & \nodata  \\ 
%\cline{1-6}
 SN 2003iq & 4928.7  & 11542.1  & \nodata  & 10486.7  & 10210.5  \\
          & 4930.7  & 10993.2  & \nodata  & 9901.3   & 9576.4   \\
          & 4940.8  & 9836.3   & 6923.9   & 8182.8   & 8145.8   \\ 
          & 4948.7  & 8495.6   & 5545.9   & 6970.3   & 7050.0   \\ 
          & 4966.7  & 7028.5   & 4214.2   & 5103.2   & 6210.8   \\ 
          & 4989.7  & 6235.4   & 3443.2   & 4616.4   & 5114.7   \\ 
          & 4996.6  & 6156.5   & 3581.6   & 4218.1   & 5568.9   \\ 
          & 5021.7  & 5814.0   & 2887.3   & 3783.4   & \nodata  \\ 
\enddata
\end{deluxetable}

\clearpage
\begin{deluxetable}{ccc}
\tablewidth{0pt}
\tablecaption{H$\beta$ to photospheric velocity ratio coefficients and dispersion 
for the {\it E96} and {\it D05} models.
\label{tab_ratio_coef}}
\tablehead{
\colhead{$j$}  & \colhead{$a_{j}({\it E96})$} & \colhead{$a_{j}({\it D05})$}}
\startdata
0 & 1.775 & 1.014       \\
1 & -1.435$\times10^{-4}$ & 4.764$\times10^{-6}$    \\
2 & 6.523$\times10^{-9}$ & -7.015$\times10^{-10}$  \\
$\sigma$ & 0.06 & 0.04 \\
\enddata
\end{deluxetable}

\clearpage
\begin{deluxetable}{cccc}
\tablewidth{0pt}
\tablecaption{SNe host galaxy and Galactic extinction adopted. 
\label{tab_reddening}}
\tablehead{
\colhead{SN}  & \colhead{$A_{V}$ ({\it OLI})}\tablenotemark{a}  
& \colhead{$A_{V}$ ({\it DES})} \tablenotemark{b}
& \colhead{$A_{V}$ (IR maps)} \tablenotemark{c} \\
\colhead{}  & \colhead{$Host$}  & \colhead{$Host$}
& \colhead{$Galactic$} 
}
\startdata
1992ba & 0.30 (0.15)  &  0.43 (0.16)  &  0.193 (0.031) \\
1999br & 0.94 (0.20)  &  0.25 (0.16)  &  0.078 (0.012) \\
1999em & 0.24 (0.14)  &  0.31 (0.16)  &  0.134 (0.021) \\
1999gi & 1.02 (0.15)  &  0.56 (0.16)  &  0.055 (0.009) \\
2002gw & 0.18 (0.16)  &  0.40 (0.19)  &  0.065 (0.010) \\
2003T  & 0.35 (0.15)  &  0.53 (0.31)  &  0.104 (0.017) \\
2003bl & 0.26 (0.15)  &  0.00 (0.16)  &  0.090 (0.014) \\
2003bn & -0.04 (0.15) &  0.09 (0.16)  &  0.215 (0.034) \\
2003ef & 0.98 (0.15)  &  1.24 (0.25)  &  0.153 (0.024) \\
2003hl & 1.72 (0.18)  &  1.24 (0.25)  &  0.241 (0.039) \\
2003hn & 0.46 (0.14)  &  0.59 (0.25)  &  0.047 (0.008) \\
2003iq & 0.25 (0.16)  &  0.37 (0.16)  &  0.241 (0.039) \\
\enddata
\tablenotetext{a}{\begin{small}  \citet{OLI08}
\end{small}}
\tablenotetext{b}{\begin{small}  \citet{DES08}
\end{small}}
\tablenotetext{c}{\begin{small} \citet{Sch98}
\end{small}}
\end{deluxetable}

\clearpage
\begin{deluxetable}{ccccc}
\tablewidth{0pt}
\tablecaption{EPM distances using the $\{VI\}$ filter subset and {\it DES} reddening. 
\label{tab_VI_DES.EPM}}
\tablehead{
\colhead{SN}  & \colhead{$D_{(E96)}$} & \colhead{$t_{0(E96)}$} &
\colhead{$D_{(D05)}$} & \colhead{$t_{0(D05)}$} \\
\colhead{}  & \colhead{$(Mpc)$} & \colhead{$(JD-2448000)$} &
\colhead{$(Mpc)$} & \colhead{$(JD-2448000)$}}
\startdata
1992ba & 16.4 (2.5)   & 883.9 (3.0)  & 27.2 (6.5)   & 879.8 (5.6)  \\
1999br & \nodata      & \nodata      & 39.5 (13.5)  & 3275.6 (7.7)  \\
1999em & 9.3 (0.5)    & 3476.3 (1.1) & 13.9 (1.4)   & 3474.0 (2.0)  \\
1999gi & 11.7 (0.8)   & 3517.0 (1.2) & 17.4 (2.3)   & 3515.6 (2.4)  \\
2002gw & 37.4 (4.9)   & 4557.9 (2.7) & 63.9 (17.0)  & 4551.7 (7.6)  \\
2003T  & 87.8 (13.5)  & 4654.2 (2.7) & 147.3 (35.7) & 4648.9 (6.1)  \\
2003bl & \nodata      & \nodata      & 92.4 (14.2)  & 4694.5 (2.0)  \\
2003bn & 50.2 (7.0)   & 4693.4 (2.7) & 87.2 (28.0)  & 4687.0 (9.0)  \\
2003ef & 38.7 (6.5)   & 4759.8 (4.7) & 74.4 (30.3)  & 4748.4 (15.6) \\
2003hl & 17.7 (2.1)   & 4872.3 (1.7) & 30.3 (6.3)   & 4865.4 (5.9)  \\
2003hn & 16.9 (2.2)   & 4859.5 (3.8) & 26.3 (7.1)   & 4853.8 (9.3)  \\
2003iq & 36.0 (5.6)   & 4909.6 (4.3) & 53.3 (17.1)  & 4905.6 (9.5)  \\
\enddata
\end{deluxetable}

\clearpage
\begin{deluxetable}{cc}
\tablewidth{0pt}
\tablecaption{Error Sources \label{tab_errors}}
\tablehead {
\colhead{Error Source}  & \colhead{Typical Error} }
\startdata
Photometry & 0.02 {\it mag}\\
SN redshift & 50 / 200 ($km$ $s^{-1}$) \tablenotemark{a}\\
%Host Galaxy peculiar motion & 300 ($km$ $s^{-1}$)\\
Foreground extinction & 0.02 {\it mag} \\
Host galaxy extinction & 0.15 {\it mag}\\
Line expansion velocity & 85 ($km$ $s^{-1}$) \\
Photospheric velocity conversion & 0.06 / 0.04 \tablenotemark{b}\\
Dilution Factors & 0.03 / 0.07 \tablenotemark{b}\\
\enddata
\tablenotetext{a}{\begin{small} Corresponds to the redshifts measured in this work and those
taken from the NED, respectively.
%We assigned an error of 50 $km$ $s^{-1}$ to the redshift
%measured in this work and of 200 $km$ $s^{-1}$ to those taken from the NED.
\end{small}}
\tablenotetext{b}{\begin{small} Corresponds to the {\it E96} and {\it D05} models, respectively. 
\end{small}}
\end{deluxetable}

\clearpage
\begin{deluxetable}{cccccc}
\tablewidth{0pt}
\tablecaption{EPM Quantities Derived for SN 1992ba using the \{VI\} filter subset,
{\it DES} reddening and the {\it D05} atmosphere models. \label{tab_SN92ba_EPM}}
\tablehead {
\colhead{JD-}  & \colhead{$T_{VI}$} & \colhead{$\theta$$\zeta_{VI}$} &
\colhead{$\zeta_{VI}$} & \colhead{$v_{phot}$} & \colhead{$\theta/vel$} \\
\colhead{2448000}  & \colhead{$(K)$} & \colhead{$(10^{15}$ $cm$ $Mpc^{-1})$} &
\colhead{} & \colhead{$(km$ $s^{-1})$} & \colhead{$(100$ $s$ $Mpc^{-1})$} }
\startdata
896.9  & 12526 (2637) &  0.0221 (0.0047) & 0.552  & 8050 & 498.3 (128.0)  \\
900.9  & 11187 (1038) &  0.0253 (0.0026) & 0.541  & 7366 & 635.2 (114.4)  \\
922.8  & 8493 (189)   &  0.0345 (0.0011) & 0.548  & 4669 & 1347.0 (203.7) \\
\enddata
\end{deluxetable}

\clearpage
\begin{deluxetable}{cccccc}
\tablewidth{0pt}
\tablecaption{EPM Quantities Derived for SN 1999br using the \{VI\} filter subset,
{\it DES} reddening and the {\it D05} atmosphere models. \label{tab_SN99br_EPM}}
\tablehead {
\colhead{JD-}  & \colhead{$T_{VI}$} & \colhead{$\theta$$\zeta_{VI}$} &
\colhead{$\zeta_{VI}$} & \colhead{$v_{phot}$} & \colhead{$\theta/vel$} \\
\colhead{2451000}  & \colhead{$(K)$} & \colhead{$(10^{15}$ $cm$ $Mpc^{-1})$} &
\colhead{} & \colhead{$(km$ $s^{-1})$} & \colhead{$(100$ $s$ $Mpc^{-1})$} }
\startdata
291.7 & 8601 (203) & 0.0113 (0.0004) & 0.546 & 4606 & 450.6 (68.6)\\
294.7 & 8398 (177) & 0.0117 (0.0004) & 0.550 & 4336 & 490.4 (74.0)\\
297.7 & 7497 (303) & 0.0134 (0.0009) & 0.577 & 4175 & 558.3 (86.8)\\
301.6 & 7700 (143) & 0.0130 (0.0004) & 0.569 & 3566 & 640.9 (94.0)\\
309.7 & 6876 (132) & 0.0160 (0.0006) & 0.612 & 3321 & 789.4 (110.5)\\
\enddata
\end{deluxetable}

\clearpage
\begin{deluxetable}{cccccc}
\tablewidth{0pt}
\tablecaption{EPM Quantities Derived for SN 1999em using the \{VI\} filter subset,
{\it DES} reddening and the {\it D05} atmosphere models. \label{tab_SN99em_EPM}}
\tablehead {
\colhead{JD-}  & \colhead{$T_{VI}$} & \colhead{$\theta$$\zeta_{VI}$} &
\colhead{$\zeta_{VI}$} & \colhead{$v_{phot}$} & \colhead{$\theta/vel$} \\ 
\colhead{2451000}  & \colhead{$(K)$} & \colhead{$(10^{15}$ $cm$ $Mpc^{-1})$} &
\colhead{} & \colhead{$(km$ $s^{-1})$} & \colhead{$(100$ $s$ $Mpc^{-1})$} }
\startdata
482.8 & 14588 (469) & 0.0321 (0.0010) & 0.574 & 11022 & 506.7 (73.0)\\
483.8 & 14349 (462) & 0.0331 (0.0011) & 0.572 & 10355 & 559.6 (81.0)\\ 
484.8 & 13986 (382) & 0.0341 (0.0009) & 0.568 & 9867  & 608.3 (88.1)\\ 
485.2 & 13810 (415) & 0.0345 (0.0011) & 0.566 & 9117  & 669.5 (97.7)\\ 
485.7 & 13550 (414) & 0.0352 (0.0011) & 0.563 & 8942  & 699.7 (102.6)\\ 
485.7 & 13544 (414) & 0.0352 (0.0011) & 0.563 & 8915  & 702.1 (103.0)\\ 
485.8 & 13479 (456) & 0.0353 (0.0012) & 0.562 & 9311  & 675.0 (99.7)\\ 
486.8 & 12812 (425) & 0.0373 (0.0013) & 0.555 & 8817  & 762.6 (113.9)\\ 
487.9 & 11985 (333) & 0.0403 (0.0013) & 0.547 & 8584  & 857.5 (128.9)\\ 
488.8 & 11587 (310) & 0.0413 (0.0013) & 0.544 & 8598  & 882.8 (133.3)\\ 
489.8 & 11352 (256) & 0.0424 (0.0011) & 0.542 & 8476  & 921.1 (138.7)\\ 
491.1 & 11077 (350) & 0.0443 (0.0016) & 0.541 & 7870  & 1040.1 (159.5)\\ 
491.2 & 11055 (358) & 0.0444 (0.0017) & 0.541 & 7824  & 1050.6 (161.4)\\ 
491.7 & 10939 (372) & 0.0453 (0.0018) & 0.540 & 7964  & 1053.3 (162.6)\\ 
492.1 & 10840 (349) & 0.0460 (0.0018) & 0.539 & 7863  & 1083.7 (166.9)\\ 
496.2 & 10264 (312) & 0.0495 (0.0019) & 0.537 & 7031  & 1311.6 (202.6)\\ 
496.7 & 10224 (301) & 0.0497 (0.0018) & 0.537 & 7172  & 1290.7 (199.0)\\ 
501.2 & 9610 (224)  & 0.0526 (0.0016) & 0.537 & 5921  & 1653.2 (252.9)\\ 
501.7 & 9386 (185)  & 0.0548 (0.0014) & 0.538 & 6107  & 1667.6 (253.4)\\ 
501.7 & 9384 (185)  & 0.0548 (0.0014) & 0.538 & 6250  & 1630.3 (247.7)\\ 
501.8 & 9362 (189)  & 0.0551 (0.0015) & 0.538 & 6506  & 1572.5 (238.9)\\ 
504.8 & 8907 (173)  & 0.0589 (0.0016) & 0.542 & 5991  & 1813.0 (274.0)\\ 
506.8 & 8655 (162)  & 0.0605 (0.0016) & 0.545 & 5691  & 1950.5 (293.2)\\ 
510.8 & 8248 (63)   & 0.0649 (0.0007) & 0.553 & 5156  & 2276.6 (334.0)\\ 
514.8 & 7819 (92)   & 0.0705 (0.0013) & 0.565 & 4877  & 2557.8 (370.3)\\ 
\enddata
\end{deluxetable}

\clearpage
\begin{deluxetable}{cccccc}
\tablewidth{0pt}
\tablecaption{EPM Quantities Derived for SN 1999gi using the \{VI\} filter subset,
{\it DES} reddening and the {\it D05} atmosphere models. \label{tab_SN99gi_EPM}}
\tablehead {
\colhead{JD-}  & \colhead{$T_{VI}$} & \colhead{$\theta$$\zeta_{VI}$} &
\colhead{$\zeta_{VI}$} & \colhead{$v_{phot}$} & \colhead{$\theta/vel$} \\
\colhead{2451000}  & \colhead{$(K)$} & \colhead{$(10^{15}$ $cm$ $Mpc^{-1})$} &
\colhead{} & \colhead{$(km$ $s^{-1})$} & \colhead{$(100$ $s$ $Mpc^{-1})$} }
\startdata
525.0 & 10112 (564) & 0.0351 (0.0027) & 0.537 & 10590 & 617.6 (103.6)\\
526.0 & 9721 (409)  & 0.0384 (0.0023) & 0.537 & 10425 & 685.1 (109.9)\\
548.9 & 7247 (183)  & 0.0607 (0.0026) & 0.589 & 6070  & 1696.8 (244.5)\\
553.9 & 7101 (236)  & 0.0618 (0.0034) & 0.597 & 5350  & 1933.8 (283.8)\\
556.9 & 6903 (295)  & 0.0646 (0.0047) & 0.610 & 5090  & 2082.6 (317.6)
\enddata
\end{deluxetable}

\clearpage
\begin{deluxetable}{cccccc}
\tablewidth{0pt}
\tablecaption{EPM Quantities Derived for SN 2002gw using the \{VI\} filter subset,
{\it DES} reddening and the {\it D05} atmosphere models. \label{tab_SN02gw_EPM}}
\tablehead {
\colhead{JD-}  & \colhead{$T_{VI}$} & \colhead{$\theta$$\zeta_{VI}$} &
\colhead{$\zeta_{VI}$} & \colhead{$v_{phot}$} & \colhead{$\theta/vel$} \\
\colhead{2452000}  & \colhead{$(K)$} & \colhead{$(10^{15}$ $cm$ $Mpc^{-1})$} &
\colhead{} & \colhead{$(km$ $s^{-1})$} & \colhead{$(100$ $s$ $Mpc^{-1})$} }
\startdata
573.1 & 9148 (135) & 0.0119 (0.0002) & 0.537 & 8194. & 269.5 (40.9)\\
576.7 & 8760 (164) & 0.0125 (0.0004) & 0.541 & 7271. & 316.9 (48.2)\\
577.7 & 8757 (208) & 0.0124 (0.0005) & 0.541 & 6895. & 332.7 (51.2)\\
585.7 & 7683 (123) & 0.0152 (0.0004) & 0.568 & 6054. & 442.2 (64.3)\\
588.8 & 7337 (97)  & 0.0165 (0.0004) & 0.583 & 5792. & 488.5 (69.1)\\
590.7 & 7275 (82)  & 0.0168 (0.0003) & 0.586 & 5324. & 537.0 (75.4)\\
\enddata
\end{deluxetable}

\clearpage
\begin{deluxetable}{cccccc}
\tablewidth{0pt}
\tablecaption{EPM Quantities Derived for SN 2003T using the \{VI\} filter subset,
{\it DES} reddening and the {\it D05} atmosphere models. \label{tab_SN03T_EPM}}
\tablehead {
\colhead{JD-}  & \colhead{$T_{VI}$} & \colhead{$\theta$$\zeta_{VI}$} &
\colhead{$\zeta_{VI}$} & \colhead{$v_{phot}$} & \colhead{$\theta/vel$} \\
\colhead{2452000}  & \colhead{$(K)$} & \colhead{$(10^{15}$ $cm$ $Mpc^{-1})$} &
\colhead{} & \colhead{$(km$ $s^{-1})$} & \colhead{$(100$ $s$ $Mpc^{-1})$} }
\startdata
667.9 & 10626 (1346) & 0.0049 (0.0007) & 0.526 & 9860 & 95.0 (20.5) \\ 
673.8 & 9835 (195)   & 0.0054 (0.0001) & 0.527 & 7314 & 139.6 (21.9)\\ 
701.7 & 7087 (126)   & 0.0084 (0.0003) & 0.596 & 4634 & 303.1 (43.2)\\ 
\enddata
\end{deluxetable}

\clearpage
\begin{deluxetable}{cccccc}
\tablewidth{0pt}
\tablecaption{EPM Quantities Derived for SN 2003bl using the \{VI\} filter subset,
{\it DES} reddening and the {\it D05} atmosphere models. \label{tab_SN03bl_EPM}}
\tablehead {
\colhead{JD-}  & \colhead{$T_{VI}$} & \colhead{$\theta$$\zeta_{VI}$} &
\colhead{$\zeta_{VI}$} & \colhead{$v_{phot}$} & \colhead{$\theta/vel$} \\
\colhead{2452000}  & \colhead{$(K)$} & \colhead{$(10^{15}$ $cm$ $Mpc^{-1})$} &
\colhead{} & \colhead{$(km$ $s^{-1})$} & \colhead{$(100$ $s$ $Mpc^{-1})$} }
\startdata
701.8 & 13712 (812) & 0.0029 (0.0002) & 0.555 & 6452 & 80.5 (12.9) \\
702.8 & 12059 (367) & 0.0034 (0.0001) & 0.540 & 7040 & 89.4 (13.8) \\ 
729.8 & 6876 (181)  & 0.0076 (0.0003) & 0.610 & 3930 & 317.1 (45.3) \\ 
735.8 & 6738 (111)  & 0.0079 (0.0002) & 0.620 & 3364 & 376.2 (51.7) \\ 
\enddata
\end{deluxetable}

\clearpage
\begin{deluxetable}{cccccc}
\tablewidth{0pt}
\tablecaption{EPM Quantities Derived for SN 2003bn using the \{VI\} filter subset,
{\it DES} reddening and the {\it D05} atmosphere models. \label{tab_SN03bn_EPM}}
\tablehead {
\colhead{JD-}  & \colhead{$T_{VI}$} & \colhead{$\theta$$\zeta_{VI}$} &
\colhead{$\zeta_{VI}$} & \colhead{$v_{phot}$} & \colhead{$\theta/vel$} \\
\colhead{2452000}  & \colhead{$(K)$} & \colhead{$(10^{15}$ $cm$ $Mpc^{-1})$} &
\colhead{} & \colhead{$(km$ $s^{-1})$} & \colhead{$(100$ $s$ $Mpc^{-1})$} }
\startdata
706.6 & 11051 (280)  & 0.0089 (0.0003) & 0.535 & 9318 & 178.6 (27.6) \\
710.8 & 10306 (1028) & 0.0099 (0.0014) & 0.532 & 8684 & 213.1 (43.7)  \\ 
729.7 & 7780 (574)   & 0.0141 (0.0018) & 0.564 & 5819 & 429.3 (83.0)  \\ 
\enddata
\end{deluxetable}

\clearpage
\begin{deluxetable}{cccccc}
\tablewidth{0pt}
\tablecaption{EPM Quantities Derived for SN 2003ef using the \{VI\} filter subset,
{\it DES} reddening and the {\it D05} atmosphere models. \label{tab_SN03ef_EPM}}
\tablehead {
\colhead{JD-}  & \colhead{$T_{VI}$} & \colhead{$\theta$$\zeta_{VI}$} &
\colhead{$\zeta_{VI}$} & \colhead{$v_{phot}$} & \colhead{$\theta/vel$} \\
\colhead{2452000}  & \colhead{$(K)$} & \colhead{$(10^{15}$ $cm$ $Mpc^{-1})$} &
\colhead{} & \colhead{$(km$ $s^{-1})$} & \colhead{$(100$ $s$ $Mpc^{-1})$} }
\startdata
780.7 & 9986 (324) & 0.0144 (0.0006) & 0.531 & 7959 & 340.6 (53.7) \\
789.7 & 9275 (296) & 0.0152 (0.0006) & 0.534 & 6241 & 455.7 (71.7) \\ 
794.6 & 8902 (260) & 0.0158 (0.0007) & 0.538 & 5574 & 527.7 (82.8) \\ 
797.6 & 9126 (332) & 0.0153 (0.0008) & 0.535 & 5284 & 540.0 (86.2) \\ 
\enddata
\end{deluxetable}

\clearpage
\begin{deluxetable}{cccccc}
\tablewidth{0pt}
\tablecaption{EPM Quantities Derived for SN 2003hl using the \{VI\} filter subset,
{\it DES} reddening and the {\it D05} atmosphere models. \label{tab_SN03hl_EPM}}
\tablehead {
\colhead{JD-}  & \colhead{$T_{VI}$} & \colhead{$\theta$$\zeta_{VI}$} &
\colhead{$\zeta_{VI}$} & \colhead{$v_{phot}$} & \colhead{$\theta/vel$} \\
\colhead{2452000}  & \colhead{$(K)$} & \colhead{$(10^{15}$ $cm$ $Mpc^{-1})$} &
\colhead{} & \colhead{$(km$ $s^{-1})$} & \colhead{$(100$ $s$ $Mpc^{-1})$} }
\startdata
879.9 & 11543 (388) & 0.0184 (0.0007) & 0.541 & 8594 & 396.5 (61.1)  \\ 
900.8 & 8060 (106)  & 0.0298 (0.0006) & 0.556 & 5291 & 1013.3 (149.1) \\ 
908.7 & 7352 (81)   & 0.0338 (0.0006) & 0.583 & 4641 & 1247.4 (176.2) \\
\enddata
\end{deluxetable}

\clearpage
\begin{deluxetable}{cccccc}
\tablewidth{0pt}
\tablecaption{EPM Quantities Derived for SN 2003hn using the \{VI\} filter subset,
{\it DES} reddening and the {\it D05} atmosphere models. \label{tab_SN03hn_EPM}}
\tablehead {
\colhead{JD-}  & \colhead{$T_{VI}$} & \colhead{$\theta$$\zeta_{VI}$} &
\colhead{$\zeta_{VI}$} & \colhead{$v_{phot}$} & \colhead{$\theta/vel$} \\
\colhead{2452000}  & \colhead{$(K)$} & \colhead{$(10^{15}$ $cm$ $Mpc^{-1})$} &
\colhead{} & \colhead{$(km$ $s^{-1})$} & \colhead{$(100$ $s$ $Mpc^{-1})$} }
\startdata
878.2 & 10539 (1143) & 0.0405 (0.0054) & 0.537 & 9467 & 796.8 (159.8) \\
888.3 & 9364 (260)   & 0.0441 (0.0016) & 0.538 & 8044 & 1020.1 (157.3) \\ 
897.9 & 8013 (171)   & 0.0514 (0.0017) & 0.558 & 6632 & 1388.0 (206.1) \\ 
900.9 & 7756 (123)   & 0.0530 (0.0014) & 0.567 & 5884 & 1589.2 (230.9) \\ 
\enddata
\end{deluxetable}

\clearpage
\begin{deluxetable}{cccccc}
\tablewidth{0pt}
\tablecaption{EPM Quantities Derived for SN 2003iq using the \{VI\} filter subset,
{\it DES} reddening and the {\it D05} atmosphere models. \label{tab_SN03iq_EPM}}
\tablehead {
\colhead{JD-}  & \colhead{$T_{VI}$} & \colhead{$\theta$$\zeta_{VI}$} &
\colhead{$\zeta_{VI}$} & \colhead{$v_{phot}$} & \colhead{$\theta/vel$} \\
\colhead{2452000}  & \colhead{$(K)$} & \colhead{$(10^{15}$ $cm$ $Mpc^{-1})$} &
\colhead{} & \colhead{$(km$ $s^{-1})$} & \colhead{$(100$ $s$ $Mpc^{-1})$} }
\startdata
928.7 & 10464 (252.) & 0.0200 (0.0006) & 0.535 & 10629 & 351.8 (54.1) \\
930.7 & 10149 (216.) & 0.0207 (0.0006) & 0.534 & 9979 & 387.8 (59.5) \\ 
940.8 & 8837 (113.)  & 0.0243 (0.0005) & 0.541 & 8135 & 551.7 (82.9) \\ 
948.7 & 8046 (106.)  & 0.0273 (0.0005) & 0.557 & 6881 & 713.5 (104.7) \\ 
\enddata
\end{deluxetable}

\clearpage
\begin{deluxetable}{ccccc}
\tablewidth{0pt}
\tablecaption{Summary of $H_0$ values. \label{tab_H0_values}}
\tablehead {
\colhead{} & \colhead{} & \colhead{$\{BV\}$} & \colhead{$\{BVI\}$} & \colhead{$\{VI\}$} }
\startdata
{\it E96/OLI}§ && 98.4 (9.2) & 100.8 (8.0) & 89.1 (6.9) \\
{\it E96/DES} & & 97.2 (8.7) & 100.5 (8.4) & 91.2 (6.7) \\
{\it D05/OLI} & & 66.2 (4.2) & 60.4 (4.1)  & 53.9 (4.3) \\
{\it D05/DES} & & 63.8 (3.9) & 59.6 (4.2)  & 52.4 (4.3) \\
\enddata
\end{deluxetable}

\clearpage
\begin{deluxetable}{ccccc}
\tablewidth{0pt}
\tablecaption{Summary of dispersions in Hubble Diagrams. \label{tab_Dispersion_values}}
\tablehead {
\colhead{} & \colhead{} & \colhead{$\{BV\}$} & \colhead{$\{BVI\}$} & \colhead{$\{VI\}$} }
\startdata
{\it E96/OLI}§ && 0.53  & 0.43 & 0.34 \\
{\it E96/DES} & & 0.50  & 0.41 & 0.37 \\
{\it D05/OLI} & & 0.57  & 0.39 & 0.36 \\
{\it D05/DES} & & 0.51  & 0.37 & 0.32 \\
\enddata
\end{deluxetable}

%%\clearpage
%%\begin{sidewaystable}
%%\centering
%%\begin{tabular}{cccccccccccccc}
%%\tableline\tableline
%%\end{tabular}
%%\caption{EPM Quantities Derived for SN 1992ba.\label{tab_SN92ba_EPM}}
%%\end{sidewaystable}

\end{document}